\documentclass[11pt,a4paper]{article}
\pdfoutput=1
\usepackage{jheppub}

\usepackage{amsmath}
\usepackage{amssymb}
\usepackage{graphicx}
\usepackage{bbm}

\newcommand{\be}{\begin{equation}}  
\newcommand{\ee}{\end{equation}}  

\newcommand{\hc}{+\,\mathrm{h.c.}}
\newcommand{\vev}[1]{\langle #1 \rangle}

\newcommand{\into}{\ensuremath{\,\rightarrow\,}}
\newcommand{\tr}{\operatorname{tr}}

\newcommand{\wh}[1]{\ensuremath{\widehat{#1}}}

\def\softsusy{{\sc SoftSusy3.2.4}}

\def\micro{{\sc micrOMEGAs2.4}}
\def\hdecay{{\sc Hdecay4.4}}

\def\mgut{M_{\rm GUT}}
\def\BR{{\rm BR}}

\title{Anatomy of maximal stop mixing in the MSSM}

\author[a]{Felix Br\"ummer,}
\author[b]{Sabine Kraml,}
\author[b]{Suchita Kulkarni,}
\affiliation[a]{Deutsches Elektronen-Synchrotron DESY, Notkestra{\ss}e 85, D-22607 Hamburg, Germany}
\affiliation[b]{Laboratoire de Physique Subatomique et de Cosmologie, UJF Grenoble 1, 
CNRS/IN2P3, INPG, 53~Avenue des Martyrs, F-38026 Grenoble, France\\}
\emailAdd{felix.bruemmer@desy.de}
\emailAdd{sabine.kraml@lpsc.in2p3.fr}
\emailAdd{suchita.kulkarni@lpsc.in2p3.fr}
\preprint{DESY-12-064, LPSC12121}

\abstract{A Standard Model-like Higgs near 125 GeV in the MSSM requires multi-TeV stop masses, or a near-maximal contribution to its mass from stop mixing. We investigate the maximal mixing scenario, and in particular its prospects for being realized it in potentially realistic GUT models. We work out constraints on the possible GUT-scale soft terms, which we compare with what can be obtained from some well-known mechanisms of SUSY breaking mediation. Finally, we analyze two promising scenarios in detail, namely gaugino mediation and gravity mediation with non-universal Higgs masses. 
}


\begin{document}
\maketitle

\clearpage 
\section{Introduction}

Recent results from ATLAS~\cite{ATLAS:2012ae} and CMS~\cite{Chatrchyan:2012tx} show intriguing hints for a Higgs boson with a mass around $124\mbox{--}126$ GeV. While the reported excess of events is only at the level of about 2--3$\,\sigma$ per experiment, the consistency between the already excluded mass range, the excess in the remaining small window, and theoretical expectations provides a strong motivation to take these hints seriously and investigate their implications.\footnote{See also the appendix of~\cite{DAgostini:2011aa} in this context.}
 
If the cause of the observed excess were the lightest Higgs boson of the MSSM, it would be rather heavy, requiring large radiative corrections to its mass from the top-stop sector. In this case either the average stop mass must be large, $M_S\equiv\sqrt{m_{\tilde t_1}m_{\tilde t_2}} \gtrsim 3$ TeV, or the stop mixing parameter $|X_t|$ must be around twice $M_S$ \cite{Djouadi:2005gj}. The latter, known as the ``maximal mixing scenario,'' is the subject of this paper.

Several recent papers, including~\cite{Hall:2011aa,Heinemeyer:2011aa,Arbey:2011ab,Arbey:2011aa,Draper:2011aa,Carena:2011aa,Cao:2012fz,Christensen:2012ei,Arvanitaki:2011ck}, have explored the implications of a heavy MSSM Higgs from a bottom-up perspective, prescribing the MSSM soft parameters at the TeV scale.\footnote{For related work in non-minimal supersymmetric models, see also \cite{Ellwanger:2011aa, Arvanitaki:2011ck,King:2012is,Kang:2012tn,Boudjema:2012cq,Vasquez:2012hn}.}
In this approach, the soft terms can simply be chosen by hand to yield maximal mixing. However, one should keep in mind one of the key motivations for low-energy supersymmetry: The supersymmetric Standard Model can naturally be extrapolated up to a very high fundamental scale. Indeed, gauge coupling unification in the MSSM points to the GUT scale $\mgut\sim 10^{16}$ GeV as the scale where it should be embedded into a more fundamental theory. It is therefore worthwhile to investigate if 
maximal mixing can result from some reasonable choice of GUT-scale parameters, what further relations between the GUT-scale soft terms this would imply, and which classes of models of high-scale SUSY breaking mediation can (or cannot) accommodate maximal mixing. Furthermore, it is clearly of interest how the GUT-scale conditions for maximal mixing affect the physical spectrum and the low-scale observables, such as Higgs cross-sections and decay rates. These subjects are addressed in the present paper.

The implications of a 125~GeV Higgs for GUT-scale MSSM scenarios have been
investigated previously
in~\cite{Baer:2011ab,Kadastik:2011aa,Buchmueller:2011ab,Cao:2011sn,
Aparicio:2012iw,Ellis:2012aa,Baer:2012uy}.\footnote{For implications
for non-minimal SUSY models with GUT-scale boundary conditions, see
\cite{Moroi:2011aa,Gunion:2012zd,Rizzo:2012rf,Chang:2012gp,Ellwanger:2012ke}.}
It was observed \cite{Baer:2011ab} (see also \cite{Baer:2011sr}) that,
in the CMSSM and in NUHM models with sizeable $m_0$,
large  $|A_0|\approx 2\,m_0$ is preferred to obtain a heavy Higgs.
Our work goes beyond these studies by providing a thorough discussion
of the prerequisites for maximal mixing in more general models,
accompanied by a detailed numerical analysis. 

This work is organized as follows: In Section \ref{sec:maxmix} we give a brief review of the maximal mixing scenario, and explain why it is non-trivial to obtain maximal mixing when running from the GUT-scale. Using semi-numerical solutions of the renormalization group equations, we derive some necessary conditions for maximal mixing. In Section \ref{sec:models} we comment on the possibility of realizing maximal mixing in several well-established classes of models of SUSY breaking mediation, namely gaugino mediation, models with strongly-coupled near-conformal hidden sectors, radion mediation in 5D models, and gauge mediation. Section \ref{sec:numerics} contains a detailed numerical analysis of a gaugino-mediated and a simple gravity-mediated model. Moreover, we comment on the case of very heavy 1st/2nd generation sfermions.
Conclusions are contained in Section \ref{sec:conclusions}. In three appendices, we give more details about the method we use to solve the MSSM renormalization group equations, and comment on fine-tuning and on the danger of introducing charge- and color-breaking minima in the scalar potential.

\section{Maximal stop mixing and the Higgs mass}\label{sec:maxmix}

We follow the sign conventions of \cite{Martin:1997ns} for soft terms, in which the MSSM superpotential is
\be
  W=y_u\,UQH_u + y_d\, DH_dQ+y_e\,EH_dL+\mu\,H_uH_d\,, 
\ee
and the SUSY-breaking Lagrangian reads
\be\begin{split}\label{susybL}
  {\cal L}_{\rm soft} =
  &-\frac{1}{2}\sum_{a=1}^3M_a\,\lambda_a\lambda_a\hc-\left(A_u y_u\,\widetilde U\widetilde Q h_u
    +A_dy_d\,\widetilde D h_d\widetilde Q+A_ey_e\,\widetilde E h_d\widetilde L\right)\hc\\
  &-B_\mu\,h_u h_d\hc+\text{scalar soft mass terms}.
\end{split}
\ee
Factorizing the trilinear couplings into Yukawa matrices $y$ and soft coefficient matrices $A$, as in Eq.~\eqref{susybL}, is convenient for our purposes. In fact, for much of what we have to say, the only relevant trilinear soft terms are those for the third generation, in particular $A_t\equiv A_{u33}$. For simplicity we will mostly assume flavor-universal soft terms in the following, in which case $A_{u,d,e}=A_0\mathbbm{1}$ at the GUT scale.

In the decoupling limit $m_{A}\gg m_Z$, taking into account the one-loop corrections from the top-stop sector, the mass of the lightest MSSM Higgs is
\begin{alignat}{3}\label{mh0sq}
  m_{h}^2&\;=\quad m_{h}^{\rm 2,tree}&\;+\;&\Delta m_{h}^{2,\text{1-loop}}\nonumber \\
  &\;=\;m_Z^2\cos^2 2\beta&\;+\;&\frac{3}{4\pi^2}\frac{m_t^4}{v^2}
  \left(\log\frac{M_S^2}{m_t^2}+\frac{X_t^2}{M_S^2}\left(1-\frac{X_t^2}{12\,M_S^2}\right)\right)\,.
\end{alignat}
Here as usual $\tan\beta=v_u/v_d$, $v=\sqrt{v_u^2+v_d^2}=174$ GeV, $m_t$ is the running top mass at the scale $m_t$, and $M_S^2=m_{\tilde t_1}m_{\tilde t_2}$ with $m_{\tilde t_{1,2}}$ the stop masses. $X_t$ is the stop mixing parameter, defined at the scale $M_S$ as
\be\label{eq:xtdef}
  X_t=A_t-\mu\cot\beta\,.
\ee
The tree-level bound $m_{h}<m_Z$ quickly saturates for $\tan\beta\gtrsim 5$. To further lift $m_{h}^2$ from $m_Z^2=(91\text{ GeV})^2$ to around $(125\text{ GeV})^2$, radiative corrections nearly as large as the tree-level value are required. In Eq.~\eqref{mh0sq} the contribution from the logarithmic term can be increased by simply raising $M_S$, but naturalness demands that the soft mass scale should be not too far above the electroweak scale. The $X_t$ contribution is easily seen to be maximized at $|X_t/M_S|\simeq\sqrt{6}=2.45$, although some studies taking two-loop effects into account suggest a maximal mixing contribution closer to $|X_t/M_S|\approx 2$ (see e.g.~\cite{Carena:2000dp} for a detailed discussion). Values larger than about $\sqrt{6}$ will however induce dangerous charge- and color-breaking minima in the scalar potential, as detailed in Appendix \ref{sec:CCB}. For these reasons, in the present paper we will focus on the range
\be\label{maxmixdef}
  1.5\,<\,\left|\frac{X_t}{M_S}\right|\,<\,2.5
\ee
which we take to define \emph{maximal stop mixing}.

If maximal mixing is to be obtained from a GUT-scale model, this places non-trivial restrictions on the GUT-scale soft terms. The $A_t$ parameter generically changes drastically during renormalization group (RG) running, because it receives large radiative corrections from gluino loops. At one loop, the dominant terms in its RG equation are
\be
  \dot A_t= \frac{3}{4\pi^2}\,|y_t|^2\,A_t+\frac{2}{3\pi^2}g_3^2\,M_3+\ldots
\ee
Sizeable gluino masses, which are now favored in the light of direct LHC search bounds, will drive $A_t$ towards large negative values at the electroweak scale (in a phase convention where $M_3$ is positive). 
The soft-breaking masses  $m^2_{Q_3}$ and $m^2_{U_3}$ entering $M_S$ 
will also receive large radiative corrections, because they carry color and because of the large top Yukawa coupling. 
The corresponding one-loop RGEs are
\begin{align}
\dot m^2_{Q_3}=&\;\frac{3}{8\pi^2}|y_t|^2\left(m_{H_u}^2+m_{Q_3}^2+m_{U_3}^2+|A_t|^2\right)-\frac{2}{3\pi^2} g_3^2|M_3|^2-\frac{3}{8\pi^2} g_2^2|M_2|^2\nonumber\\
&+\frac{3}{8\pi^2}|y_b|^2\left(m_{H_d}^2+m_{Q_3}^2+m_{D_3}^2+|A_b|^2\right)+\ldots\,,\\
\dot m^2_{U_3}=&\;\frac{3}{4\pi^2}|y_t|^2\left(m_{H_u}^2+m_{Q_3}^2+m_{U_3}^2+|A_t|^2\right)-\frac{2}{3\pi^2} g_3^2|M_3|^2+\ldots
\end{align}
It is evident that a sizeable $M_3$ will also have a large effect on the scalar soft masses.

To better quantify the effects of RG running, let us assume universal GUT-scale gaugino masses $M_{1/2}$ 
as predicted by many GUT models, and universal trilinears $A_0$. 
The most relevant parameters are actually $M_3$, $M_2$ and $A_t$, so
we are effectively imposing $M_2(\mgut)=M_3(\mgut)$.\footnote{For simplicity of notation, in the following we will use hatted symbols such as $\wh m_{H_u}^2$, $\wh \mu$,  etc.~to denote GUT-scale boundary values of some of the running MSSM parameters. The exceptions are $M_{1/2}\equiv\wh M_1=\wh M_2=\wh M_3$ and $A_0\equiv\wh A_t=\wh A_b=\wh A_\tau$ by definition and, where appropriate, a universal scalar mass $m_0\equiv m_0(\mgut)$.}
We can solve the two-loop RGEs \cite{Martin:1993zk} semi-numerically and express the electroweak-scale mixing parameter as a function of the GUT-scale soft terms (see Appendix \ref{sec:RGEs}). For $\mgut=2\cdot 10^{16}$ GeV, $M_S=1$ TeV and $\tan\beta=20$ we obtain
\be\begin{split}\label{xt4}
X_t^4\approx\; & 9.4\,M_{1/2}^4- 7.5\,A_0\,M_{1/2}^3+ 2.2\, A_0^2\,M_{1/2}^2 -0.3\, A_0^3\,M_{1/2} 
\\& + 1.1\, M_{1/2}^3\,\wh \mu- 0.7\, A_0\,M_{1/2}^2\,\wh \mu\,.
\end{split}
\ee
Terms with coefficients $<0.2$ have been suppressed. Similarly, we have
\be\begin{split}\label{mqmu}
M_S^4=\left.m_{U_3}^2m_{Q_3}^2\right|_{M_S}\approx &\;8.7\, M_{1/2}^4  + 2.5\, M_{1/2}^2\,\wh m_{U_3}^2+ 1.7\, M_{1/2}^2\,\wh m_{Q_3}^2+ 1.2\, A_0\,M_{1/2}^3\\
&-0.4\, A_0^2\,M_{1/2}^2  - 0.9\, M_{1/2}^2\,\wh m_{H_{u}}^2  
+0.8\, \wh m_{U_3}^2\,
\wh m_{Q_3}^2\,.\end{split}
\ee
The coefficients in these equations vary at most by about $10\%$ in the range $5<\tan\beta<40$. The only exception are the coefficients involving $\wh \mu$ in Eq.~\eqref{xt4}, which grow as $1/\tan\beta$ for smaller $\tan\beta$ as is evident from Eq.~\eqref{eq:xtdef}. Explicitly, for the same parameters but with $\tan\beta=5$ we obtain
\be\begin{split}\label{xt4_smalltb}
X_t^4\approx\; & 9.2\,M_{1/2}^4- 7.2\,A_0\,M_{1/2}^3+ 2.1\, A_0^2\,M_{1/2}^2 -0.3\, A_0^3\,M_{1/2} 
\\& + 4.3\, M_{1/2}^3\,\wh \mu- 2.5\, A_0\,M_{1/2}^2\,\wh \mu+ 
 0.5\,A_0^2\,M_{1/2}\,\wh \mu +0.8\,M_{1/2}^2\,|\wh \mu|^2+0.3\,A_0\,M_{1/2}\,|\wh\mu|^2\,,
\end{split}
\ee
and
\be\begin{split}\label{mqmu_smalltb}
M_S^4\approx &\;8.7\, M_{1/2}^4  + 2.5\, M_{1/2}^2\,\wh m_{U_3}^2+ 1.6\, M_{1/2}^2\,\wh m_{Q_3}^2+ 1.1\, A_0\, M_{1/2}^3\\
&-0.4\, A_0^2\,M_{1/2}^2  - 1.0\, M_{1/2}^2\,\wh m_{H_{u}}^2  
+0.8\, \wh m_{U_3}^2\,
\wh m_{Q_3}^2\,.
\end{split}
\ee

From these expressions one can read off some conditions on the GUT-scale soft terms under which maximal mixing results. We will make the extra assumptions that no soft term is hierarchically larger than the gaugino mass (such that all terms which we neglected in Eqns.~\eqref{xt4}--\eqref{mqmu_smalltb} are indeed subdominant)\footnote{Very large scalar soft masses for the first two generations may be an interesting and natural alternative scenario for maximal mixing, see e.g.~\cite{Baer:2012uy2,SPtalk,Badziak:2012xx}
and our Section \ref{sec:esusy}.}
 and that all scalar soft masses-squared except possibly $\wh m_{H_{u}}^2$ and $\wh m_{H_{d}}^2$ are positive.\footnote{The relatively large coefficients of the $M_{1/2}^2 \wh m_{U_3}^2$ and $M_{1/2}^2 \wh m_{Q_3}^2$ terms in Eqns.~\eqref{mqmu} and \eqref{mqmu_smalltb} have led the authors of \cite{Dermisek:2006ey} to propose tachyonic GUT-scale soft masses for the scalars of the third generation. This would render $M_S$ small and thus allow for a sizeable $|X_t/M_S|$ ratio. In this paper however we prefer to restrict ourselves to models where sfermion masses are positive at all scales, to avoid possible
complications from introducing additional vacua to the scalar potential.} It is convenient to classify all possibilities by the largest GUT-scale soft parameter among those appearing in Eqns.~\eqref{xt4}--\eqref{mqmu_smalltb}.

\begin{figure}\centering
   \includegraphics[width=90mm]{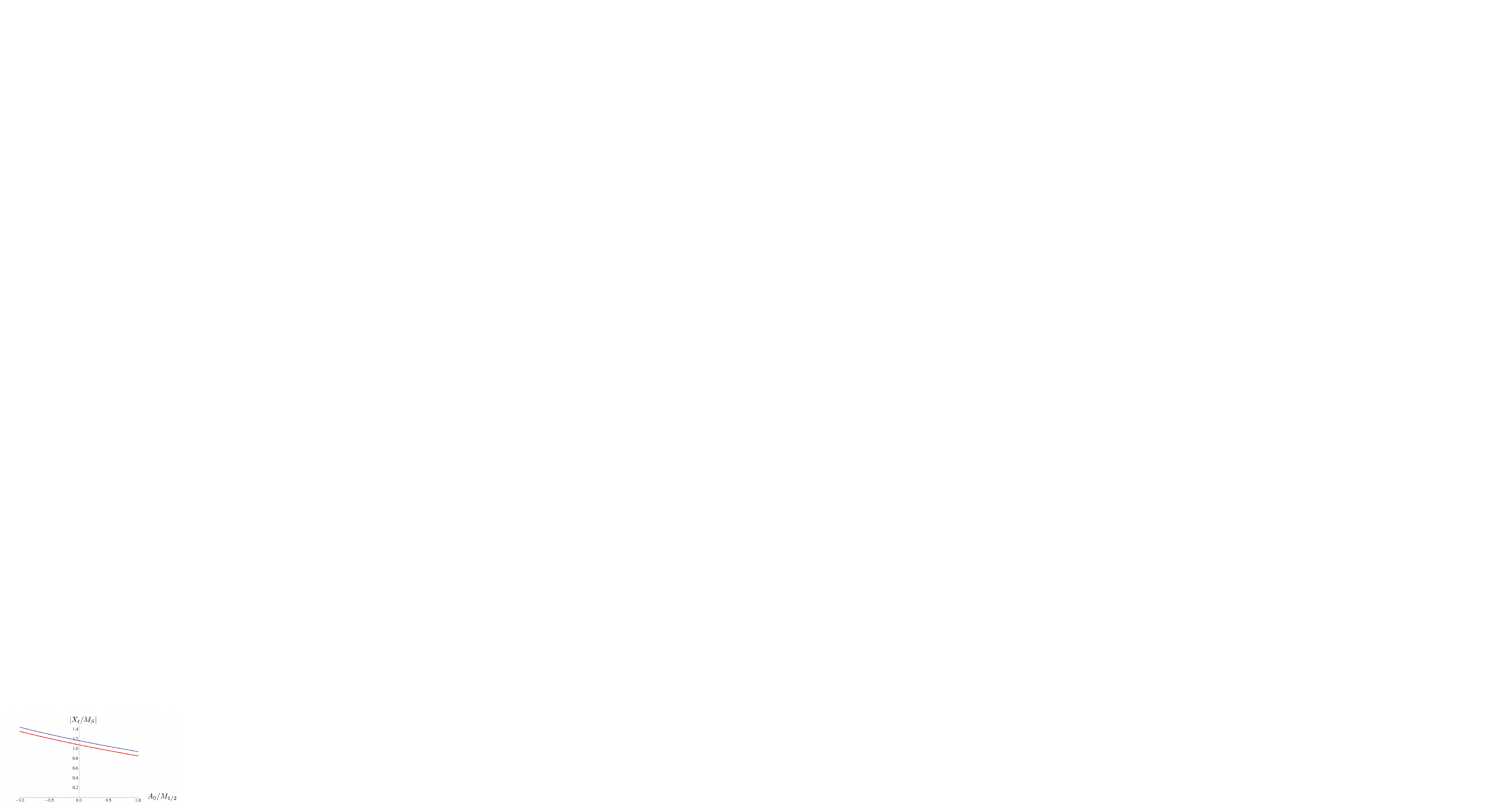}
\caption{\label{fig:mixM12} $|X_t/M_S|$ as a function of $A_0/M_{1/2}$, with $M_{1/2}$ the largest GUT-scale soft parameter. The upper, violet curve is for $\tan\beta=5$, while the lower, red curve is for $\tan\beta=20$. The other soft masses are $\widehat m_{H_u}^2=\widehat m_{H_d}^2=|\widehat \mu|^2=M_{1/2}^2$ and $\wh m_{Q_3}^2=\wh m_{U_3}^2=\wh m_{D_3}^2=0$, in order to maximize $X_t$ in Eq.~\eqref{xt4} while minimizing $M_S$ in Eq.~\eqref{mqmu}.  Even with this optimal choice of parameters, there is no maximal mixing.}
\end{figure}

\begin{figure}\centering
   \includegraphics[width=60mm]{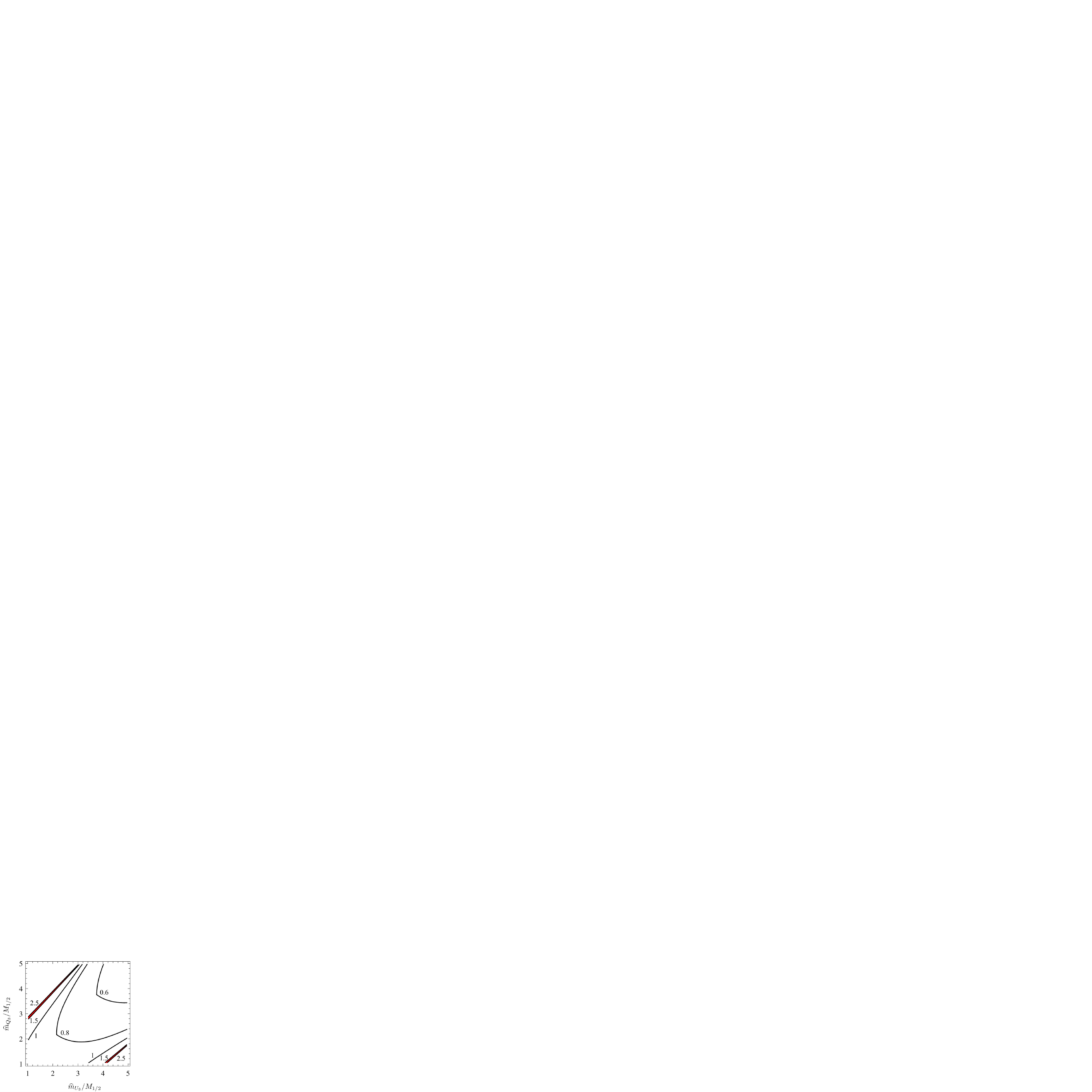} $\qquad$
   \includegraphics[width=60mm]{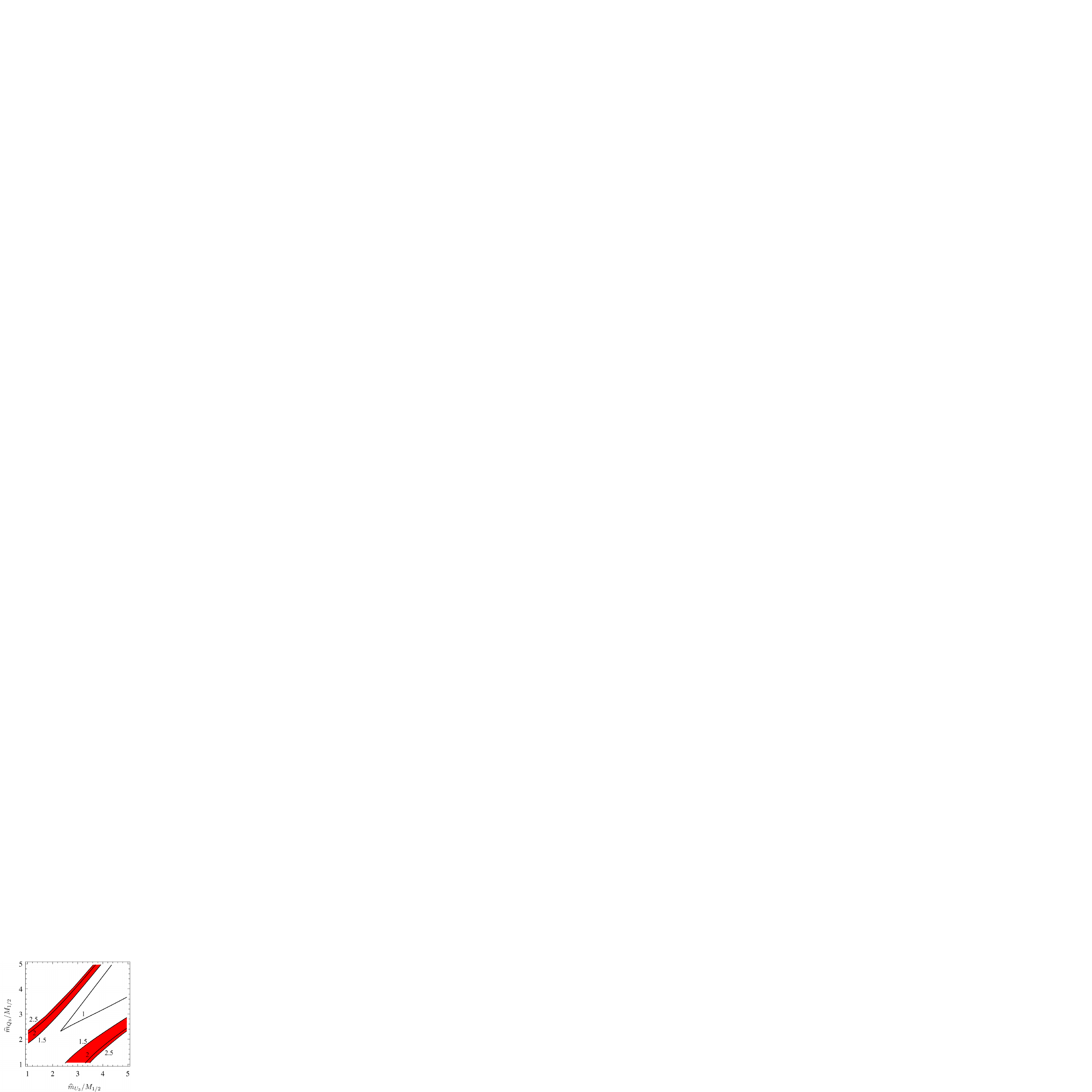}\\[5mm]
   \includegraphics[width=60mm]{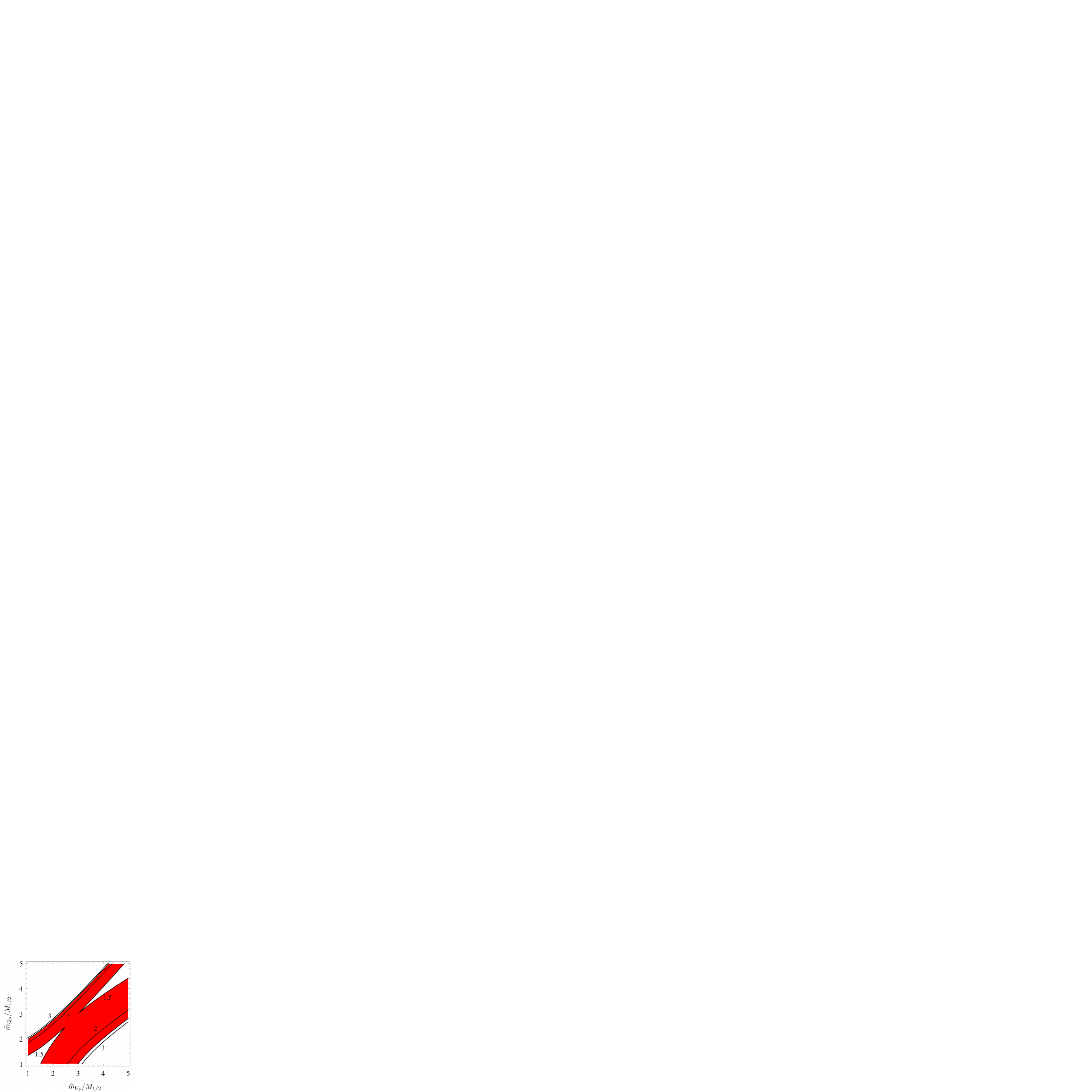} $\qquad$
   \includegraphics[width=60mm]{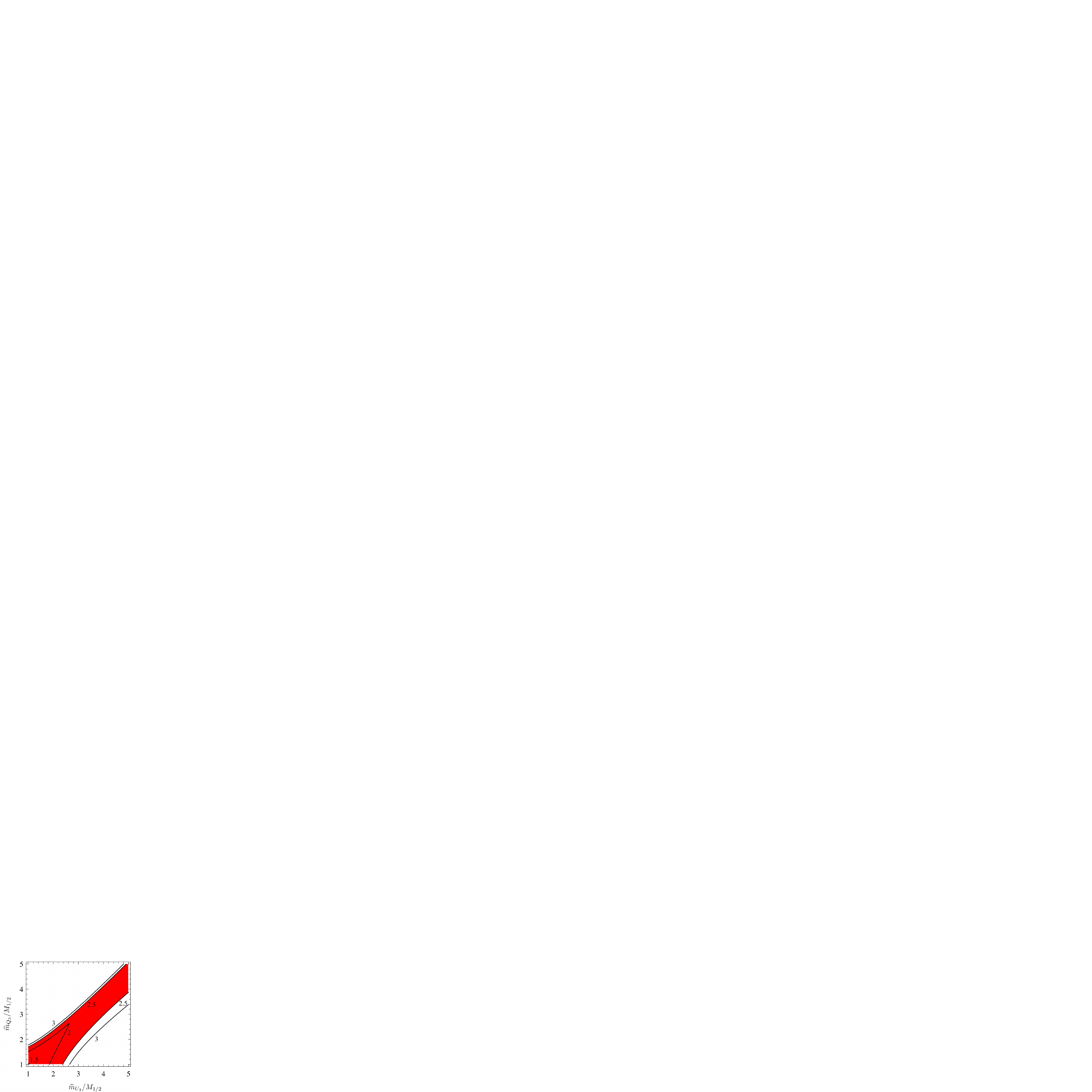}
\caption{Contours of $|X_t/M_S|$ as a function of $\wh m_{Q_3}/M_{1/2}$ (vertical axes) and of $\wh m_{U_3}/M_{1/2}$ (horizontal axes), for $\tan\beta=20$ and for $A_0=(0,\,-1,\,-1.5,\,-2)\times\max(\wh m_U,\wh m_Q)$ 
in the top left, top right, bottom left, bottom right panels, respectively. 
The other soft masses are $\wh\mu=\wh m_{H_u}=\wh m_{H_d}=\max(\wh m_{U_3},\wh m_{Q_3})$. 
Maximal mixing occurs in the red areas, which shrink to nearly zero as $|A_0|$ is reduced (cf.\ top left plot).
\label{fig:mixUQ} }
\end{figure}

\begin{figure}\centering
  \includegraphics[width=60mm]{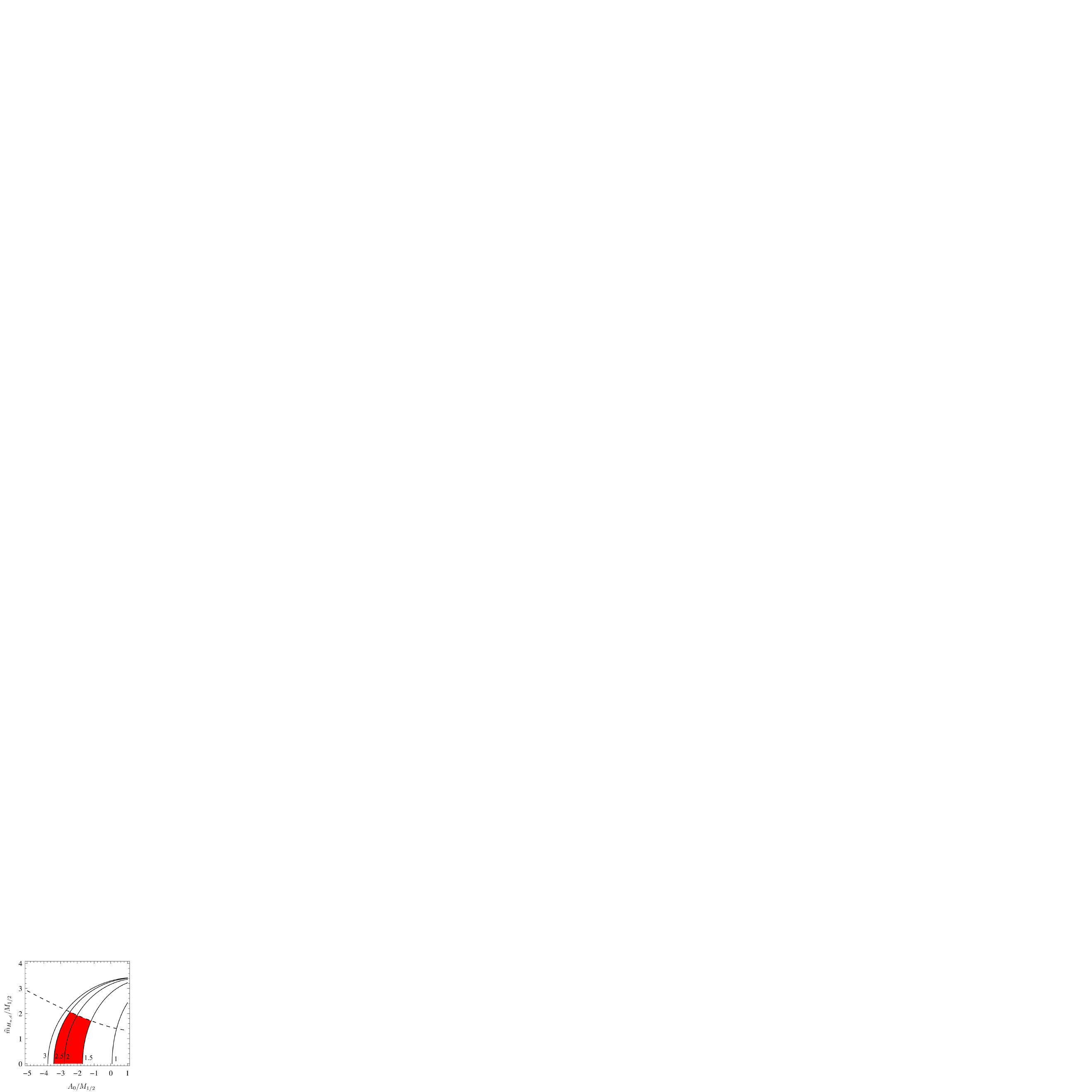}$\qquad$
  \includegraphics[width=60mm]{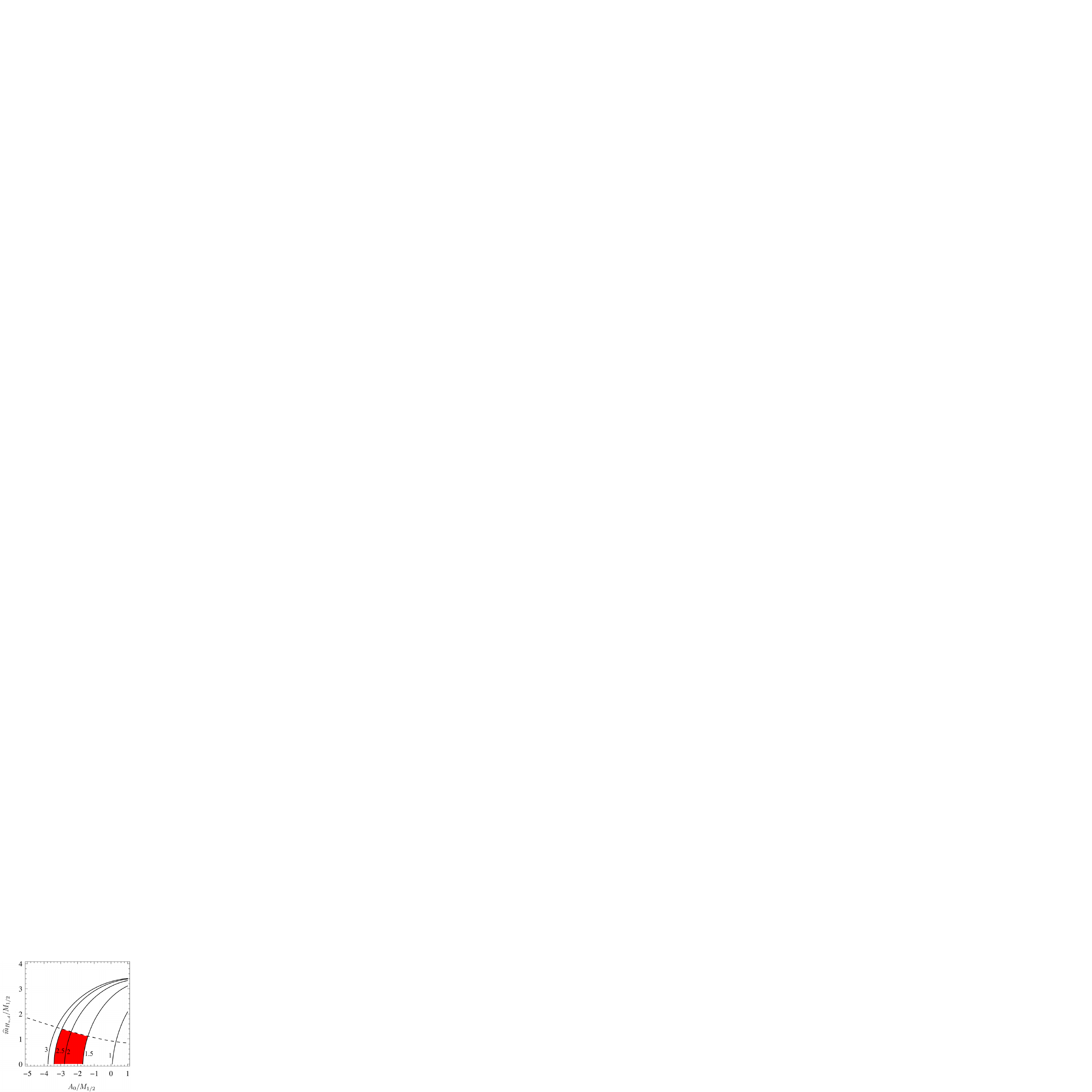}
\caption{Contours of $|X_t/M_S|$ as a function of $A_0/M_{1/2}$ and of $\wh m_{H_u}/M_{1/2}$, with $\wh m_{H_d}=\wh m_{H_u}$ for simplicity and with negligible $\wh m_U^2$ and $\wh m_Q^2$. Left panel: $\wh\mu=0$, right panel: $\wh\mu=\wh m_{H_u}$. The dashed line indicates the border of the electroweak symmetry breaking region according to Eq.~\eqref{mzsq}. In the red region electroweak symmetry is broken and stop mixing is maximal.\label{fig:mixH} }
\end{figure}

\begin{itemize}
\item If the largest GUT-scale soft parameter is $M_{1/2}$, then maximal mixing is excluded. Specifically, for $A_0> -M_{1/2}$ and ($\wh m_{H_{u}},\,\wh m_{H_{d}},\,\wh m_{Q_3},\,\wh m_{U_3},\,\wh \mu)<M_{1/2}$, we find
$\left|X_t/M_S\right|< 1.4$, as is depicted in Fig.~\ref{fig:mixM12}.
\item If the sfermion masses are universal, specifically $\wh m_{Q_3}^2=\wh m_{U_3}^2\equiv m_0^2$, then maximal mixing does not allow for $m_0$ to be the largest soft parameter. However, maximal mixing is possible if either $\wh m_{Q_3}^2$ or $\wh m_{U_3}^2$ is large and the other is small. In that case a necessary condition on the spectrum is
\be
  \wh m_{Q_3(U_3)}\gtrsim 2\,M_{1/2}
\ee
and sizeable negative $A_0$ is strongly preferred (even though there remains a tiny slice of parameter space where $A_0$ can be zero, if all other parameters are chosen optimally). When deviating from the optimal case of sizeable $\wh m_{H_{u}}$ and negligible $\wh m_{U_3(Q_3)}$, the required $\wh m_{Q_3(U_3)}:M_{1/2}$ ratio can grow very large. This is illustrated in the top row of Fig.~\ref{fig:mixUQ}.
\item $|A_0|$ can easily be the largest soft parameter if $A_0$ is negative. In the limit that all other soft terms are negligible, we find that values of
\be
  A_0\approx -(1\mbox{--}\,3)\,\max\left(M_{1/2},\wh m_{Q_3},\wh m_{U_3}\right)
\ee
are generic in situations with maximal mixing, particularly if the scalar masses are unified or negligible. This is easily understood in the light of the large coefficients of the $A_0$ terms in Eqns.~\eqref{xt4} and \eqref{mqmu}. It is illustrated in Fig.~\ref{fig:mixH} for the case of negligible $\wh m_{U_3}^2$ and $\wh m_{Q_3}^2$, and in the bottom row of Fig.~\ref{fig:mixUQ} for the case of dominant sfermion masses $\wh m_{Q_3,U_3}>M_{1/2}$.
\item If $\wh m_{H_{u}}^2$ is positive, and $\wh m_{H_{u}}$ is the largest GUT-scale soft parameter, then maximal mixing seems possible, at first sight, even without significant $A_0$ contributions as can be seen from Fig.~\ref{fig:mixH}. This is because of the negative-sign $M_{1/2}^2\wh m_{H_{u}}^2$ contribution in Eq.~\eqref{mqmu}. However, maximal mixing in this case requires a moderate hierarchy,
\be
  \wh m_{H_{u}}\gtrsim 3\,M_{1/2}\,,
\ee
which becomes more pronounced if $\wh m_{Q_3}$ and $\wh m_{U_3}$ are non-negligible or if $A_0$ is positive, and weaker if $A_0$ is negative. Such a hierarchy is in conflict with electroweak symmetry breaking, as can be understood from the equivalent formula for the electroweak symmetry breaking (EWSB) order parameter $m_Z$ (here quoted for $\tan\beta=20$):
\be\begin{split}\label{mzsq}
m_Z^2&\approx-2\left.\left(|\mu|^2+m_{H_u}^2\right)\right|_{M_S}\\
&=0.2\,A_0^2 - 0.7\,A_0\,M_{1/2} + 2.9\,M_{1/2}^2 - 2.1\,|\wh\mu|^2 - 1.3\,\wh m_{H_{u}}^2 + 0.7\, \wh m_{Q_3}^2+ 0.8\,\wh m_{U_3}^2
\end{split}
\ee
In RGE language, the gaugino masses (specifically the gluino mass) are typically responsible for driving $m_{H_u}^2$ negative at the electroweak scale, thus triggering electroweak symmetry breaking. If the GUT-scale value for $m_{H_u}^2$ is too large, and $M_{1/2}$ is too small, electroweak symmetry is not broken, because the RHS of Eq.~\eqref{mzsq} remains negative. Sizeable $\wh m_{Q_3}$ or $\wh m_{U_3}$, or sizeable negative $A_0$ can remedy this, but only the latter is favorable for maximal mixing. 
\end{itemize}

\begin{figure}\centering
  \includegraphics[width=60mm]{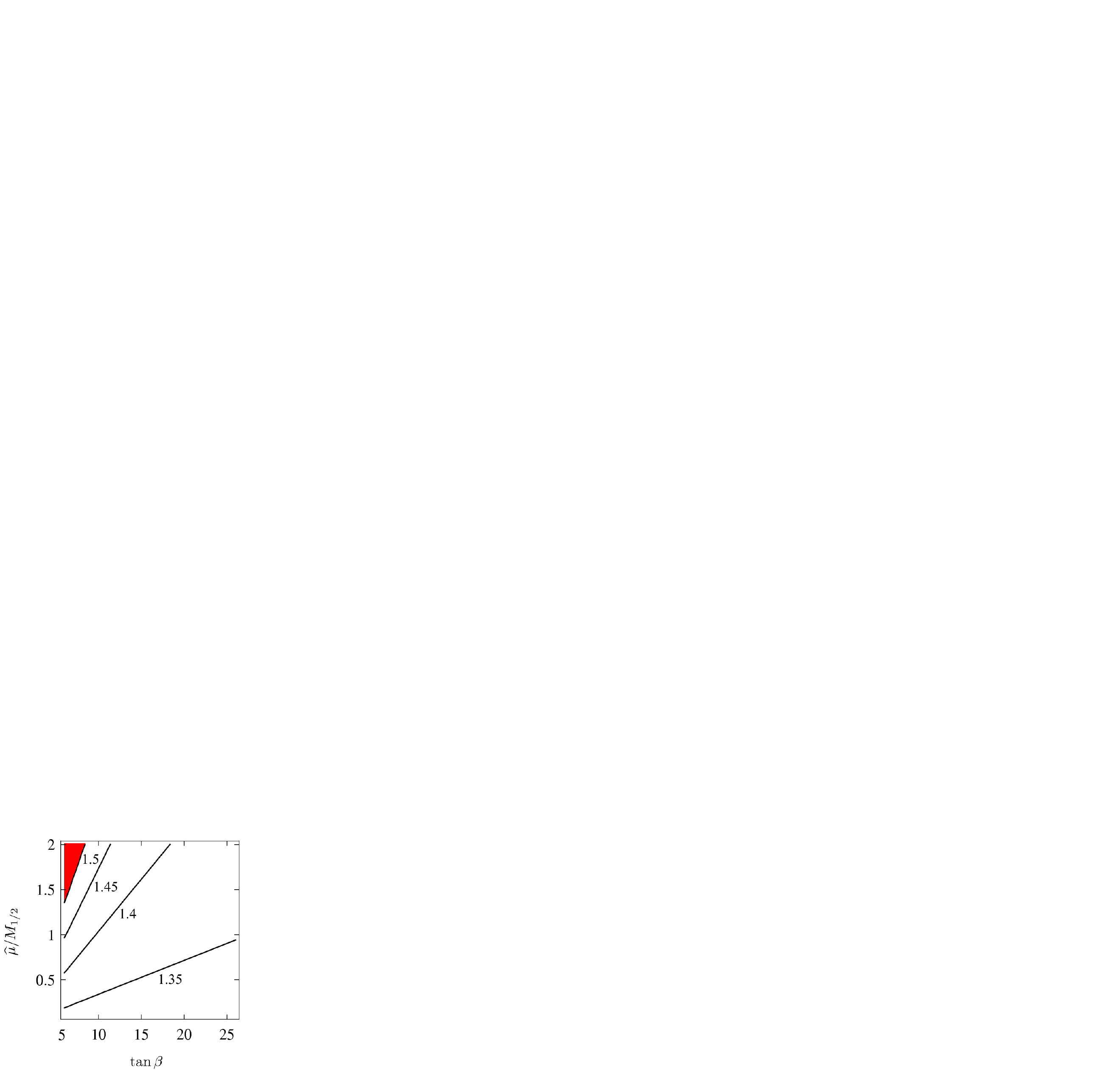}$\qquad$
  \includegraphics[width=60mm]{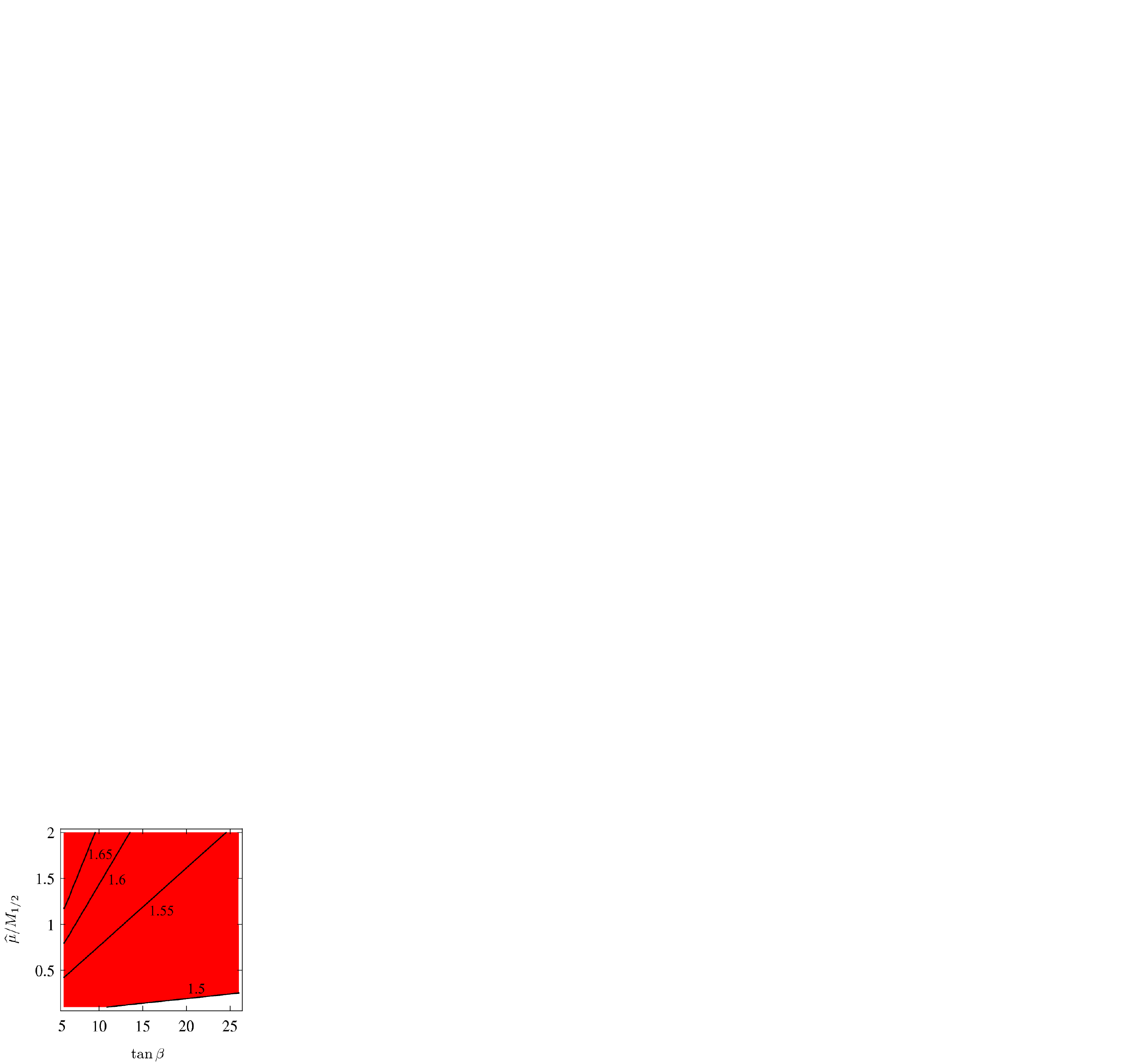}
\caption{Contours of $|X_t/M_S|$ as a function of $\wh\mu/M_{1/2}$ and $\tan\beta$, for $\wh m_{H_u}=\wh m_{H_d}=M_{1/2}$ and $\wh m_{Q_3}=\wh m_{U_3}=0$. Left panel: $A_0/M_{1/2}=-1$, right panel: $A_0/M_{1/2}=-3/2$. Clearly, at small $\tan\beta$, large $\mu$ can lead to maximal mixing in regions which are otherwise at the borderline. \label{fig:mixmutb} }
\end{figure}

Figs.~\ref{fig:mixUQ} and \ref{fig:mixH} are based on $\tan\beta=20$, cf.~Eqns.~\eqref{xt4} and \eqref{mqmu}. The situation remains the same qualitatively also for smaller $\tan\beta$. For $\tan\beta\approx 5$, relatively large $\wh\mu$ can however slightly widen the allowed regions for maximal mixing, since $X_t=A_t-\mu/\tan\beta$ has a stronger $\mu$-dependence if $\tan\beta$ is small. The effect of $\wh\mu$ on maximal mixing at smaller $\tan\beta$ is also illustrated in Fig.~\ref{fig:mixmutb}. We do not consider values of $\tan\beta<5$, since they no longer maximize the tree-level Higgs mass Eq.~\eqref{mh0sq}.

To summarize, generically, maximal mixing in a GUT-scale model with unified gaugino masses at $\tan\beta\gtrsim 5$ requires 
\begin{center}\emph{large negative $A_0\approx -(1\mbox{--}\,3)\,\max\left(M_{1/2},\wh m_{Q_3},\wh m_{U_3}\right)$}. \end{center}
This is a non-trivial requirement on any GUT-scale model. In addition, it is beneficial but not strictly necessary for maximal mixing if there are \begin{center}\emph{positive up-type Higgs soft masses $\wh m_{H_{u}}^2$,}\end{center}
and 
\begin{center}
\emph{small third-generation soft masses $\wh m_{U_3}^2$, $\wh m_{Q_3}^2$},
\end{center}
as compared to the gaugino mass.

The resulting $X_t/M_S$ at the electroweak scale will necessarily be negative. All this we have deduced from the semi-numerical solution of the RGEs underlying Eqns.~\eqref{xt4}, \eqref{mqmu}, \eqref{xt4_smalltb}, and \eqref{mqmu_smalltb}; see Appendix \ref{sec:RGEs} for more details. The picture is confirmed by parameter space scans performed with \softsusy, whose results are shown in Section \ref{sec:numerics}. 

\clearpage
\section{Models}\label{sec:models}

We assume $F$-term SUSY breaking in some hidden sector, mediated to the visible sector
by messenger states which can be supersymmetrically integrated out at some high scale $M$. If $X=F\theta^2$ is the Goldstino background field which parametrizes SUSY breaking, the lowest-order operator inducing a gaugino mass takes the form
\be\label{L12}
{\cal L}_{M_{1/2}}=c_{1/2}\,\int d^2\theta\frac{X}{M}\tr W^\alpha W_\alpha\hc
\ee
As before we assume gaugino mass unification, i.e.~a GUT-preserving $F$-term VEV.

If the hidden sector couples to the Higgs, we can have the Giudice-Masiero operator
\be\label{Lmu}
{\cal L}_\mu=c_{\mu}\,\int d^4\theta\,\frac{X^\dag}{M} H_u H_d\hc
\ee
which induces a $\mu$ term, and the operators
\be\label{LmHsq}
{\cal L}_{m_H^2}=\int d^4\theta \frac{X^\dag X}{M^2}\left(c_{H_u}\,|H_u|^2+c_{H_d}\,|H_d|^2\right)
\ee
which contribute to the Higgs soft masses. The operators
\be\label{LA}
{\cal L}_A=\int d^4\theta \frac{X}{M}\left(c_{A_u}\,|H_u|^2+c_{A_d}\,|H_d|^2\right)\hc
\ee
are often neglected because they can be absorbed by a holomorphic field redefinition $H_{u,d}\into H_{u,d}\left(1+c_{A_{u,d}}X/M\right)$. In order to cleanly separate hidden and visible sectors, we instead keep these operators explicit, noticing that they also induce soft Higgs masses, and (together with the Yukawa couplings) flavor-universal trilinear $A$-terms, and (together with ${\cal L}_{\mu}$) a $B_\mu$ term. If $c_{A_u}$ and $c_{A_d}$ were set to zero by the above field redefinition, these soft terms would instead arise from the induced superpotential terms and the change in the $c_{H_{u,d}}$ coefficients in Eq.~\eqref{LmHsq}. Finally, there are the operators 
\be\label{LBmu}
{\cal L}_{B_\mu}=c_{B_\mu}\,\int d^4\theta \frac{X^\dag X}{M^2}H_u H_d\hc
\ee
which also induce a $B_\mu$ term. 

If the hidden sector couples to matter fields, the equivalents of Eq.~\eqref{LmHsq} and \eqref{LA} with Higgs fields replaced by matter fields can be present.They will induce soft masses and flavor non-universal $A$-terms. The latter can also arise from superpotential operators of the form
\be
\int d^2\theta\frac{X}{M}\left(QH_uU+Q H_d D+L H_d E\right)\hc
\ee
We will assume that the $\mu$/$B_\mu$ problem is solved; in particular there are no unacceptably large contributions to $\mu$ and $B_\mu$ as would arise from the renormalizable terms
\be
\int d^2\theta\left(\tilde\mu H_u H_d+X H_u H_d\right)\hc
\ee

From the discussion in Section \ref{sec:maxmix} it is clear that, in order to realize maximal mixing, models with vanishing or suppressed direct couplings to matter fields are preferred. The operators of Eqns.~\eqref{L12}--\eqref{LBmu} are then all that is needed to fix the high-scale soft masses. Furthermore, the model should allow for $c_{1/2}$ and $c_{A_u}$ to be chosen such that the resulting $A_t\;:\:M_{1/2}$ ratio is around $-(1\mbox{--}\,3)$. SUSY breaking scenarios with suppressed couplings of the hidden sector to matter fields are not generic, because the scalar masses cannot easily be forbidden by symmetry (by contrast, suppressed gaugino masses are easily obtained but phenomenologically undesirable). We will list a number of well-known examples, and comment on the implications of our study.

\subsection{Gaugino(-Higgs) mediation}

Gaugino-mediated supersymmetry breaking in its minimal form
is defined by the gaugino masses being the only non-vanishing terms at the mediation scale. In other words, only the operators of Eq.~\eqref{L12} are present at the scale $M$; the only MSSM fields to couple directly to the hidden sector are the gauginos. This was originally motivated by 5D models \cite{Kaplan:1999ac, Chacko:1999mi, Schmaltz:2000gy} in which the 
hidden sector and the chiral (Higgs and matter) superfields of the MSSM were separated in an extra dimension, with only the MSSM gauge fields coupling to both. Deconstructed models \cite{Csaki:2001em,Cheng:2001an}, models with Seiberg duality \cite{Green:2010ww}, and models with strongly coupled near-conformal hidden sectors (see below) may also give rise to gaugino mediation.

Independently of maximal mixing, the minimal scenario is not viable phenomenologically because it is missing a $\mu$ term. Realistic extensions therefore require that also the Higgs fields should be directly coupled to the hidden sector. In the 5D picture, the Higgs and gauge fields would be bulk fields, while the MSSM matter fields would be localized on one brane and the hidden sector on the other. Generically, all operators in Eqns.~\eqref{L12}-\eqref{LBmu}
are then present at the scale $M$, and maximal mixing can be naturally accommodated.
In fact, all three conditions listed at the end of Section \ref{sec:maxmix} are satisfied if $c_{A_u}\simeq -2\, c_{1/2}$.

It should be mentioned that the mediation scale $M$ can be parametrically lower than $M_{\rm GUT}$ in gaugino-mediated models. For significantly lower $M$, running effects become less important, and the condition for maximal mixing eventually approaches Eq.~\eqref{maxmixdef} with the boundary values of $A_t$ and $\mu$ at the scale $M$ substituted, and with $m_{Q_3}$ and $m_{U_3}$ well approximated at leading-log order.

\subsection{Scalar sequestering}\label{HSD}

Scalar sequestering \cite{Roy:2007nz,Murayama:2007ge} is a more restrictive version of gaugino-Higgs mediation. The hidden sector is assumed to be close to a strongly-coupled conformal fixed point over a large range of energies, with $X$ a composite operator satisfying $\Delta_{X^\dag X}-2\Delta_X>0$. This implies that \cite{Murayama:2007ge,Perez:2008ng}\footnote{In \cite{Craig:2009rk} it was shown that, strictly speaking, these relations do not hold generically but depend on additional assumptions about the hidden sector.}, at the scale $M$ where the theory exits the strong coupling regime,
\be \label{lod}
c_{1/2}\,\sim\,c_{A_{u,d}}\,\sim\,c_\mu\;\gg\;(c_{H_{u,d}}-|c_{A_{u,d}}|^2-|c_\mu|^2)\,\sim\, (c_{B_\mu}-c_\mu(c_{A_u}+c_{A_d}))
\ee 
The operators responsible for matter soft masses also end up being suppressed. Eventually the dominant soft
terms at the scale $M$ are $M_{1/2}$, $A_0$, $\wh m_{H_{u,d}}^2$ and $\wh \mu$, which are all of the same order, while the combination
\be \label{dhmm}
  \wh m_{H_{u,d}}^2+\wh \mu^2\,\sim\,\wh B_\mu
\ee
and the matter soft terms are suppressed.

Our analysis is applicable for $M$ close to $M_{\rm GUT}$. The phenomenological consequences of the condition 
Eq.~\eqref{dhmm} were discussed in detail in \cite{Brummer:2010gh}. It was shown that 
for $\mu>0$, $\wh B_\mu$ almost always has the same sign as $\wh A_t$ 
(and the opposite one if $\mu<0$). Moreover, small or vanishing $\wh B_\mu$ 
requires also small $\wh A_t$,  in apparent conflict with maximal mixing, cf.\ Fig.~3 in \cite{Brummer:2010gh}.

\subsection{Radion mediation / Scherk-Schwarz SUSY breaking}

In 5D models, supersymmetry can be broken by giving an $F$-term to the radion multiplet. This multiplet hosts the 5D gravitational degree of freedom which corresponds to the compactification radius. Radion mediation is equivalent to breaking SUSY by the Scherk-Schwarz mechanism, and generalizes to modulus-mediated SUSY breaking in superstring models. Of particular interest are models whose compactification scale is close to $M_{\rm GUT}$, since in that case boundary conditions can be used to break the grand-unified symmetry down to the MSSM. Minimal models of this kind give rise to specific GUT-scale soft term patterns, so it is natural to ask if they can also accommodate maximal mixing.

The role of the operator $X$ in Eqns.~\eqref{L12} to \eqref{LBmu} is then played by $X=T\,M/(2 R)$, where $M$ is identified with the 5D Planck mass, and $T$ is the radion multiplet with $\vev T=R+F^T\theta^2$. For the simplest model with a flat $S^1/\mathbb{Z}_2$ extra dimension, the gaugino masses are (see e.g.~\cite{Marti:2001iw})
\be
  M_{1/2}=\frac{F^T}{2R}\,,
\ee
and the $A$-terms depend on the localizations of the matter fields and on the origin of the Higgs field. They are given by the sum of the contributions from the Higgs and the matter fields involved in the respective trilinear coupling. Roughly speaking, in gauge-Higgs unified models where the Higgs originates from the 5D gauge multiplet, the Higgs contributes
\be
  \Delta\wh A_t=-\frac{F^T}{2R}
\ee
but bulk matter fields $Q_3$ and $U_3$ originating from 5D hypermultiplets each contribute
\be
  \Delta\wh  A_t=+\frac{F^T}{2R}\,
\ee
leading to $\wh A_t=+M_{1/2}$ in the most naive model with unlocalized $Q_3$ and $U_3$. Localizing
$Q_3$ and $U_3$ towards one of the branes brane allows to reduce their contribution, but in the potentially interesting limit in which they are completely brane-localized (which would leave us with $\wh A_t=-F^\omega=-M_{1/2}$ from the Higgs contributions),
the Yukawa couplings vanish. In models without gauge-Higgs unification, with the Higgs coming from a bulk hypermultiplet, it gives a wrong-sign contribution
\be
  \Delta\wh  A_t=+\frac{F^T}{2R}\,.
\ee

In short, maximal mixing does not occur in minimal radion-mediated models. While in realistic models the exact sfermion soft terms depend on the modelling of the matter sector, and more elaborate examples may allow for maximal mixing in principle, we were unable to find an example. For instance, for the model of \cite{Brummer:2009ug} we find that $\wh A_t/M_{1/2}$ is bounded such that it can never become large and negative, $\wh A_t/M_{1/2}\gtrsim -0.72$. Furthermore, the third-generation soft masses are typically comparable and of the order of the gaugino mass, so large ratios which might still allow for maximal mixing (see the upper row of Fig.~\ref{fig:mixUQ}) do not appear. In conclusion, maximal mixing seems not to be a generic feature of radion-mediated SUSY breaking in 5D.

\subsection{Gauge mediation}

Gauge-mediated supersymmmetry breaking, by definition, encompasses models whose hidden sector decouples from the MSSM as the MSSM gauge couplings are switched off \cite{Meade:2008wd}. Similar as in gaugino-mediated models (with which there is indeed some overlap), $\mu$ is missing in pure gauge mediation, and needs to be generated by additional Higgs-hidden sector couplings. This generically induces a too large $B_\mu$, a problem which needs to be solved in realistic models, or overcome with very special soft mass patterns \cite{Csaki:2008sr, DeSimone:2011va}.

Pure gauge-mediated models also predict vanishing $A$-terms at leading order at the mediation scale. A priori it is not clear that this prediction is maintained when the model is extended such as to solve the $\mu/B_\mu$ problem, but it was shown in \cite{Komargodski:2008ax} that generically this is indeed the case (although Higgs-messenger couplings in the superpotential can be used to obtain large negative $A$-terms while leaving the $\mu/B_\mu$ problem unsolved \cite{Kang:2012ra}). Therefore models of gauge-mediated supersymmetry breaking do not lead to maximal mixing.

The mediation scale $M$ is often taken to be far below $M_{\rm GUT}$ in gauge-mediated models (for exceptions, see e.g.~\cite{Hiller:2008sv,Brummer:2011yd}). However, this generically does not improve the prospects for achieving maximal mixing (see also \cite{Draper:2011aa}). 
The implications of a 124--126 GeV CP-even Higgs boson for minimal gauge mediation have been investigated in detail in \cite{Ajaib:2012vc}, one of the conclusions being that ``the majority of the sparticle masses are in the several to multi-TeV range''.   

\clearpage
\section{Numerical analysis}\label{sec:numerics}

Let us now illustrate the impact of maximal mixing by means of parameter space scans of two models. 
The first is gaugino mediation, and the second is a more generic gravity-mediated model with 
non-universal Higgs masses (NUHM) and universal sfermion masses $m_0$. In the NUHM case, we will 
distinguish between $m_0<M_{1/2}$ and $m_0>M_{1/2}$; gaugino mediation can be regarded as a special NUHM scenario with $m_0=0$. In addition, we will also briefly survey a NUHM scenario where maximal mixing is generated from large sfermion masses for the first two generations.

We use \softsusy~\cite{Allanach:2001kg} for the spectrum calculation, 
\micro~\cite{Belanger:2001fz} for low-energy observables, and 
\hdecay~\cite{Djouadi:1997yw} for computing Higgs decays. 
Besides radiative EWSB and the absence of tachyons, we require the following phenomenological 
constraints to be satisfied: 
\begin{itemize}
   \item {\bf Mass limits:} $m_{\tilde\chi^\pm_1}>103$~GeV,  $m_{\tilde\tau_1}>92$~GeV, $m_{\tilde e_{L,R}}>100$~GeV,  
   $m_{\tilde t_1, \tilde b_1}>100$~GeV, $m_{\tilde g}>500$~GeV, and $m_h>115$~GeV 
            to avoid the most stringent constraints from collider searches, as well as 
   \item {\bf Flavor constraints:}  $2.87\times 10^{-4}<{\rm BR}(B\to X_s\gamma)<4.23\times 10^{-4}$ \cite{Asner:2010qj} and 
             ${\rm BR}(B_s\to \mu^+\mu^-)<5.4\times 10^{-9}$ (i.e.\ the LHCb limit 
\cite{LHCb:moriond} augmented by a 20\% theoretical uncertainty) to be compatible with recent $B$-physics results. 
\end{itemize}
We furthermore compute the dark matter relic density $\Omega h^2$ and the SUSY contribution $\Delta a_\mu$ 
to the muon anomalous magnetic moment, but do not impose any restrictions on them a priori. 

We will also discuss the implications of maximal mixing for a possible Higgs signal near a mass of 125 GeV. 
To this end, we approximate the signal strength for a given final state $X$,  
relative to the Standard Model expectation for the same Higgs mass, as 
\begin{equation}
   R(X) \equiv 
   \frac{ \left[\sigma(gg\to h)\, \BR(h\to X)\right]_{\rm MSSM} }{ \left[\sigma(gg\to h)\, \BR(h\to X)\right]_{\rm SM}} \approx 
   \frac{ \left[\Gamma(h\to gg)\, \BR(h\to X)\right]_{\rm MSSM} }{ \left[\Gamma(h\to gg )\, \BR(h \to X)\right]_{\rm SM}} \,. 
\end{equation} 

\noindent
This is justified because differences in $\sigma(gg\to h)$ versus $\Gamma(h\to gg)$ should largely cancel out when taking the  taking the MSSM/SM ratio.
It is important to note that SUSY contributions can lead to modifications in both Higgs production and Higgs decays as compared to the SM. 
The effective $ggh$ coupling is dominated by the top-quark loop, while the effective $h\gamma\gamma$ coupling is dominated by the contribution from $W$ bosons with a subdominant contribution of the opposite sign from top quarks. Both couplings can receive a large contribution from third-generation sfermions, in particular from stops~\cite{Djouadi:1998az}. In case of no stop mixing, the light stop loop interferes constructively with the top loop, while the interference is destructive in the case of large mixing. In the first case $hgg$ rate will increase, while in the latter case it will decrease. The contrary is true for the $h\gamma\gamma$ couplings. Moreover, weakly interacting particles such as charginos and sleptons will contribute only to the $h\gamma\gamma$ coupling. In particular, light staus that are strongly mixed can enhance the $h \to \gamma\gamma$ rate~\cite{Carena:2011aa}. 
Another important effect is the enhancement or suppression of $h\to b\bar b$, which can significantly change the overall branching ratios. In fact an enhancement of the $\gamma\gamma$ signal, i.e.\ $R(\gamma\gamma)>1$, is often due to a suppression of BR($h\to b\bar b$).

\subsection{Gaugino mediation}  \label{sec:gm}

In the gaugino-mediated model the non-vanishing soft masses at the GUT scale are $M_{1/2}$, $A_0$, $\wh\mu$, $\wh{B_\mu}$, $\wh m_{H_u}^2$ and $\wh m_{H_d}^2$. Given $m_Z$, the last four parameters can be traded against $\tan\beta$, $\mu$, and the pseudoscalar mass $m_A$. Our input parameters are therefore $M_{1/2}$ and $A_0$ at $\mgut$, together with 
$\tan\beta$, $\mu$ and $m_A$ at the electroweak scale.  We choose $\mu>0$ and 
perform flat random scans letting $M_{1/2}$, $\mu$ and $m_A$ vary up to 2~TeV; 
$\tan\beta$ is allowed to vary between 1 and 60.  
Regarding $A_0$, we scan over two different intervals, $|A_0/M_{1/2}|\le 3$ and $A_0\in [-4,0]$~TeV. 
Of 256k valid points from these scans, 158k remain after the basic mass limits and the flavor constraints listed above. 
Of these, 10k points have $m_h=123\mbox{--}127$~GeV. 

\begin{figure}[t]\centering
  \includegraphics[width=80mm]{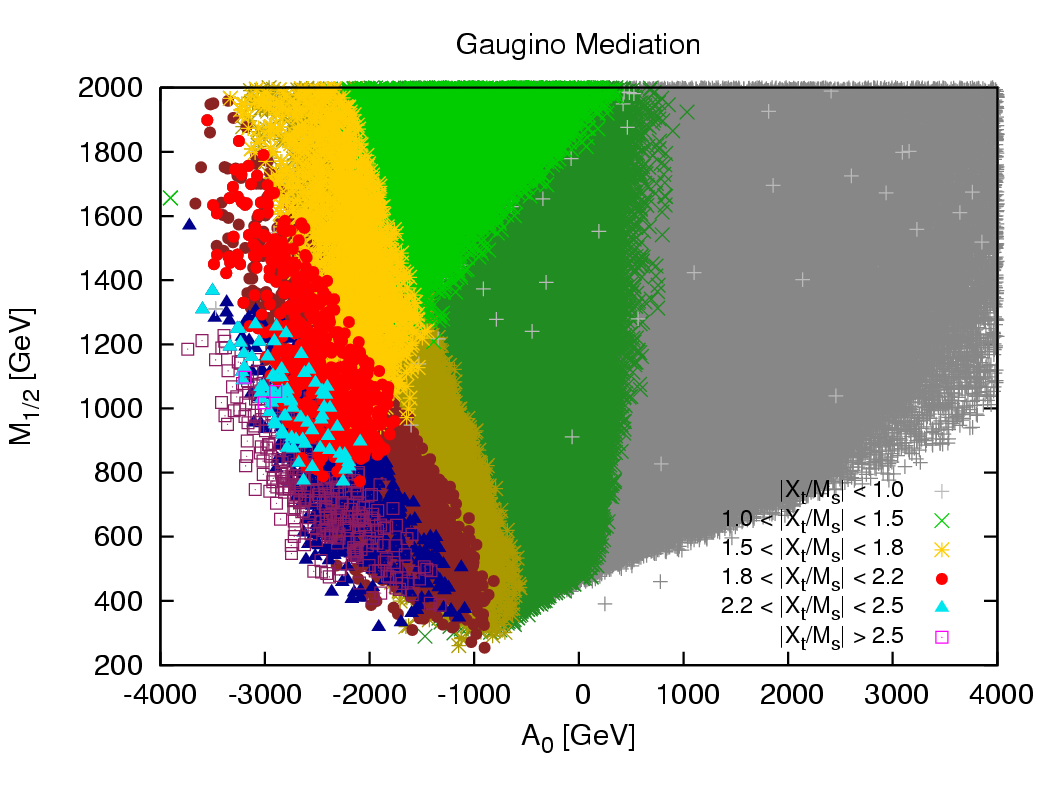}  
  \caption{Projected parameter space in the gaugino mediation model. The amount of stop mixing, $X_t/M_{S}$, is shown by a color code. Brighter (darker) shaded points of each color have 
    $m_h>123$ ($m_h<123$) GeV. Evidently, maximal mixing requires $A_0/M_{1/2}\simeq -1$ to $-3$. 
    \label{fig:gm-A0-Mhf}}
\end{figure}

Figure~\ref{fig:gm-A0-Mhf} shows a projection of the scanned parameter space in the $A_0$ versus $M_{1/2}$ plane, 
with the amount of stop mixing visualized by a color code. For each color, marking a certain interval of stop mixing, 
points with $m_h>123$ ($m_h<123$) GeV are shown in brighter (darker) shades. 
Evidently, maximal mixing (yellow, red and blue points) requires $A_0/M_{1/2} \simeq -1$ to $-3$. 
As can also be seen, a SM-like $h$ with mass $m_h>123$~GeV 
requires large mixing with a large negative $A_0$, or very large $M_{1/2}$. 
Even with $|X_t/M_S|\approx 2$, a Higgs near 125 GeV requires $M_{1/2} \gtrsim 750$~GeV;  
for $|X_t/M_S|\sim 1$, $M_{1/2}$ is pushed up to 2~TeV.

\begin{figure}[t]\centering
  \includegraphics[width=75mm]{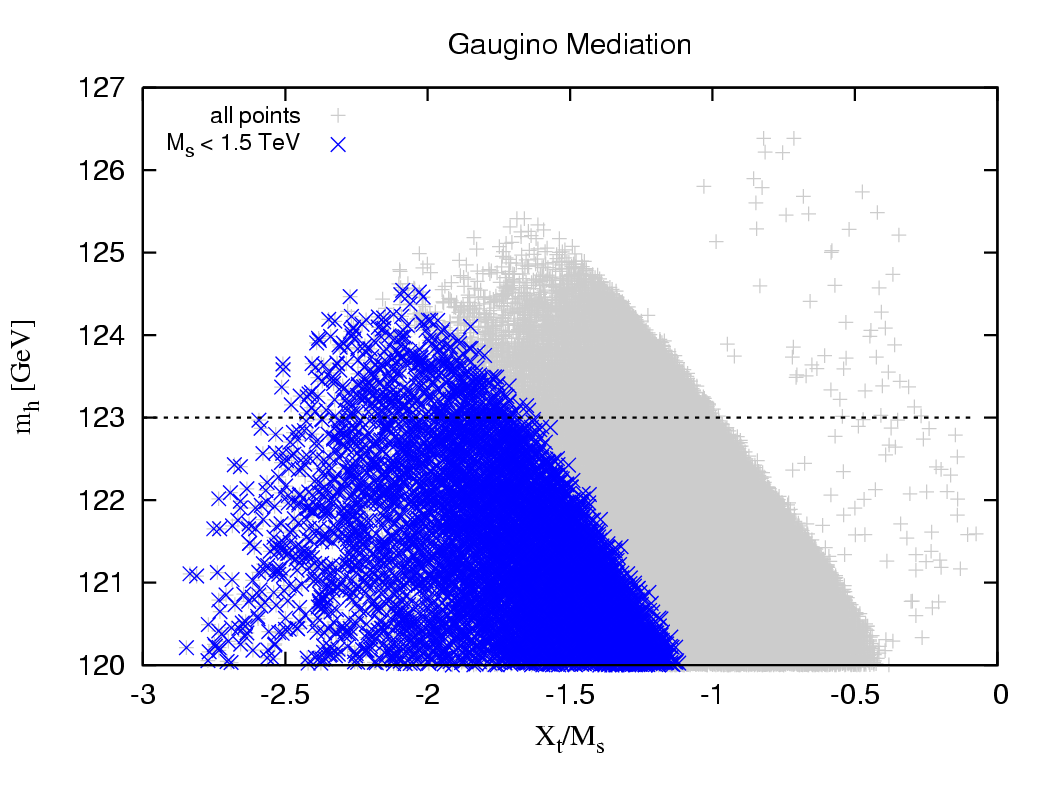}  
  \includegraphics[width=75mm]{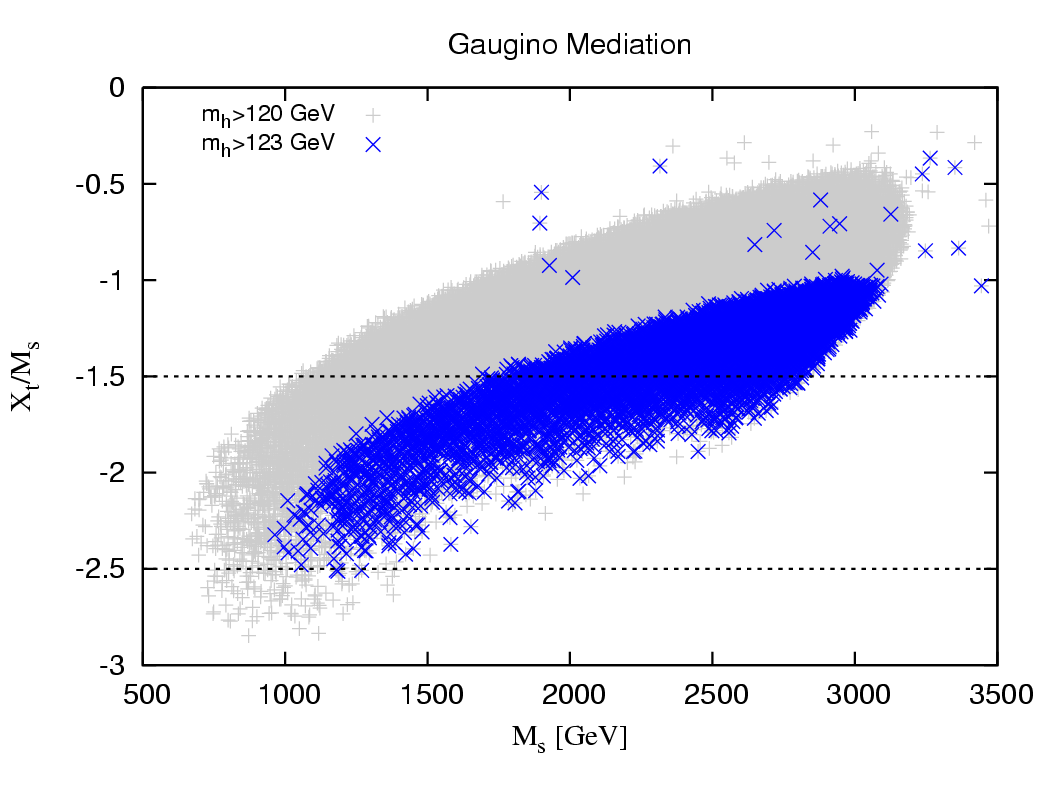}  
  \caption{Higgs mass $m_h$ versus stop mixing $X_t/M_S$ (left) and $X_t/M_S$ versus $M_S$ (right) in the gaugino mediation model. 
  \label{fig:gm-mixing}}
\end{figure}

The correlation of $m_h$ with the stop mass scale and mixing is further illustrated in Fig.~\ref{fig:gm-mixing}. We see that light stops with $M_S\lesssim 1.5$~TeV indeed require maximal mixing, $-2.5\lesssim X_t/M_S\lesssim -1.5$, for having a Higgs mass compatible with the ATLAS and CMS excesses. We also see that while in principle it is easy to have $M_S<1$~TeV as preferred by naturalness, once we impose $m_h>123$~GeV the stop mass scale is pushed to a TeV and above; if we demand $M_S\approx 1$~TeV in addition to $m_h=123\mbox{--}127$~GeV, this requires $-2.5\lesssim X_t/M_S\lesssim -2$, see the RHS plot in Fig.~\ref{fig:gm-mixing}.
 
\begin{figure}[t]\centering
  \includegraphics[width=75mm]{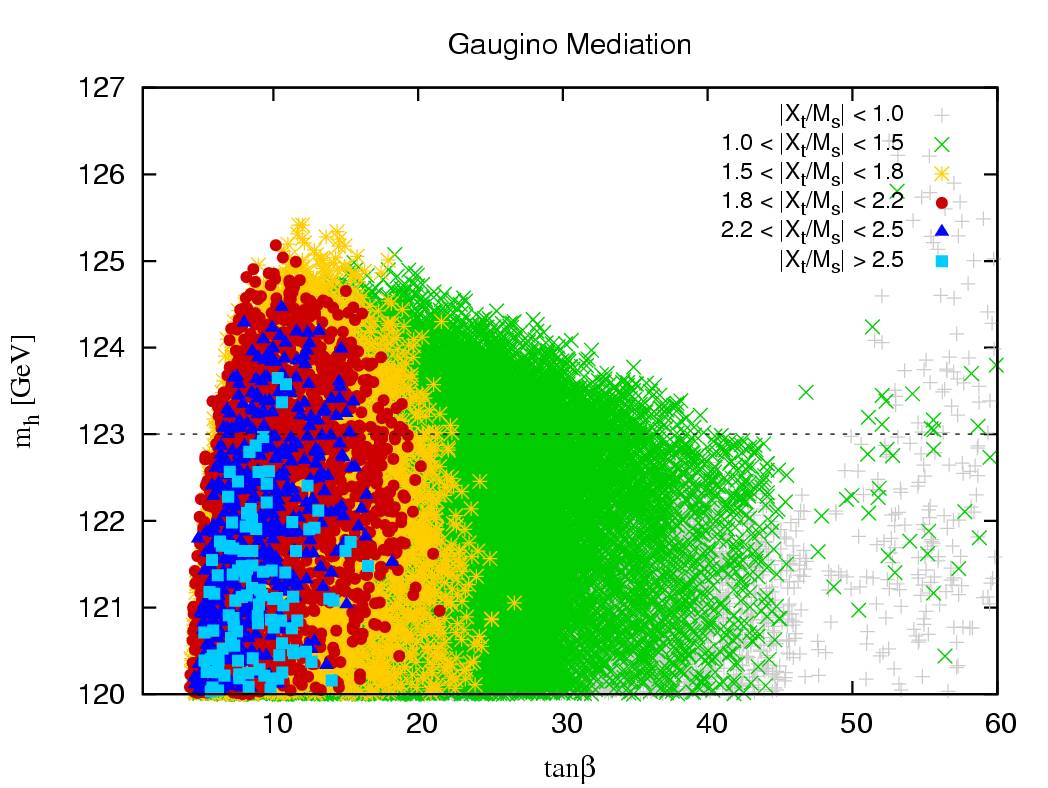}  
  \includegraphics[width=75mm]{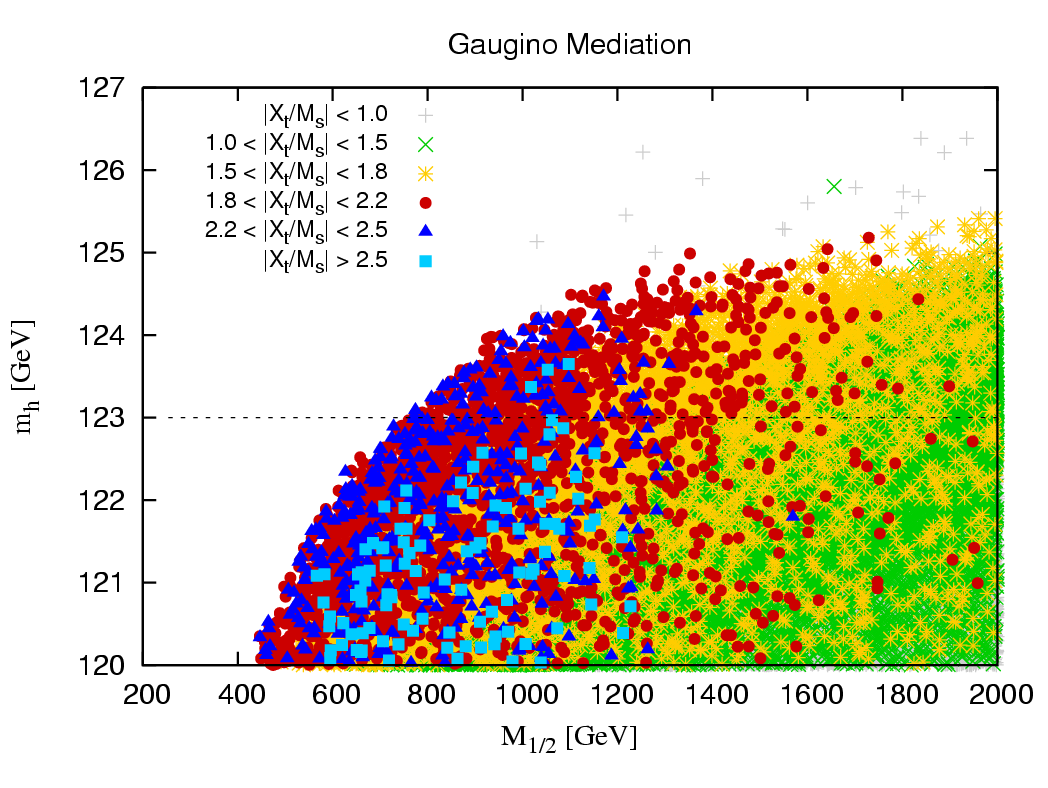}\\
  \includegraphics[width=75mm]{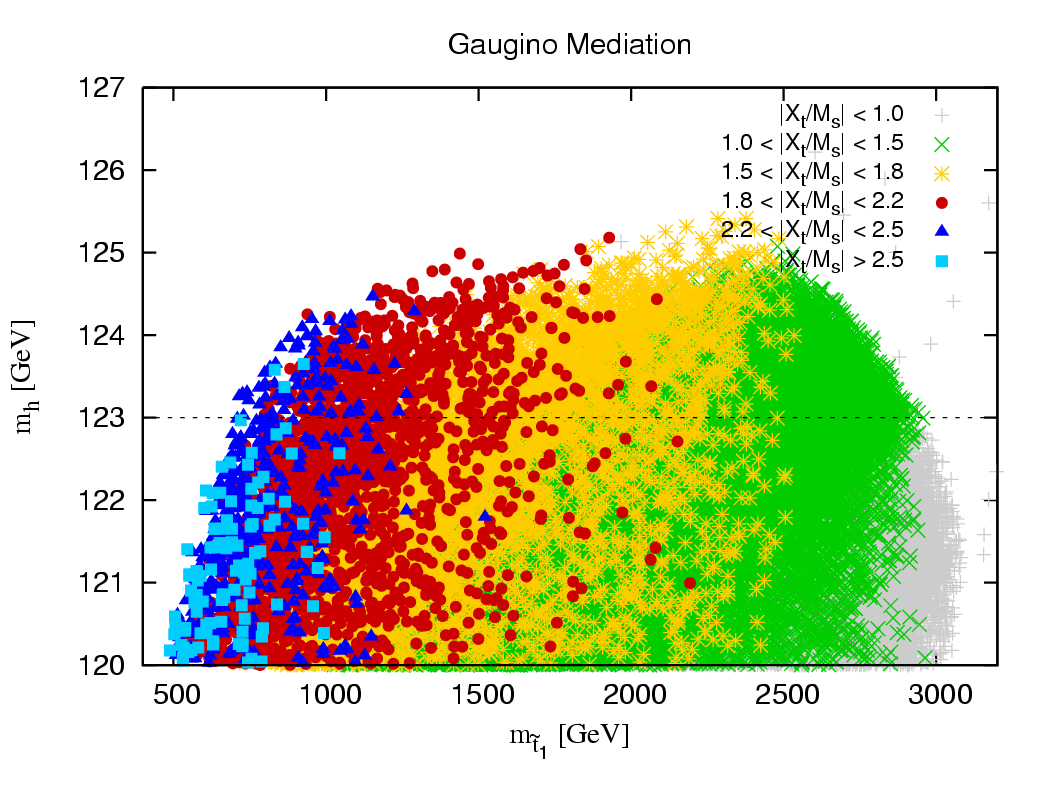}  
  \includegraphics[width=75mm]{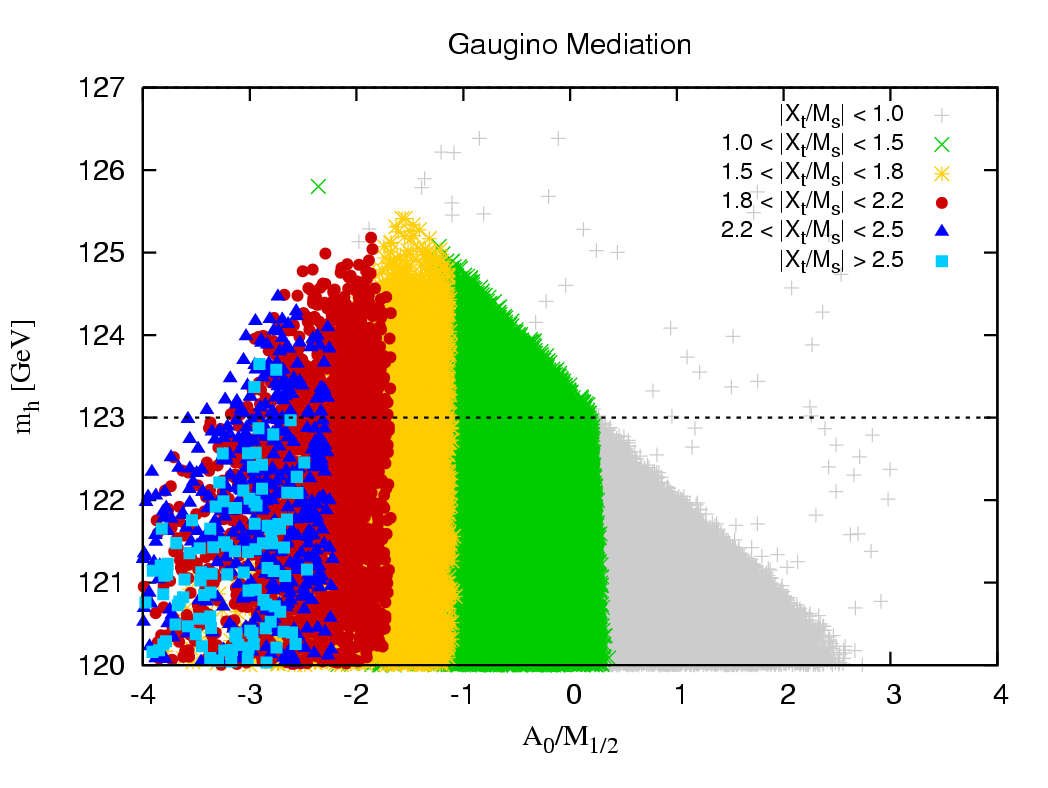}
\caption{Lightest Higgs mass dependence on various quantities, in the gaugino mediation model. 
Top row: Large $m_h$ favors not too large $\tan\beta$, especially when demanding a lighter spectrum, 
and $M_{1/2}\gtrsim 750$ GeV (i.e.\ $m_{\tilde g}\gtrsim 1.7$~TeV). 
Bottom row: $\tilde t_1$ masses below 1~TeV can still be reconciled with large $m_h$, but only at maximal mixing. 
If the gluino is light enough to be seen at the LHC, maximal mixing is also strongly favored. 
\label{fig:gm-mh} }
\end{figure}

\begin{figure}[t]\centering
  \includegraphics[width=75mm]{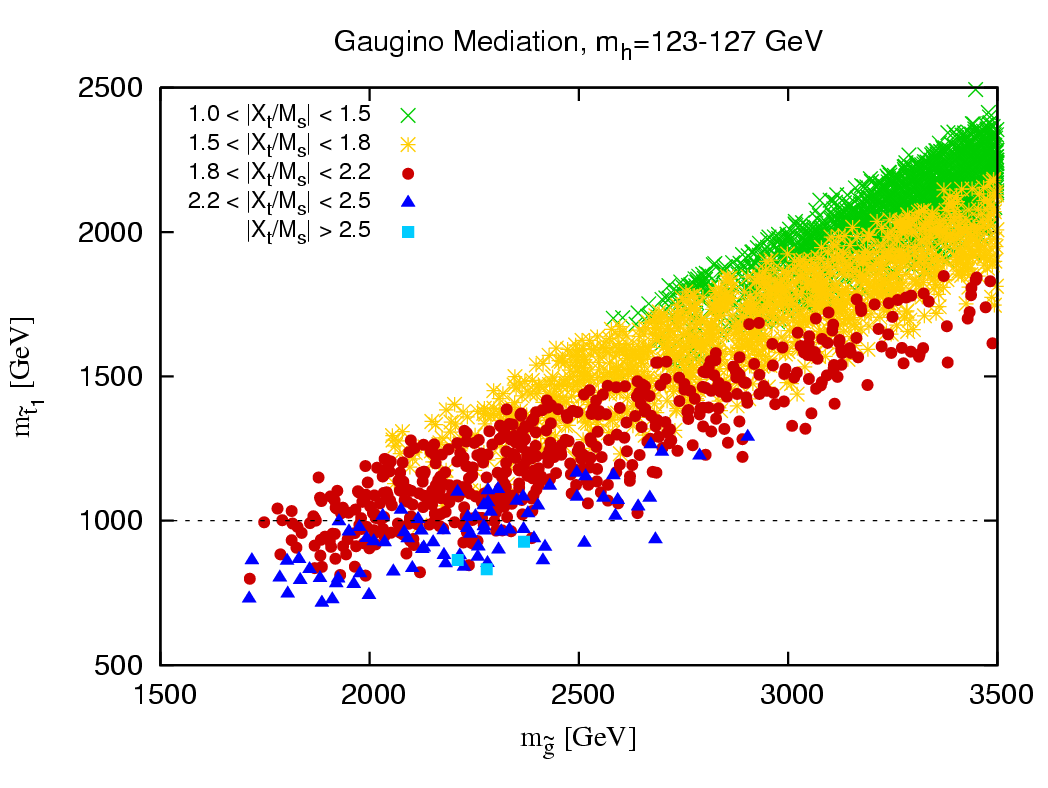}  
  \includegraphics[width=75mm]{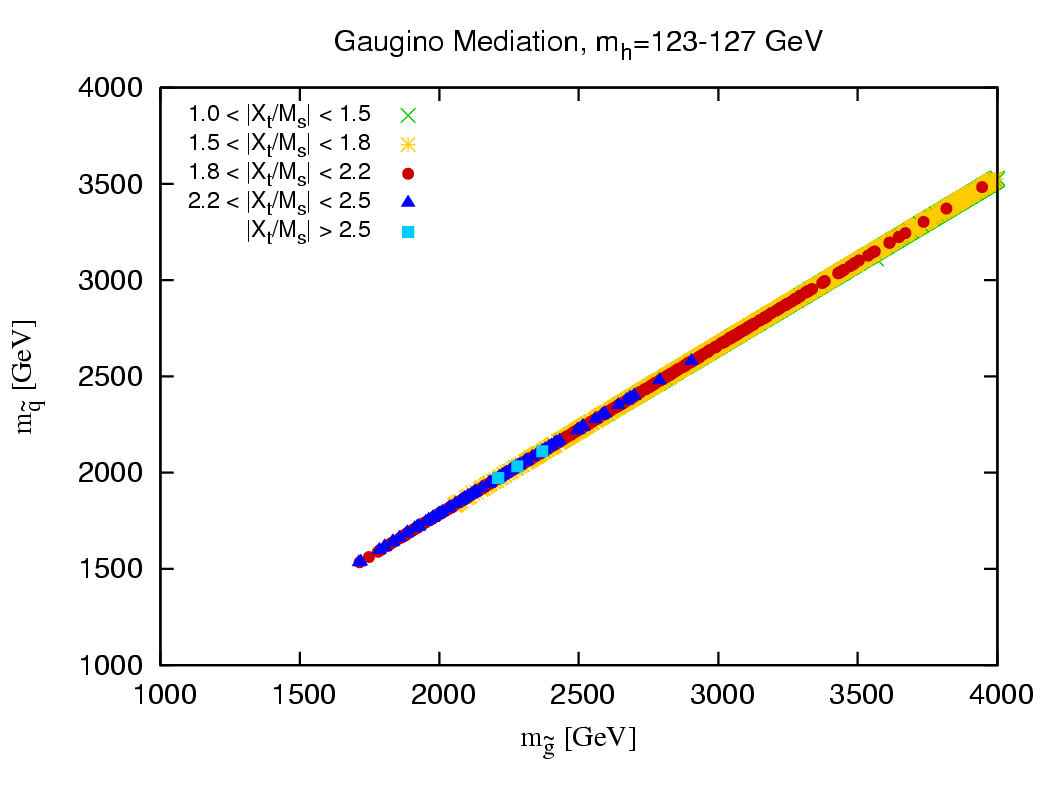}  
  \caption{Scatter plots of points with $m_h=123\mbox{--}127$~GeV in the gaugino mediation model, 
  on the left in the ${\tilde t_1}$ versus ${\tilde g}$ mass plane, on the right in the 
  ${\tilde q}$ versus ${\tilde g}$ mass plane. 
  Since $m_0\equiv 0$, $m_{\tilde q}\lesssim m_{\tilde g}$ throughout the parameter space. 
  \label{fig:gm-gluino}}
\end{figure}

The $m_h$ dependence on various quantities ($\tan\beta$, $M_{1/2}$, $m_{\tilde t_1}$ and $A_0/M_{1/2}$) 
is shown explicitly in Fig.~\ref{fig:gm-mh}. Maximal mixing clearly prefers not too large $\tan\beta<25$, while $m_h>123$~GeV needs $M_{1/2} \gtrsim 750$~GeV. The highest $h$ mass is achieved for $A_0/M_{1/2}\approx -1.5$, but stops are heavy in this case, above 2~TeV. 
Although large $m_h$ still allows for $\tilde t_1$ masses below 1~TeV 
in case of maximal mixing, overall a Higgs in the $123\mbox{--}127$~GeV mass range points towards a 
heavy spectrum. 
The correlations between gluino, stop and 1$^{st}$/2$^{nd}$ generation squark masses are shown in Fig.~\ref{fig:gm-gluino}. 
Remarkably, the latest LHC bounds of $m_{\tilde g,\tilde q}\gtrsim 1.4$~TeV \cite{atlas5fbi,cms5fbi} 
for CMSSM-like scenarios with $m_{\tilde g}\simeq m_{\tilde q}$ are automatically evaded. 
In fact, requiring $m_h>123$~GeV, we find $m_{\tilde t_1}\gtrsim 715$~GeV, 
$m_{\tilde q}\gtrsim 1.5$~TeV and $m_{\tilde g}\gtrsim 1.7$~TeV in gaugino mediation. 
Moreover, for $\tan\beta\le 50$ we find a maximal $h$ mass of around $125$~GeV (which is however subject to a 1--2~GeV theoretical uncertainty).

\begin{figure}[t]\centering
  \includegraphics[width=75mm]{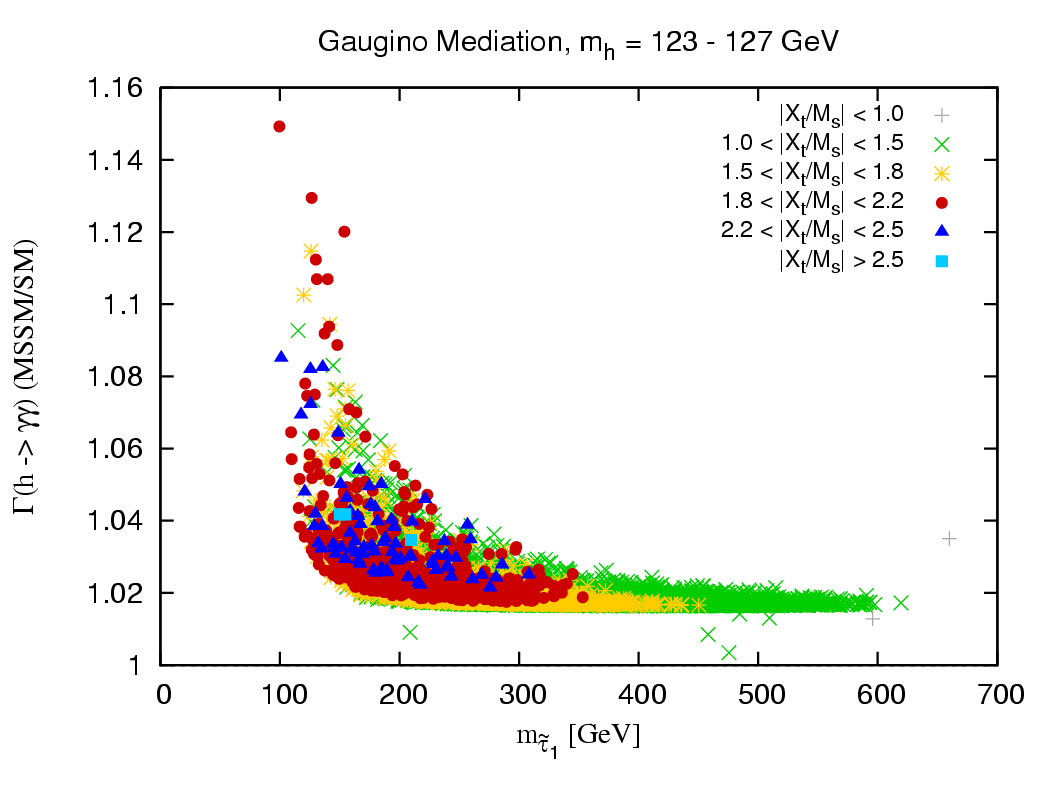} 
  \includegraphics[width=75mm]{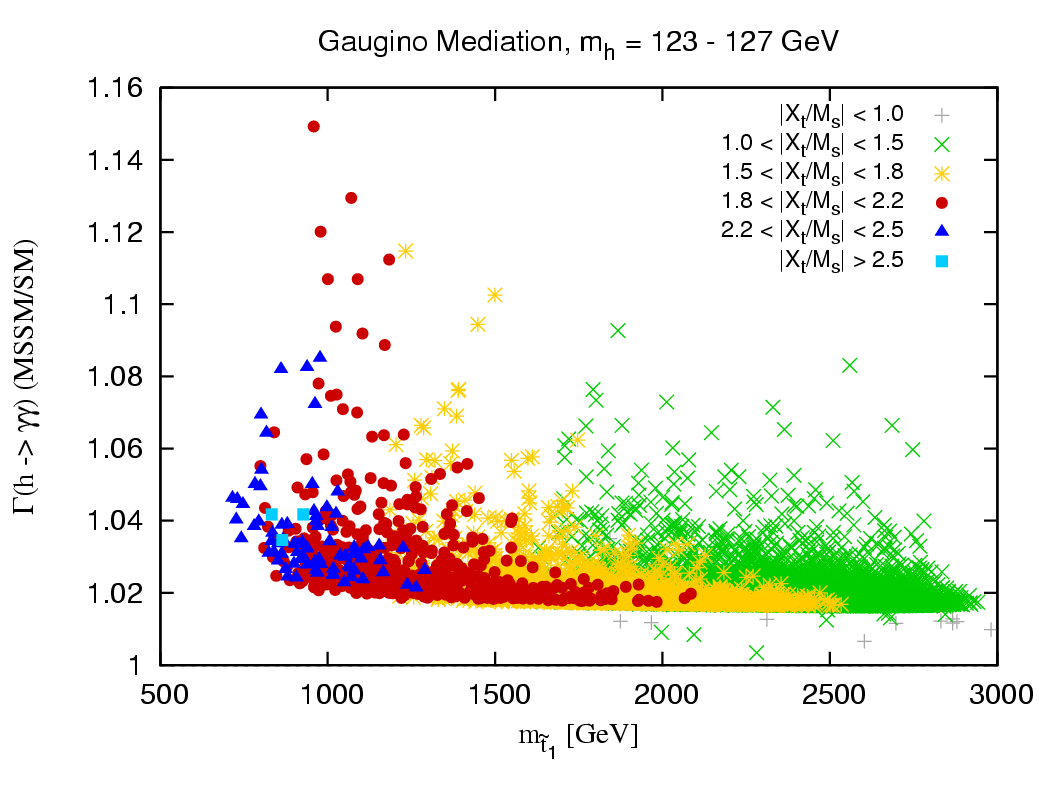} 
  \includegraphics[width=75mm]{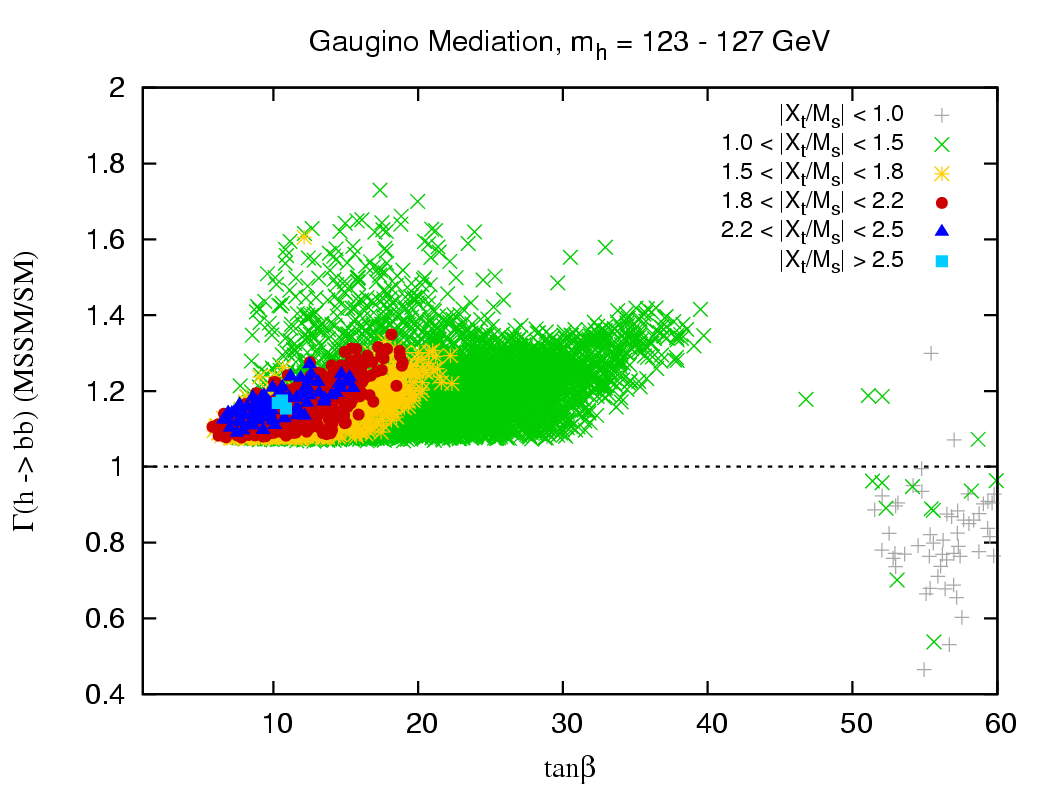} 
  \includegraphics[width=75mm]{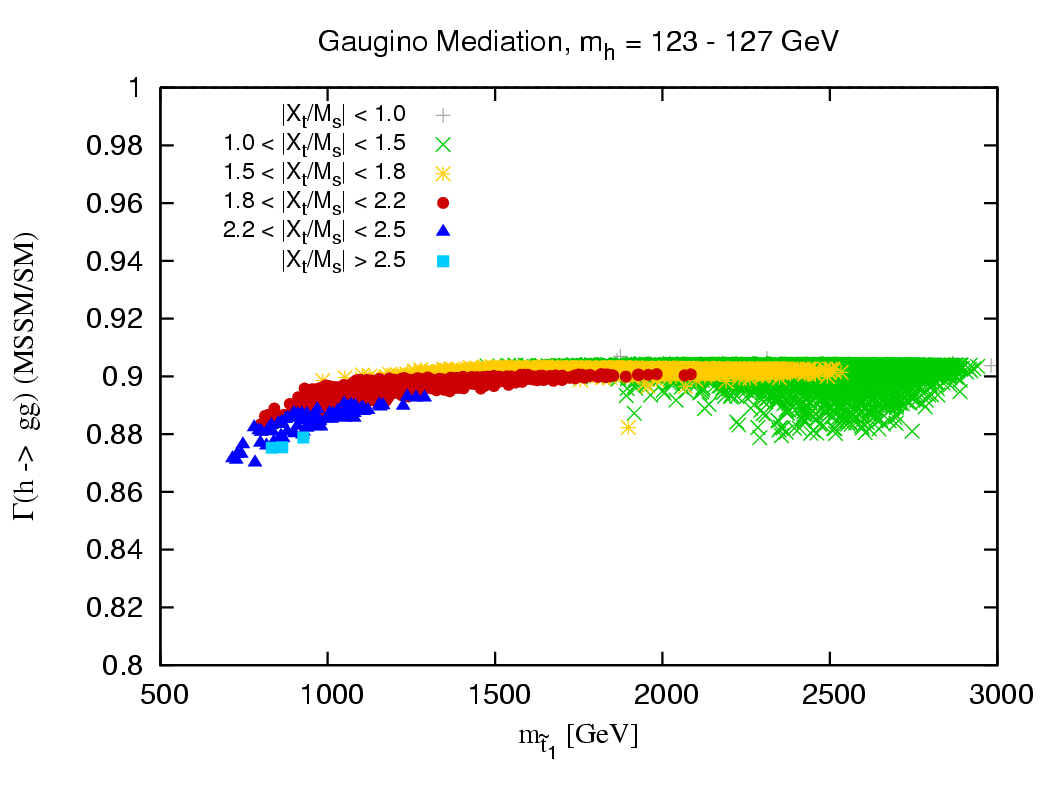} 
\caption{Partial widths of $h\to\gamma\gamma$, $h\to b\bar b$ and $h\to gg$, 
relative to SM expectations. \label{fig:gm-hwidths} }
\end{figure}

\begin{figure}[t]\centering
  \includegraphics[width=75mm]{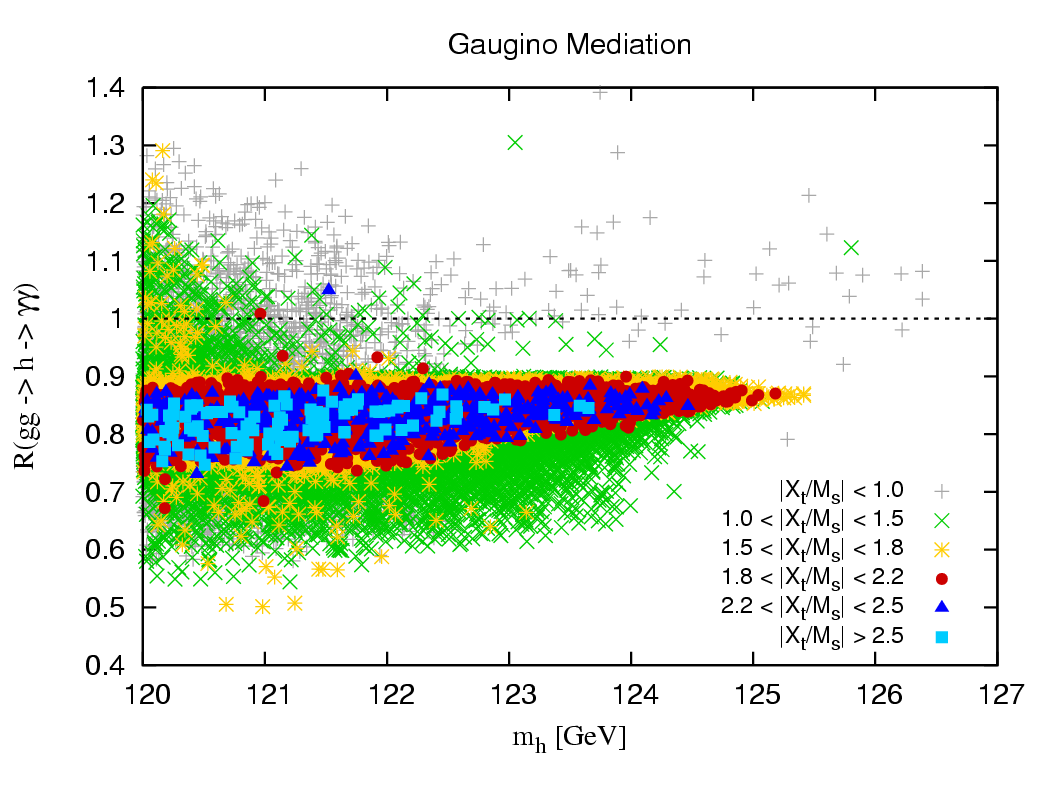} 
  \includegraphics[width=75mm]{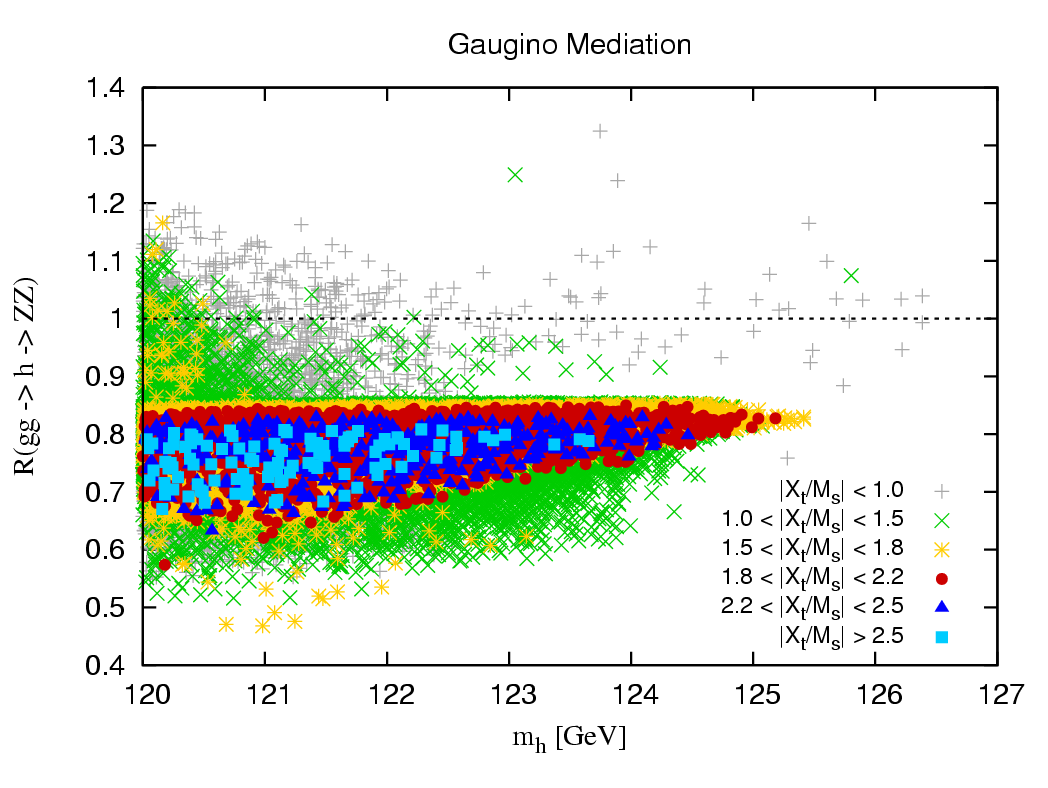}
\caption{In the gaugino mediation model with maximal mixing, the $gg\to h\into\gamma\gamma$ rate is slightly below the Standard Model one (left), mostly as a result of the reduced gluon fusion production cross section. On the right, we show the $ZZ$ signal strength. The few points with a signal strength $R>1$ at $m_h>123$~GeV have very large $\tan\beta$
\label{fig:gm-hsignal}.  }
\end{figure}

Owing to the vanishing sfermion soft masses, $m_0=0$, over most of the gaugino mediation parameter space, the LSP is the lighter stau $\tilde\tau_1$.\footnote{More precisely, to obtain a consistent cosmological picture the stau should in that case be the next-to-LSP, and the true LSP a gravitino or axino~\cite{Berger:2008ti}.} 
(92\% of the points satisfying mass limits, flavor constraints and $m_h=123\mbox{--}127$~GeV have a $\tilde\tau_1$ LSP, while 7\% have a neutralino LSP.)
Hence one might expect that the $gg\to h\to\gamma\gamma$ rate be enhanced by the light $\tilde\tau_1$ contribution~\cite{Carena:2011aa}. While the $h\to\gamma\gamma$ partial width is indeed enhanced for light staus\footnote{Note that $m_h>123$~GeV leads to $m_{\tilde\tau_1}>99$~GeV in our dataset.} 
(and light stops with large mixing), this is mostly compensated by the suppression of the $h\to gg$ partial width due to the $\tilde t$ loop contribution, and by the larger total $h$ width (mostly due to larger $\Gamma(h\to b\bar b)$). As a result, the $gg\to h\to\gamma\gamma$ signal strength is typically 80--90\% of that in the SM. 
A similar suppression arises for the $ZZ$ final state. 
This is illustrated in Figs.~\ref{fig:gm-hwidths} and \ref{fig:gm-hsignal}. 
The few points with a signal strength $R>1$ at $m_h>123$~GeV feature very large $\tan\beta=52$--$60$, and a reduced $h\to b\bar b$ rate.

\begin{figure}[t]\centering
  \includegraphics[width=75mm]{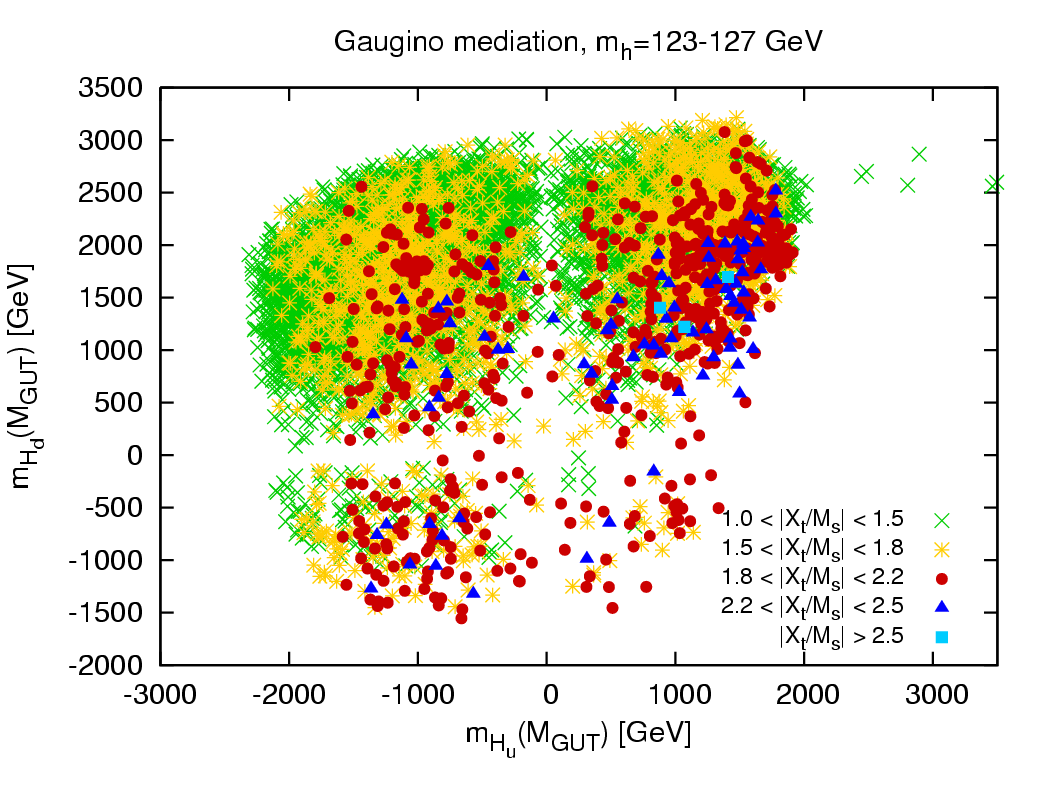} \\
  \includegraphics[width=75mm]{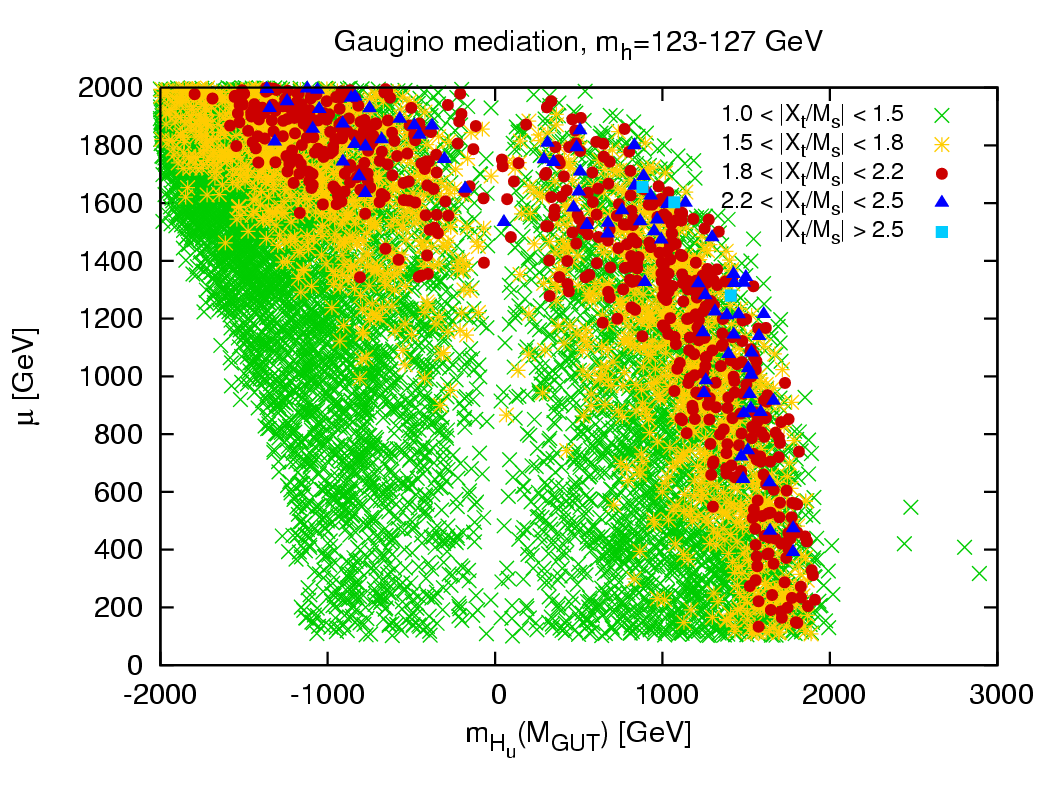} \\
  \includegraphics[width=75mm]{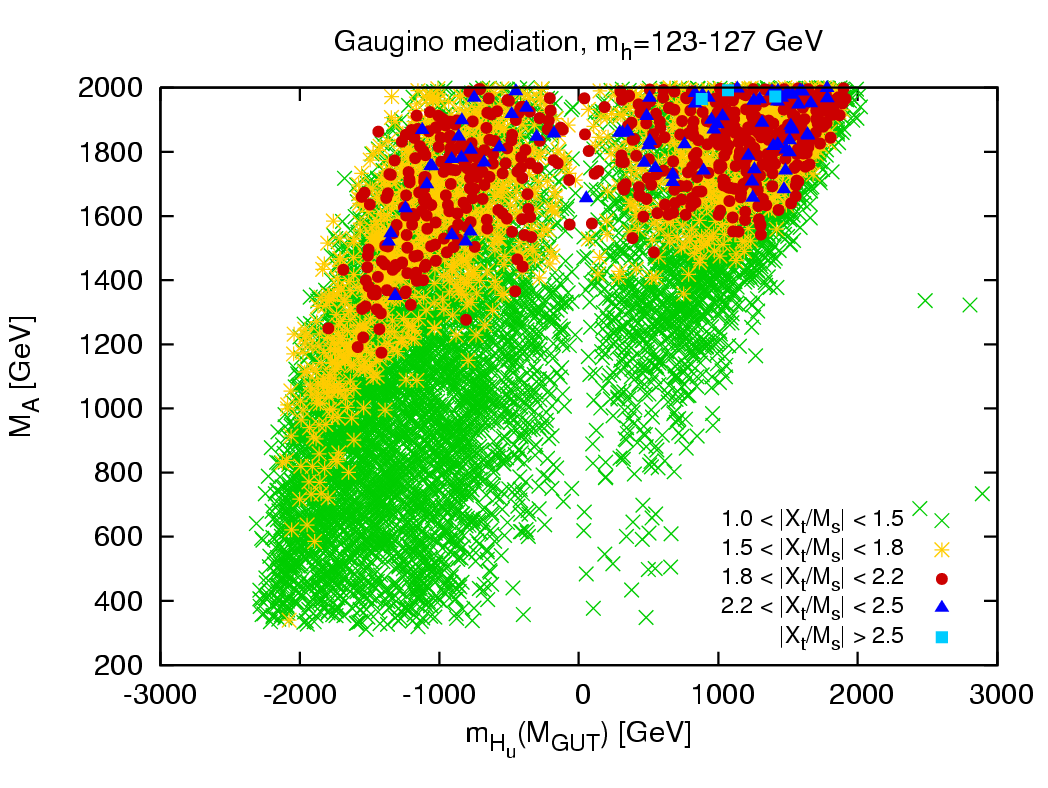} 
\caption{Correlations between GUT-scale Higgs soft masses $\wh m_{H_u}$ and $\wh m_{H_d}$, 
and weak-scale Higgs mass parameters $\mu$ and $m_A$.
    \label{fig:gm-mHu-mHd}}
\end{figure}

Finally, in Fig.~\ref{fig:gm-mHu-mHd}, we illustrate the implications of maximal mixing and a heavy CP-even Higgs for the GUT-scale Higgs soft masses $\wh m_{H_u}$, $\wh m_{H_d}$, and for the weak-scale Higgs mass parameters $\mu$ and $m_A$. (We use the convention $m_{H_{u,d}}\equiv {\rm sign}(m_{H_{u,d}}^2)\times\sqrt{|m_{H_{u,d}}^2|}$). As can be seen, positive up-type Higgs soft masses $\wh m_{H_{u}}^2$ are preferred, in particular in case of small $\mu$ (as preferred by fine-tuning). This confirms our expectations from section~\ref{sec:maxmix}. Moreover, the pseudoscalar Higgs mass $m_A$ prefers to be large in the maximal mixing case, well above current limits.

\clearpage
\subsection{NUHM model}  \label{sec:nuhm}

Let us now turn to the gravity-mediated model with non-universal Higgs soft masses (NUHM). 
The parameters are the same as in the gaugino-mediation case, but for non-vanishing soft masses for squarks and sleptons at $M_{\rm GUT}$. For simplicity, we take the latter two to be universal at the GUT scale. (As discussed in section~\ref{sec:maxmix}, the scalar masses which actually affect maximal mixing are mainly $m_{Q_3}$ and $m_{U_3}$.) 
We scan over $M_{1/2}$, $A_0$, $\tan\beta$, $\mu$ and $m_A$ as in the previous subsection, allowing however $|A_0|$ up to 6~TeV. In addition, we let $m_0$ vary from 0 to 5~TeV. 
Over most of the parameter space, the non-vanishing $m_0$ makes the leptons heavier than the $\tilde\chi^0_1$. We thus keep only points with a neutralino LSP, without however restricting $\Omega h^2$. We will discuss the $\tilde\chi^0_1$ relic density at the end of this subsection.

\begin{figure}\centering
  \includegraphics[width=75mm]{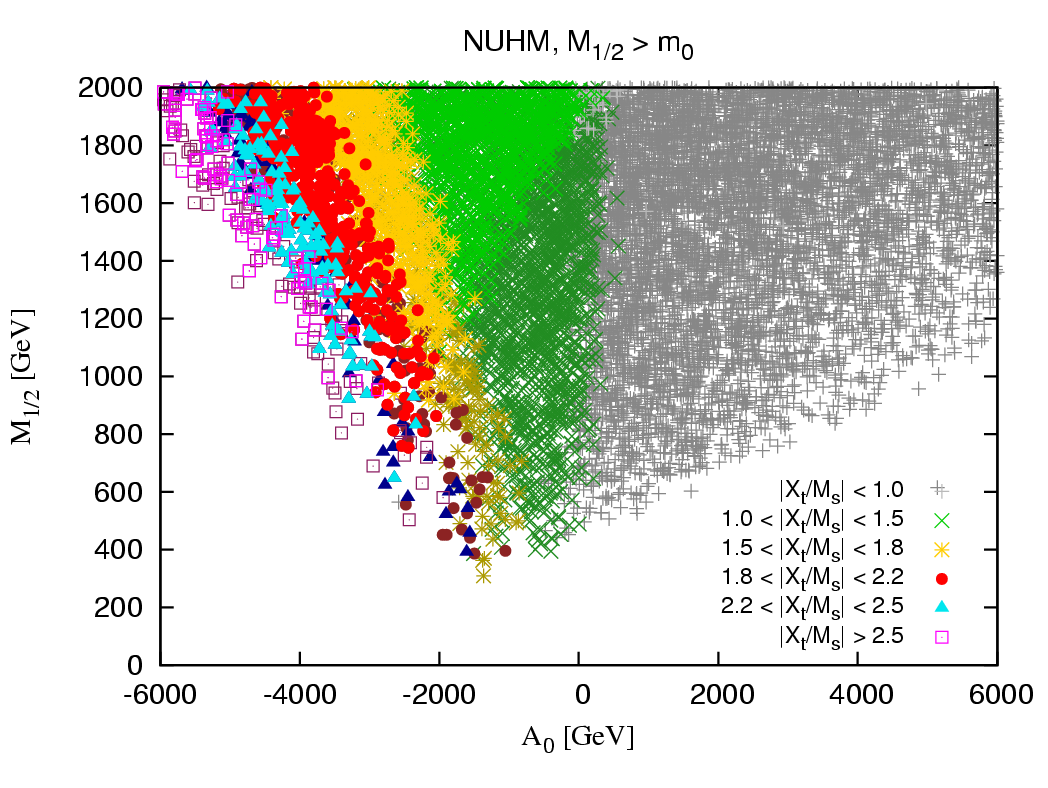}
  \includegraphics[width=75mm]{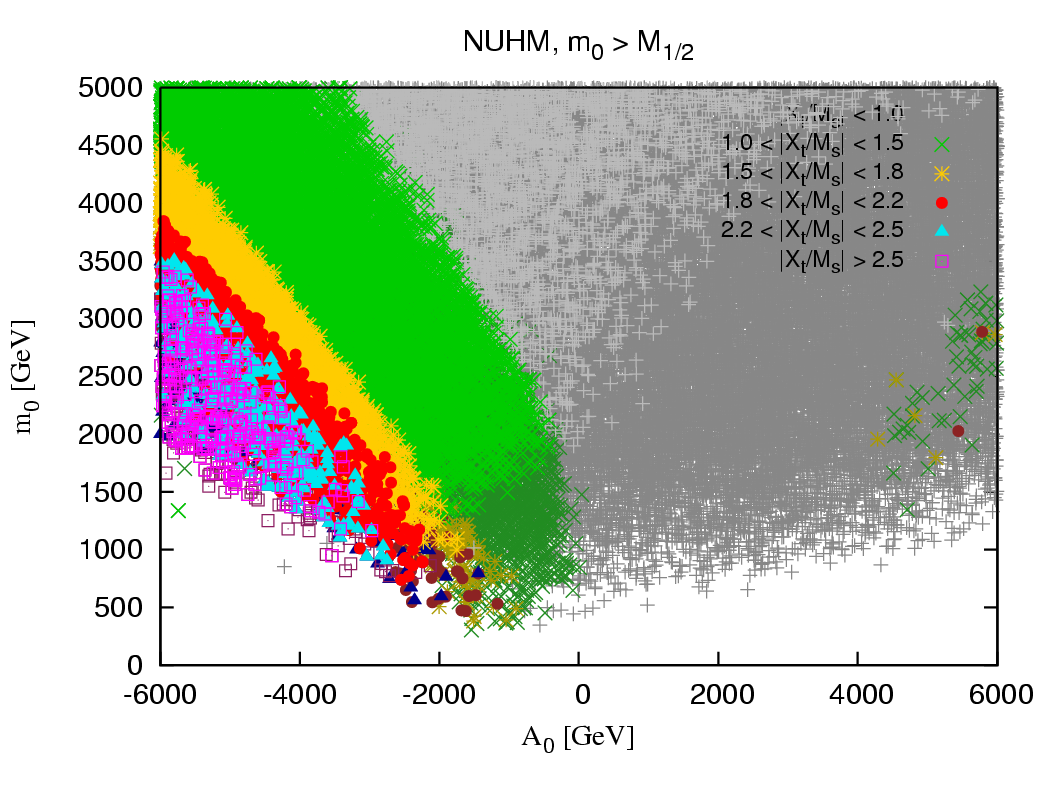}
\caption{Projected parameter space in the NUHM model, analogous to Fig.~\ref{fig:gm-A0-Mhf}. We distinguish $M_{1/2}>m_0$ (left) and $m_0>M_{1/2}$ (right). The amount of stop mixing, $X_t/M_{S}$, is indicated by a color code, with brighter (darker) points of each color having $m_h>(<)\,123$ GeV. It is apparent that negative $A_0$ is necessary for maximal mixing, with the ratio $A_0/{\rm max}(M_{1/2},m_0)\approx -1$ to $-3$. 
\label{fig:nuhm-A0-Mhf}}
\end{figure}

\begin{figure}\centering
  \includegraphics[width=75mm]{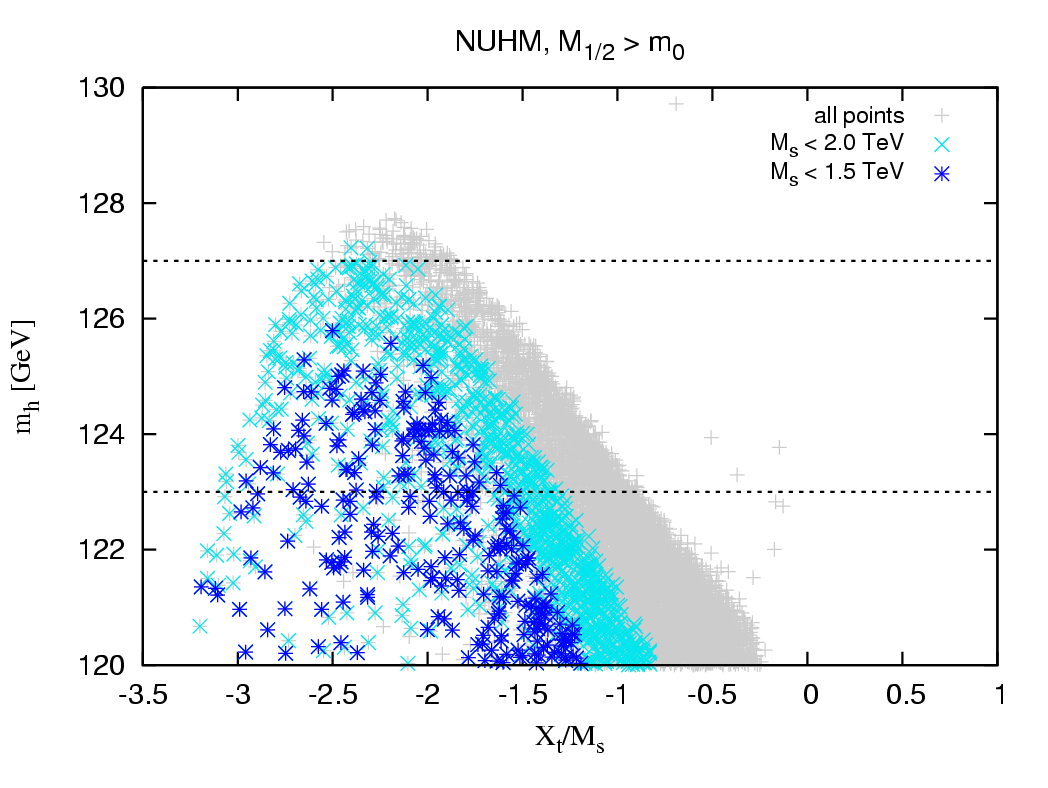} 
  \includegraphics[width=75mm]{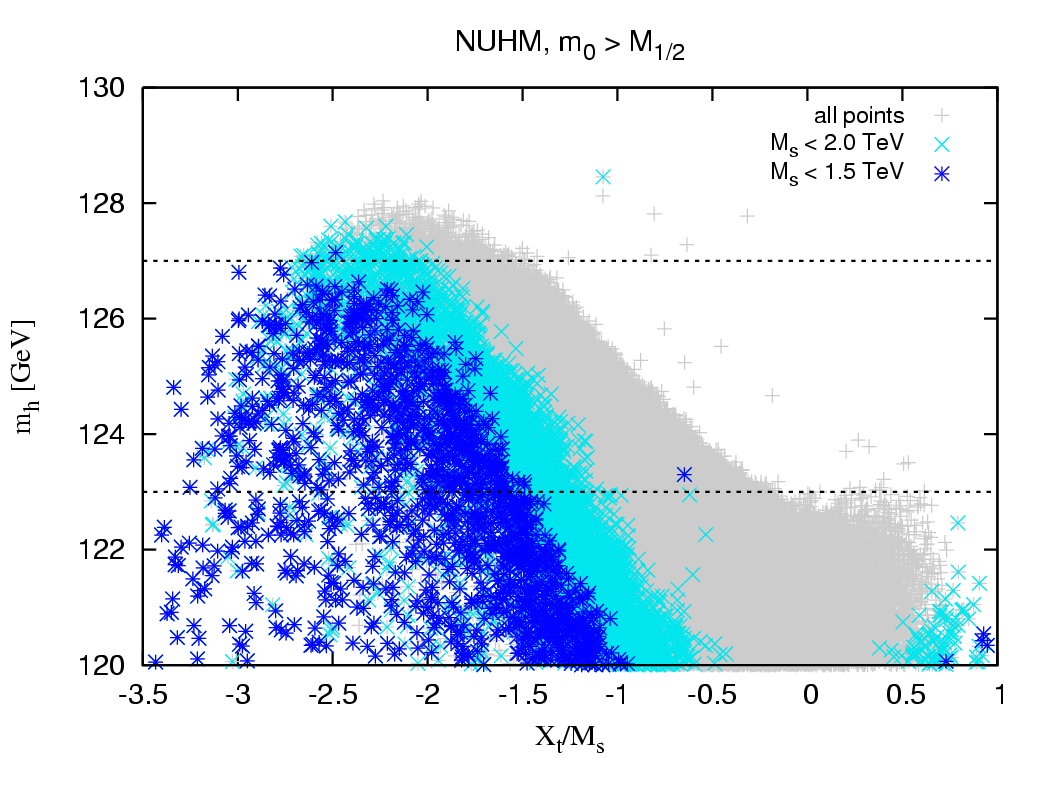}  \\
  \includegraphics[width=75mm]{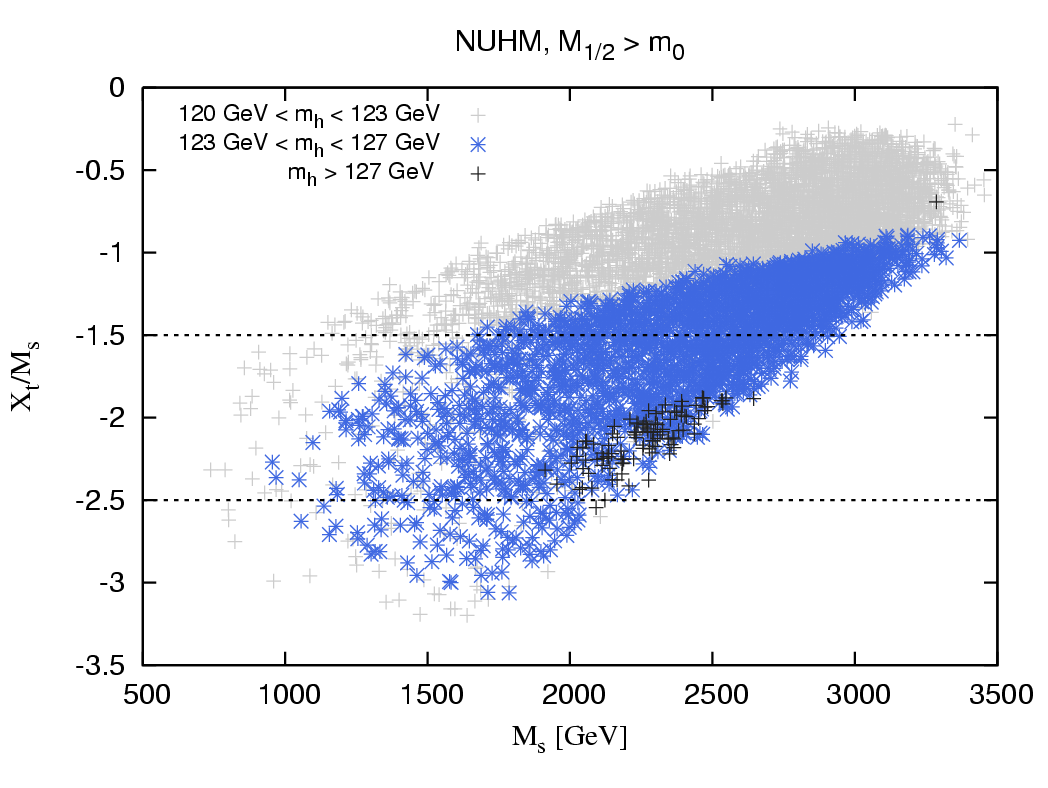}  
  \includegraphics[width=75mm]{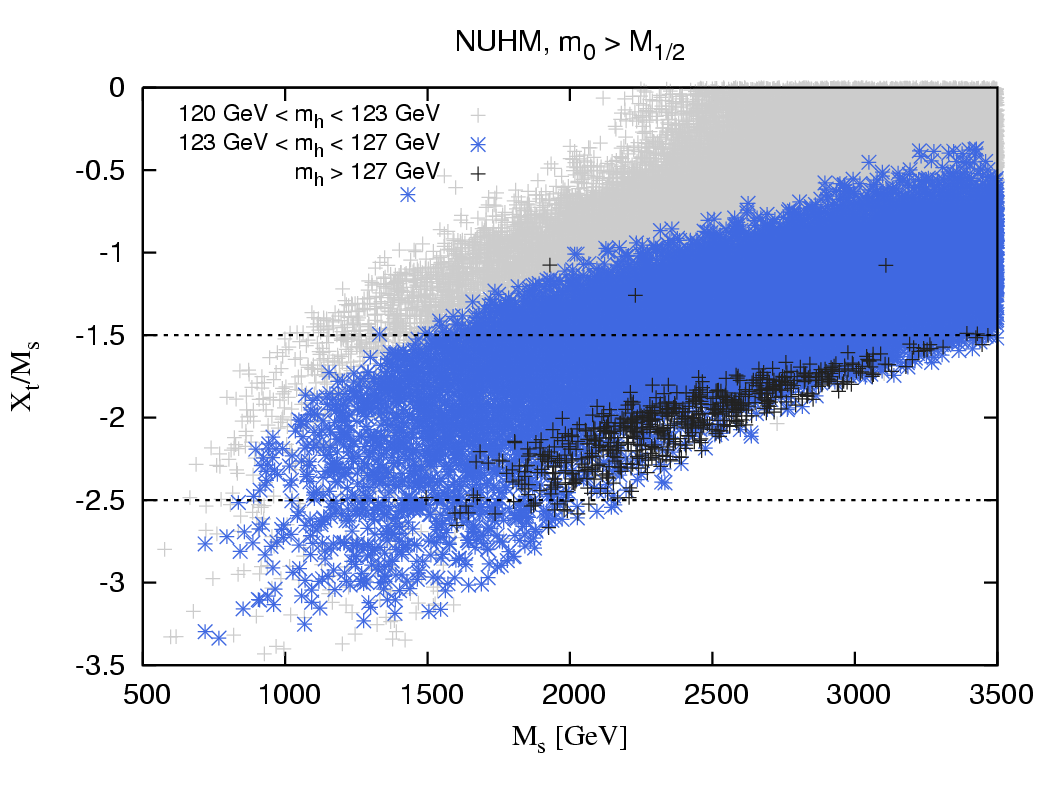}  
  \caption{Higgs mass $m_h$ versus stop mixing $X_t/M_S$ (top row) and $X_t/M_S$ versus $M_S$ (bottom row) in the NUHM model. 
  \label{fig:nuhm-mixing}}
\end{figure}

\begin{figure}\centering
  \includegraphics[width=75mm]{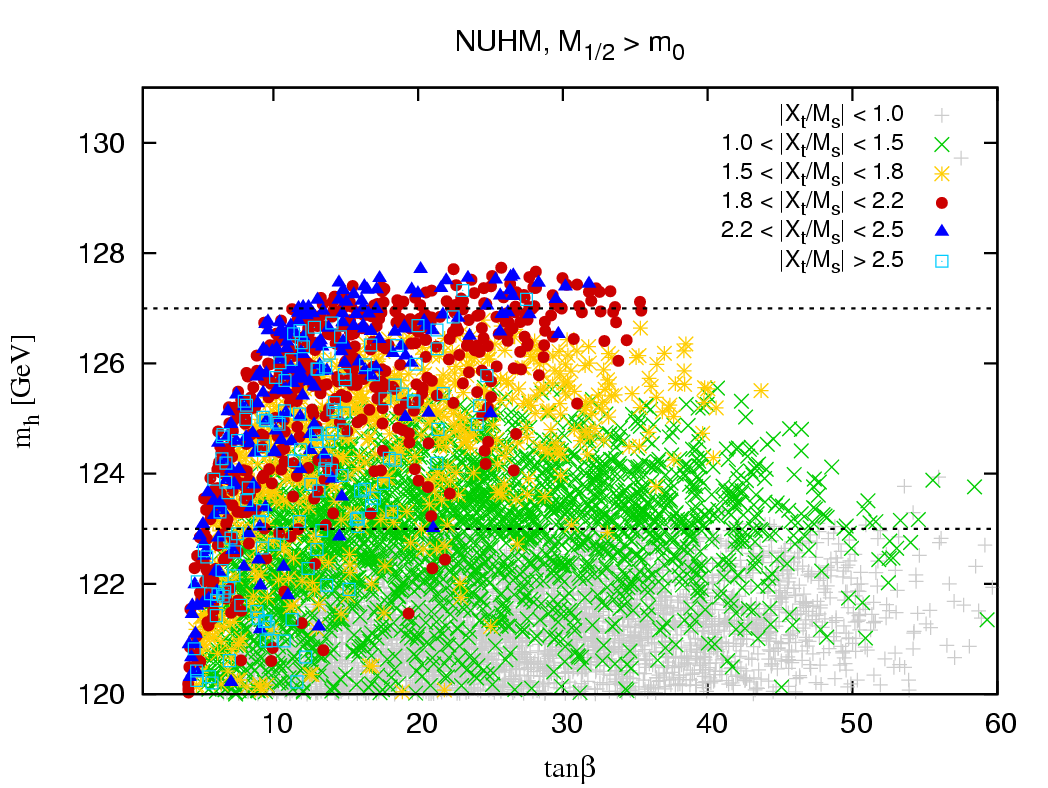} 
  \includegraphics[width=75mm]{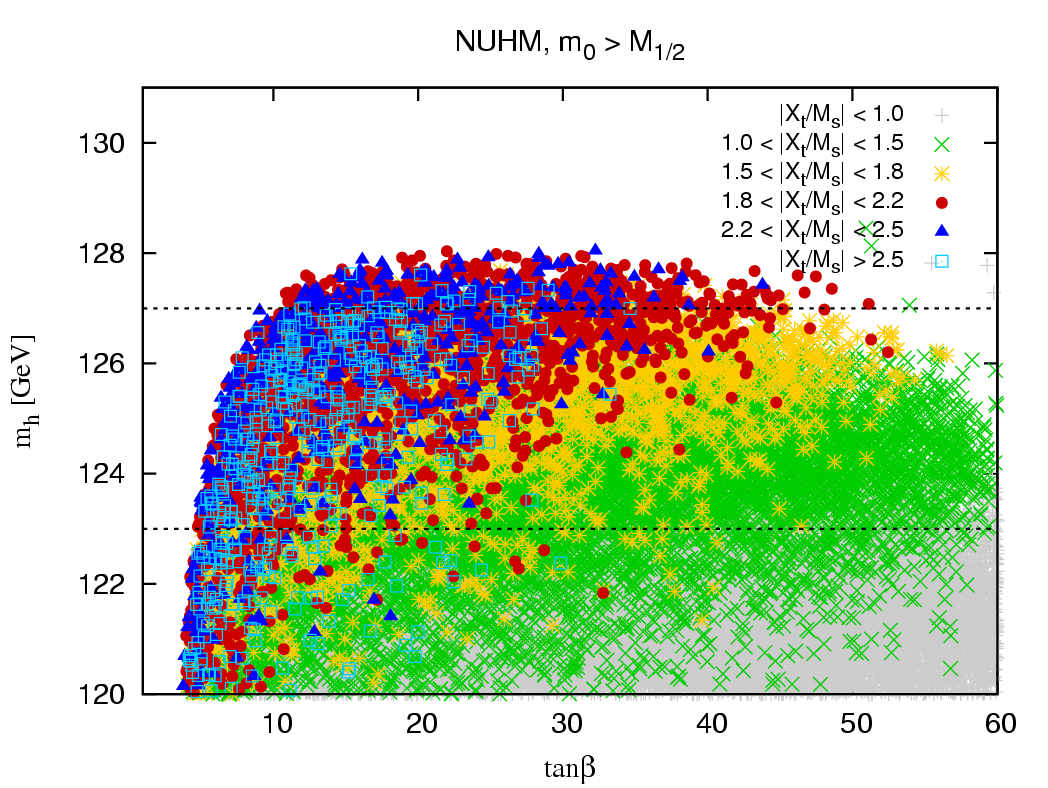}  \\
  \includegraphics[width=75mm]{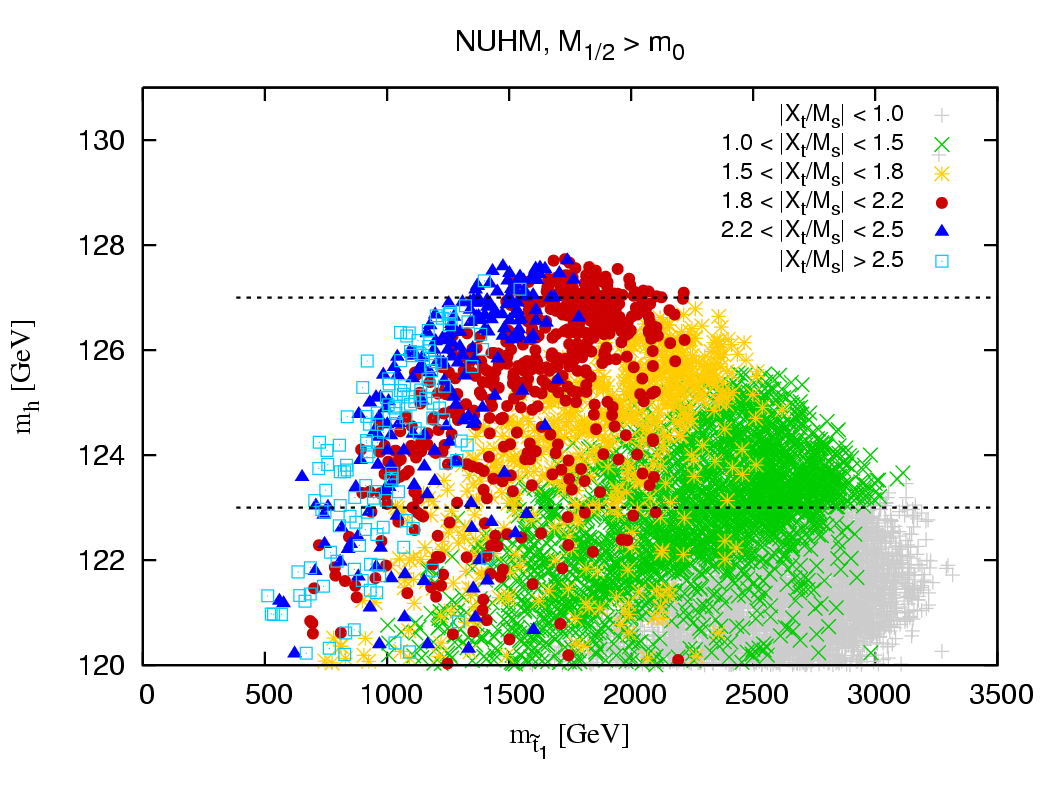} 
  \includegraphics[width=75mm]{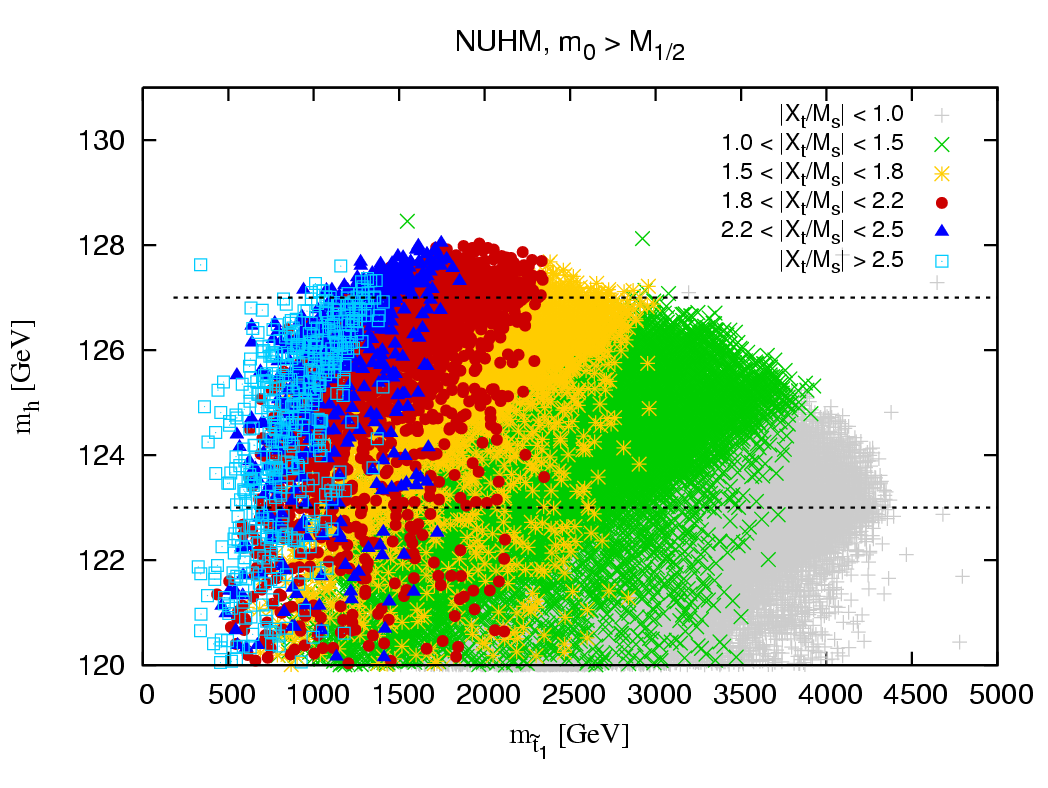}  
  \caption{Dependence of $m_h$ on $\tan\beta$ (top row) and on $m_{\tilde t_1}$ (bottom row) in the NUHM model. 
  As $m_0$ increases, larger $\tan\beta$ values are consistent with maximal mixing. Moreover, 
$\tilde t_1$ masses below 1~TeV can easily be reconciled with large $m_h\sim 125$~GeV if $m_0$ and the stop mixing are large enough. 
  \label{fig:nuhm-mh}}
\end{figure}

Figure~\ref{fig:nuhm-A0-Mhf} shows a projection of the scanned NUHM parameter space, analogous to Fig.~\ref{fig:gm-A0-Mhf} in the gaugino mediation model. As discussed in section~\ref{sec:maxmix}, it makes a difference whether $M_{1/2}$ or $m_0$ is the largest soft mass. In Fig.~\ref{fig:nuhm-A0-Mhf} and following figures, we hence distinguish between the two cases $M_{1/2}>m_0$ and $m_0>M_{1/2}$. 
As expected, maximal mixing requires a ratio between $A_0$ and ${\rm max}(M_{1/2},m_0)$ of about $-1$ to $-3$. 
Moreover, a heavy  ($m_h>123$~GeV) MSSM Higgs requires large mixing with a large negative $A_0$ --- the tip of the scatter plot being again around $A_0\approx-2$~TeV --- or an overall heavy spectrum. A difference to the case with vanishing $m_0$ is that now much larger negative values of $A_0$ still give a valid spectrum. Another interesting difference is that with increasing $m_0$, larger values of $|X_t/M_S|$ become consistent with a heavy $h$. Indeed, for large $m_0$, the lowest $M_S$ giving $m_h=123\mbox{--}127$~GeV is found for $X_t/M_S\approx -3$ to $-3.5$. 
On the other hand, as previously mentioned such large values of $|X_t/M_S|$ give rise to dangerous charge- and color-breaking minima in the scalar potential, so although SOFTSUSY gives a valid spectrum the corresponding points should not be trusted to be phenomenologically viable. 
The correlations between $m_h$, $M_S$ and amount of stop mixing are shown in  Fig.~\ref{fig:nuhm-mixing}. Furthermore, Fig.~\ref{fig:nuhm-mh} shows the dependence of $m_h$ on $\tan\beta$ and $m_{\tilde t_1}$, with $X_t/M_S$ 
again indicated by a color code. We see that at large $m_0$, 
maximal mixing is possible also for large $\tan\beta$. As a side remark we note that the maximal $h$ mass in these scans is about 128~GeV, consistent with the current 95\% CL limit. 

\begin{figure}\centering
  \includegraphics[width=75mm]{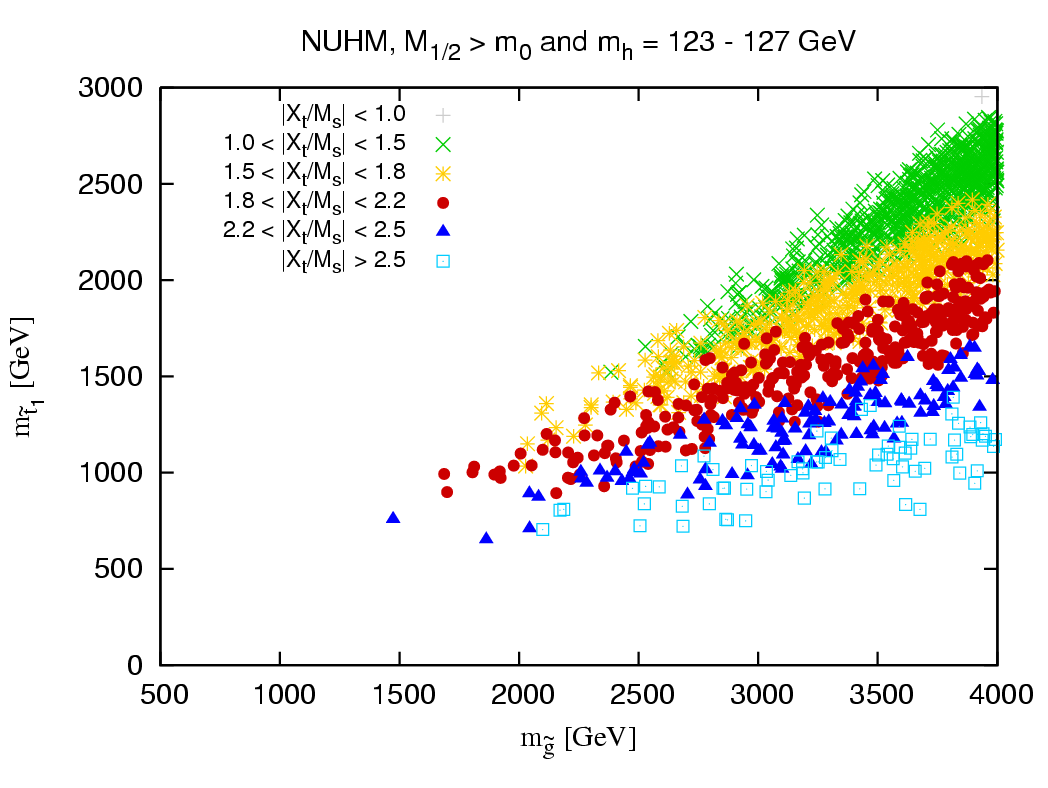} 
  \includegraphics[width=75mm]{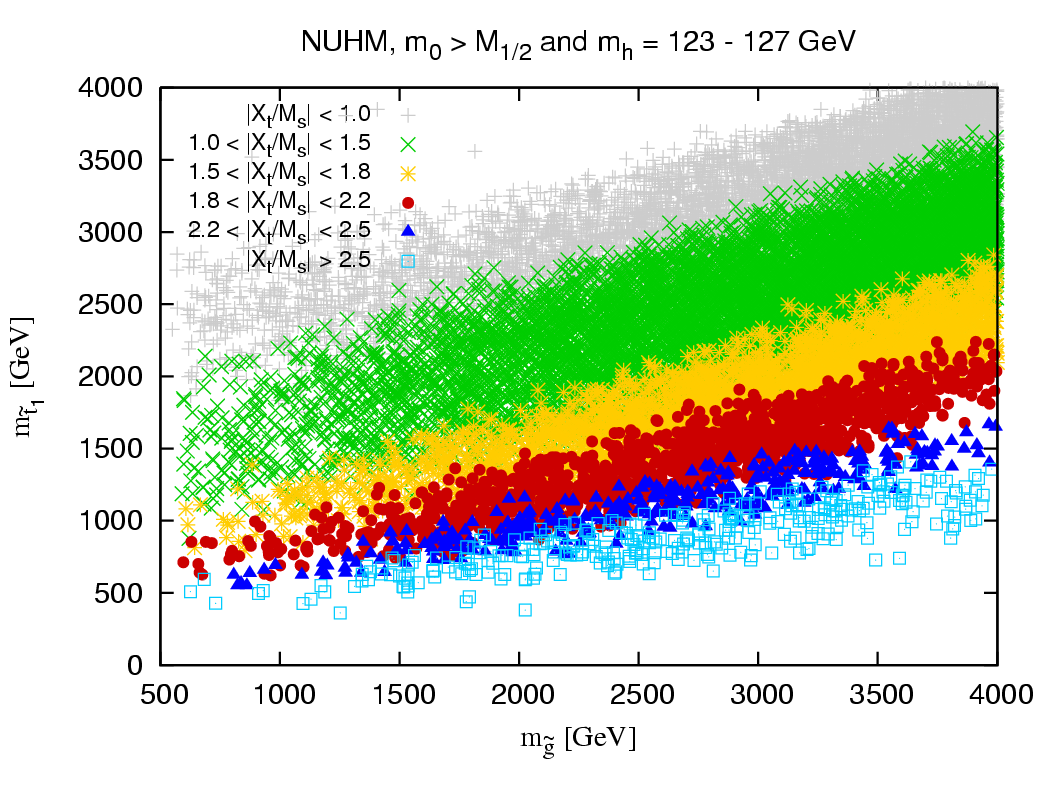}  \\
  \includegraphics[width=75mm]{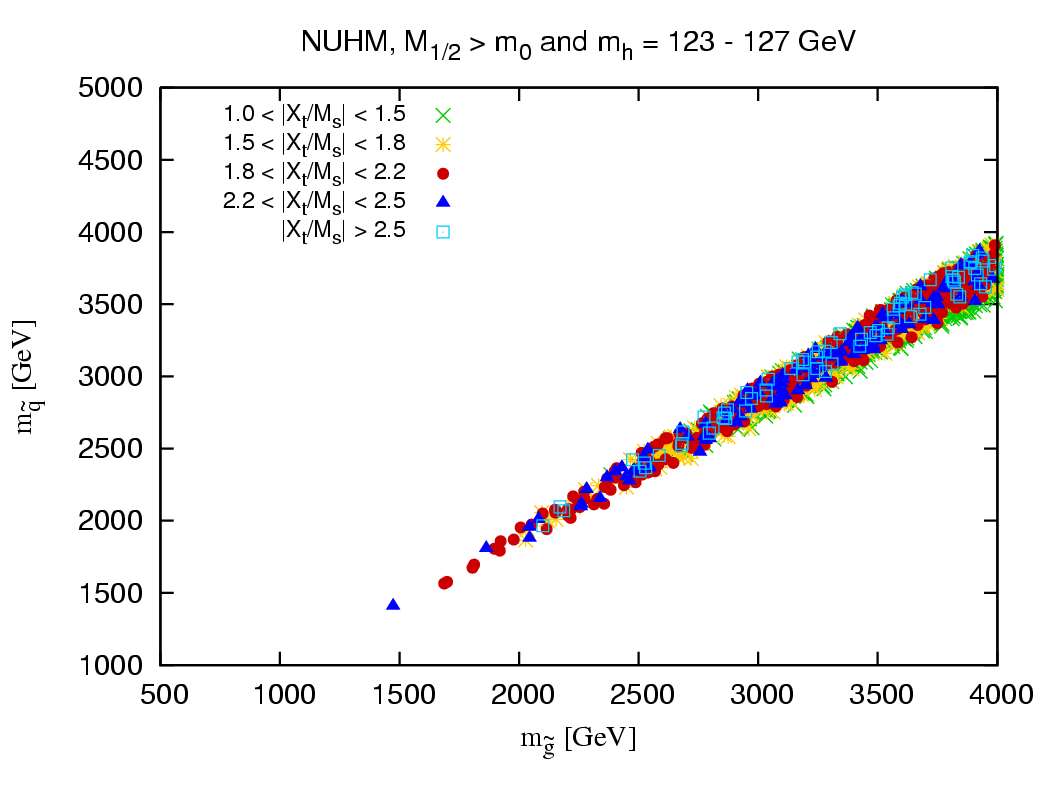} 
  \includegraphics[width=75mm]{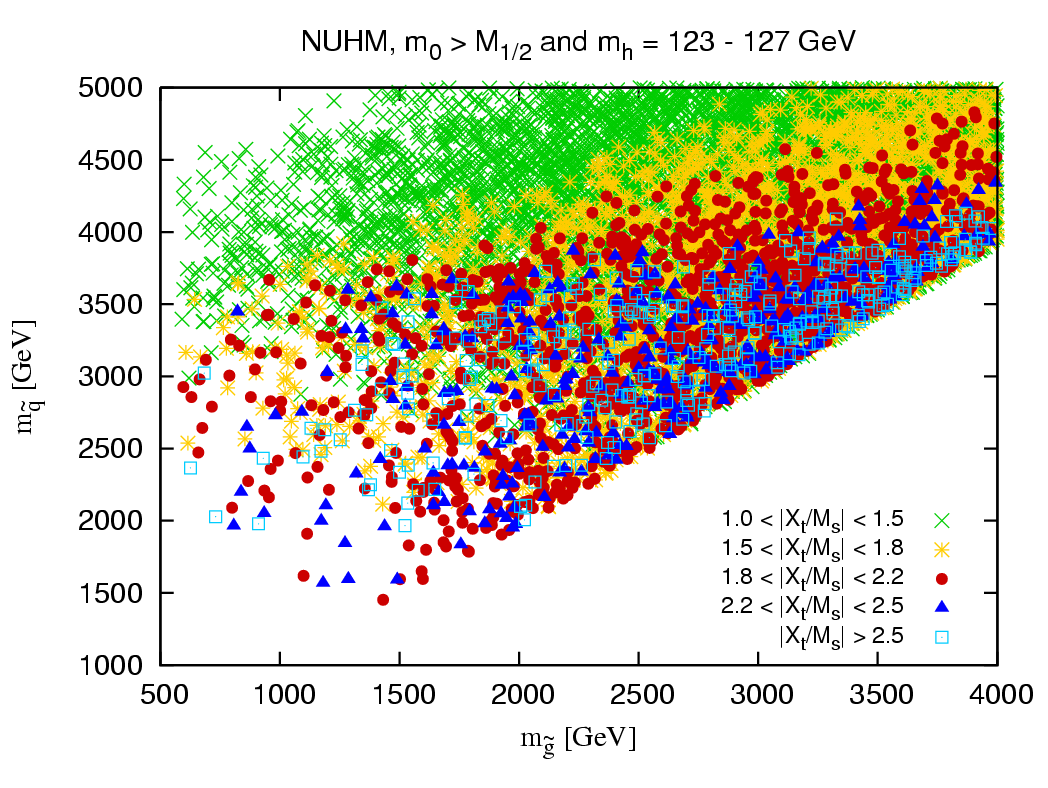} 
  \caption{Scatter plots of points with $m_h=123\mbox{--}127$~GeV in the NUHM model, 
  in the top row ${\tilde t_1}$ versus ${\tilde g}$ masses, in the bottom row 
  ${\tilde q}$ versus ${\tilde g}$ masses. 
  \label{fig:nuhm-mgluino}}
\end{figure}

Consequences for LHC SUSY searches are illustrated in Fig.~\ref{fig:nuhm-mgluino}. 
For $M_{1/2}>m_0$, we still find $m_{\tilde q}\simeq m_{\tilde q}\gtrsim 1.5$~TeV. 
For $m_0>M_{1/2}$, however, gluinos can be as light as 500--600~GeV (with stops being light and maximally mixed, and $m_h$ in the desired range). First and second generation squarks need to be heavy in this case, around 2--3~TeV, as can be seen in the bottom-right panel in Fig.~\ref{fig:nuhm-mgluino}.

\begin{figure}\centering
  \includegraphics[width=75mm]{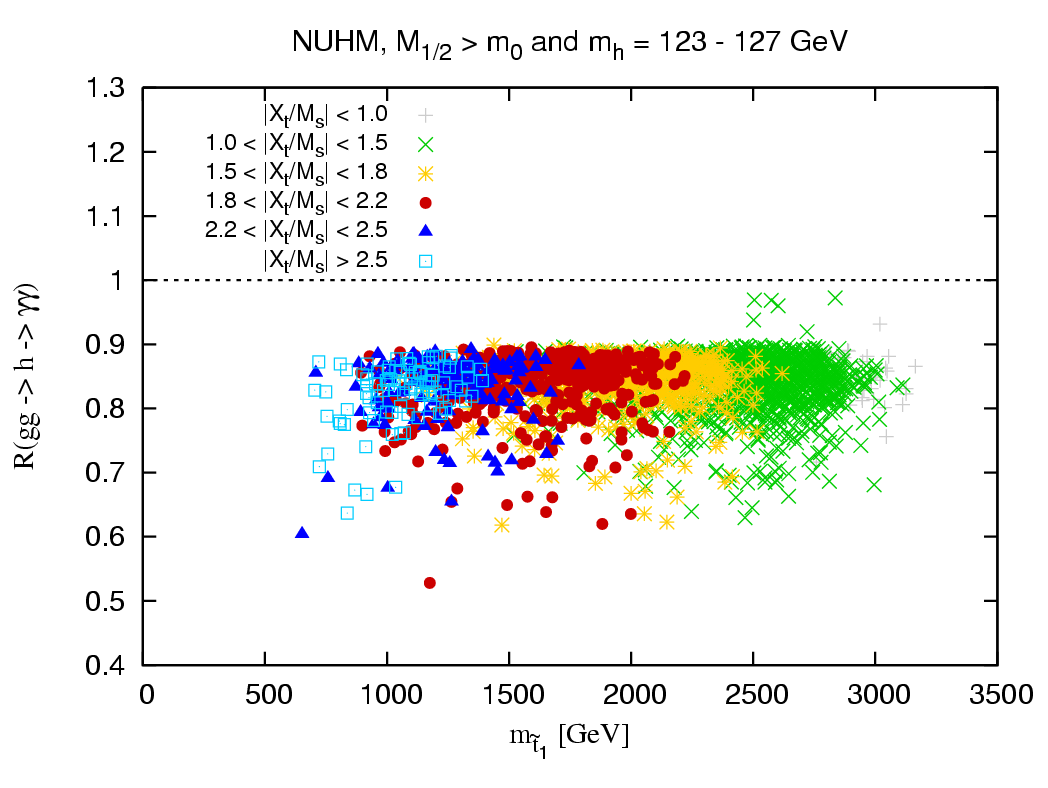} 
  \includegraphics[width=75mm]{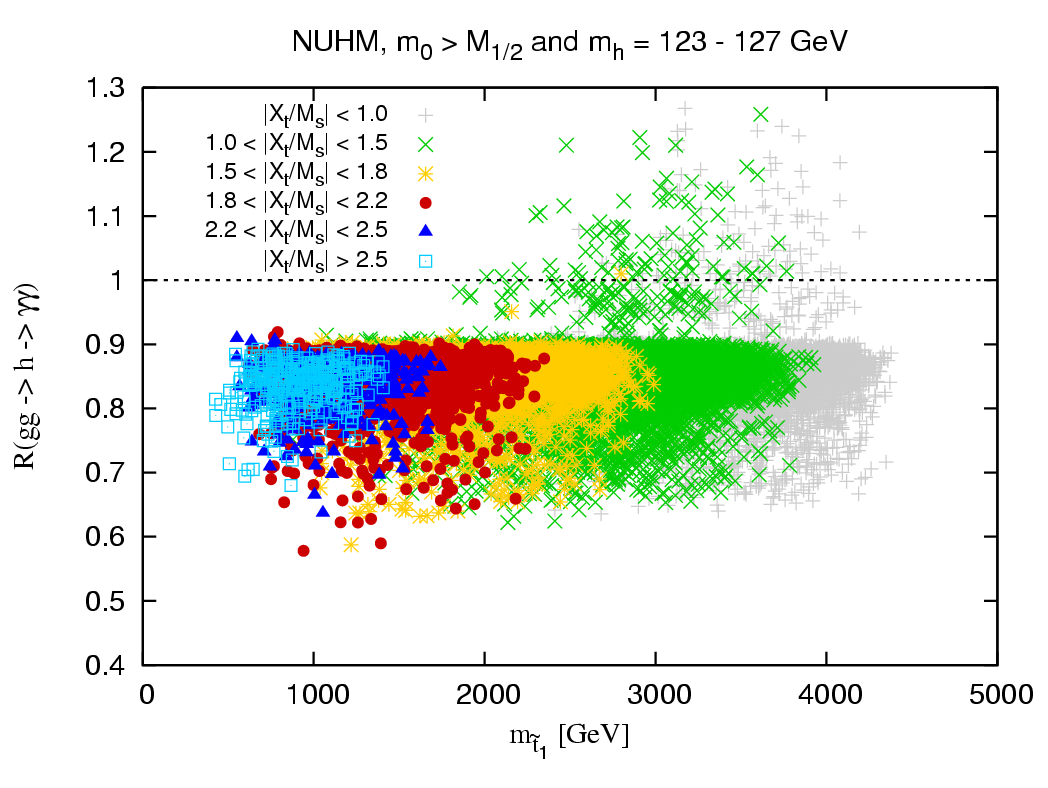} 
  \includegraphics[width=75mm]{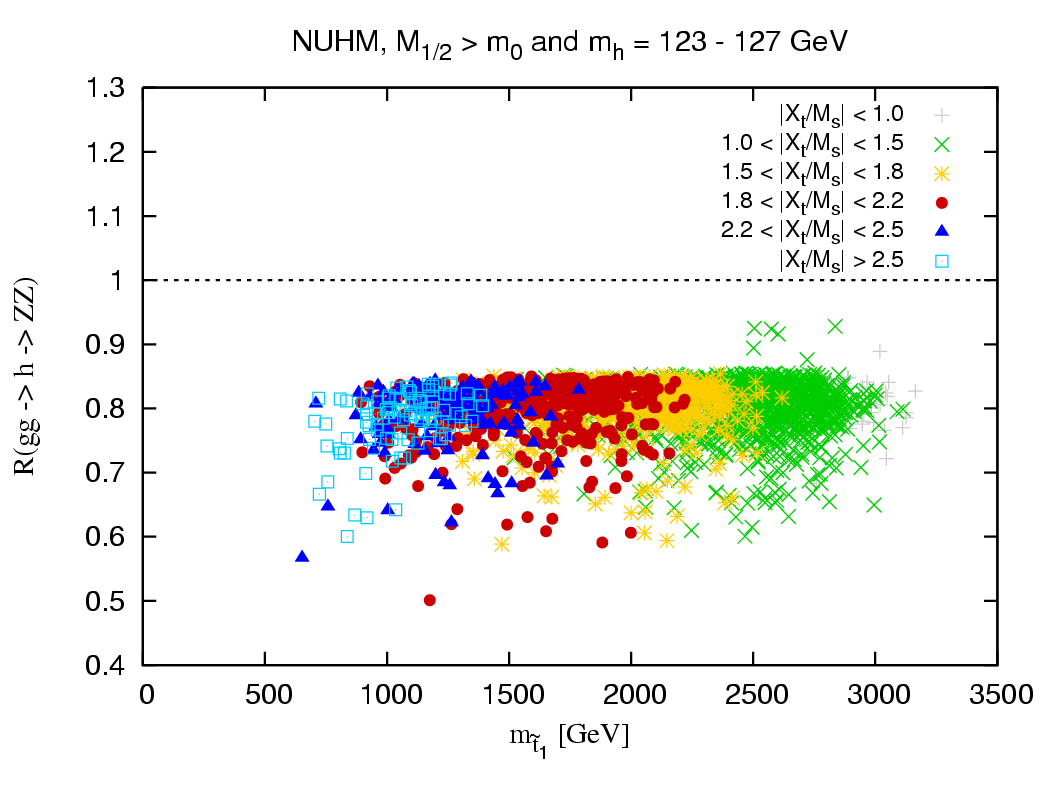}
  \includegraphics[width=75mm]{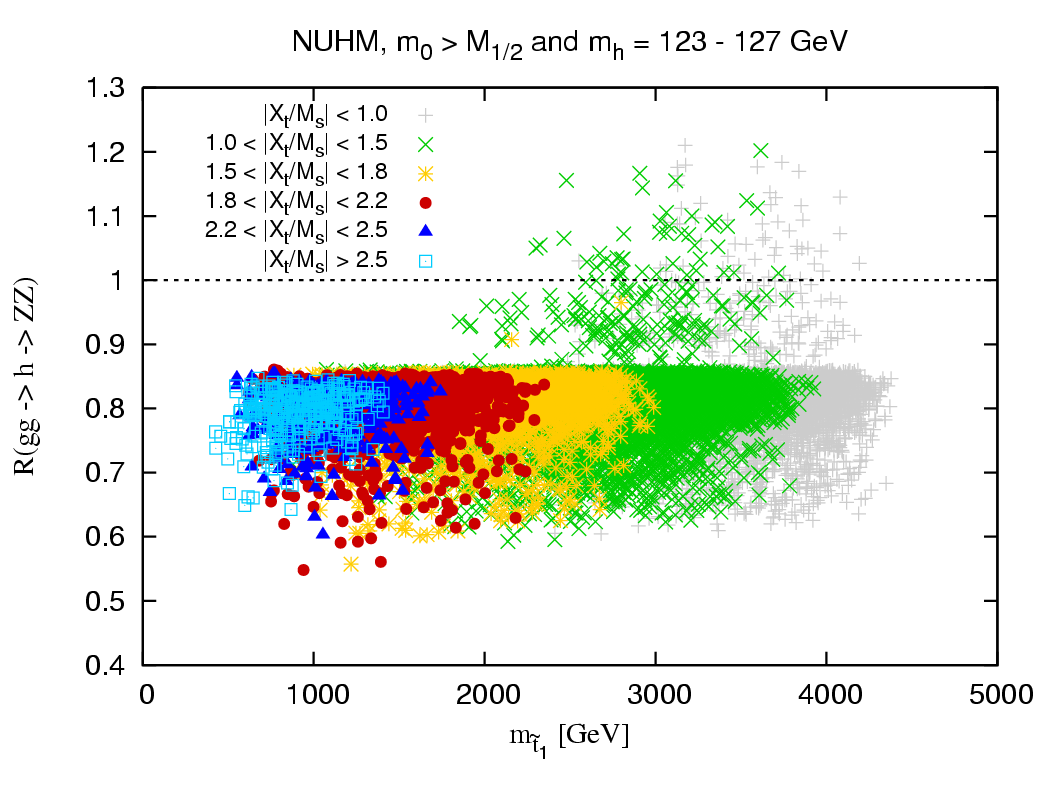}
\caption{In the NUHM model with maximal mixing and $m_h=123\mbox{--}127$~GeV, the $gg\to h\into\gamma\gamma$ and $ZZ$ rates are only about 60--90\% of those in the Standard Model one, mostly as a result of the reduced gluon fusion production cross section. An enhancement ($R>1$) only occurs for small stop mixing and when the $h\to b\bar b$ decay mode is suppressed at large $\tan\beta$ and large $m_A$.
\label{fig:nuhm-hsignal}.  }
\end{figure}

Expectations for the Higgs signal in the $\gamma\gamma$ and $ZZ$ channels are shown in Fig.~\ref{fig:nuhm-hsignal}. We observe that for a Higgs mass in the desired range, 
$R(\gamma\gamma)$ and $R(ZZ)\lesssim 0.9$, rather independently of the stop mass (a decoupling effect of heavy stops can however be seen in the lower boundary of $R$ at a given $X_t/M_S$). Higgs signal strengths close to or above 1 occur for $m_0\gtrsim M_{1/2}$; they require heavy stops with small mixing, combined with large $\tan\beta$ and large $m_A$, such that the $h\to b\bar b$ rate is  suppressed.

\begin{figure}\centering
  \includegraphics[width=75mm]{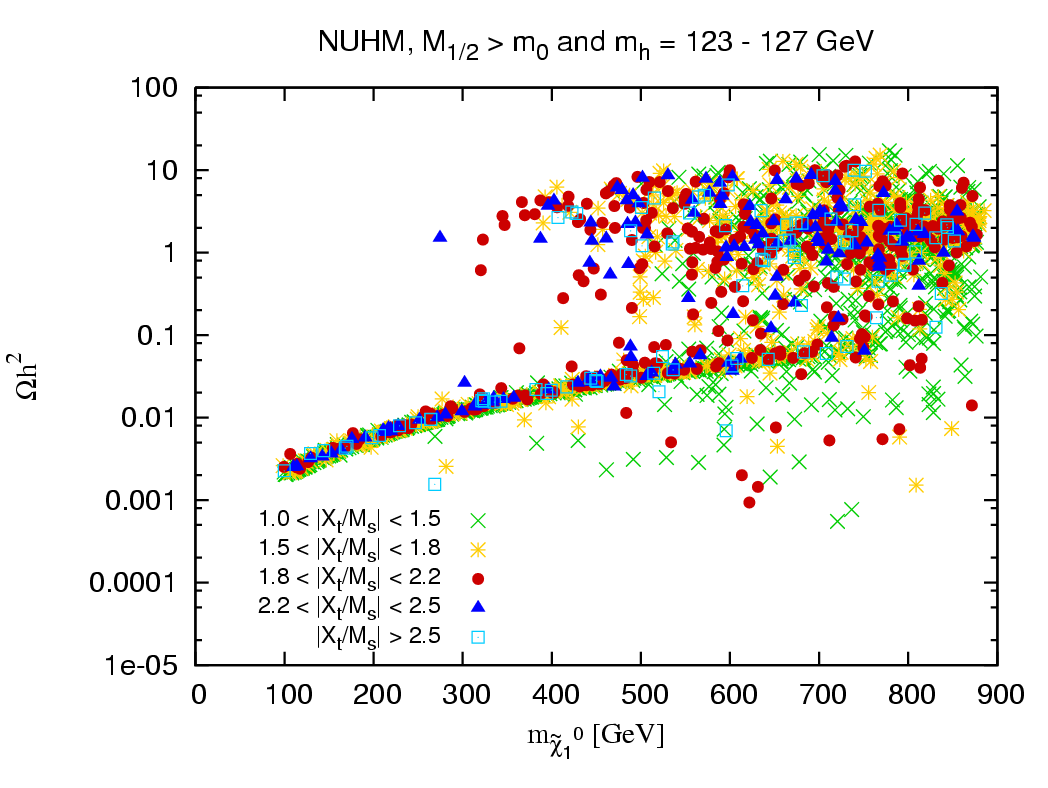} 
  \includegraphics[width=75mm]{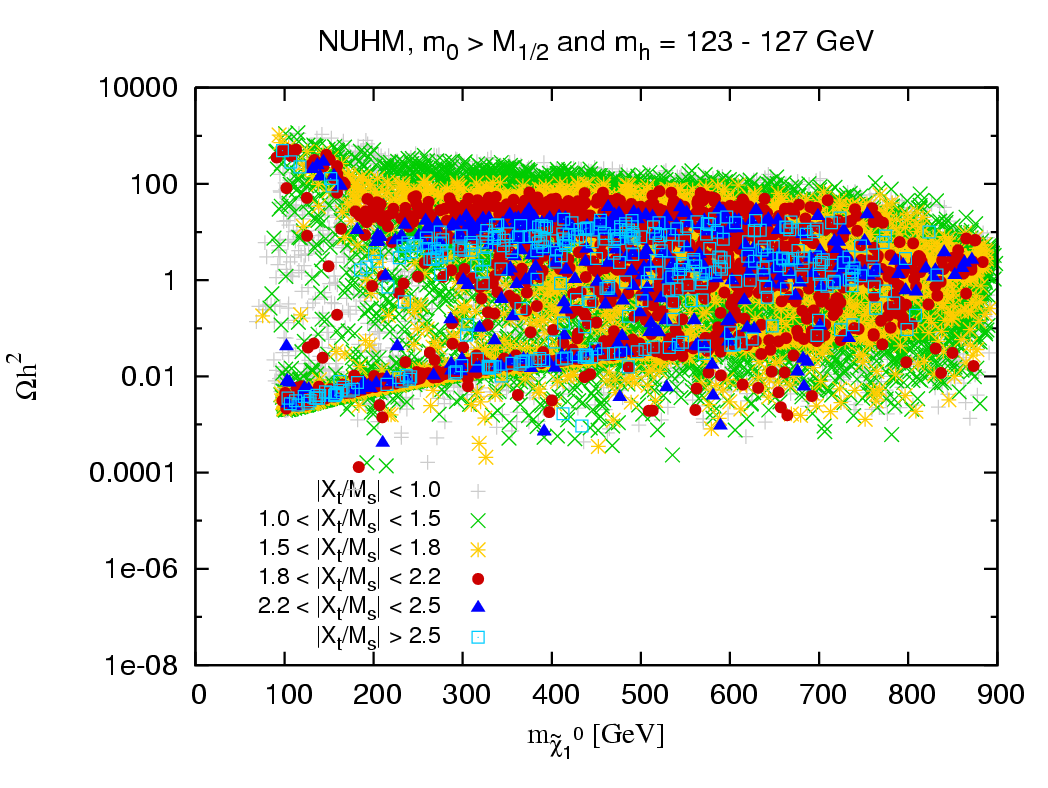}  
  \caption{Dark matter relic density versus neutralino LSP mass in the NUHM model, 
on the left for $M_{1/2}>m_0$, on the right for $m_0>M_{1/2}$. All points have $m_h=123\mbox{--}127$~GeV.
  \label{fig:nuhm-omega}}
\end{figure}

Finally, although it does not directly have to do with maximal mixing, 
let us consider the question of neutralino dark matter. 
The relic density of the neutralino LSP is plotted versus the neutralino mass in Fig.~\ref{fig:nuhm-omega}. Interestingly, for $M_{1/2}>m_0$, a large fraction (45\%) of the points have $\Omega h^2<0.135$, and overall the relic density does not exceed 20. The points with very low $\Omega h^2$ typically feature a higgsino-like LSP, which makes the scenario difficult to detect at the LHC~\cite{Baer:2011ec,Bobrovskyi:2011jj}. In the remaining cases, when $\mu$ is large, $\Omega h^2$ is low because of co-annihilations. For $m_0>M_{1/2}$, the situation is quite different, and we find the ``usual" MSSM picture with $\Omega h^2$ ranging from $10^{-5}$ to $10^{3}$. Roughly 2\% of the points have $0.09<\Omega h^2<0.135$. An example of a ``perfect'' point the sense of light stops, maximal mixing, $m_h=125$~GeV and $\Omega_{\tilde\chi^0_1} h^2=0.1$ is given in Table~\ref{tab:benchmarks}.

\clearpage 
\subsection{Split generations, inverted sfermion-mass hierarchy }  \label{sec:esusy}

So far we have considered only the case that there is no large hierarchy between the three generations of sfermions. An interesting alternative, motivated also by the absence of any signal of new physics in the flavor sector, is the case of an inverted sfermion-mass hierarchy, with squarks and sleptons of the first two generations being very heavy (in the multi-TeV range) while the third generation and the gauginos are light, of the order of 1~TeV. Together with the requirement of small $\mu$, this is often referred to as ``effective SUSY'' or ``natural SUSY'' in the literature. 

\begin{figure}\centering
  \includegraphics[width=75mm]{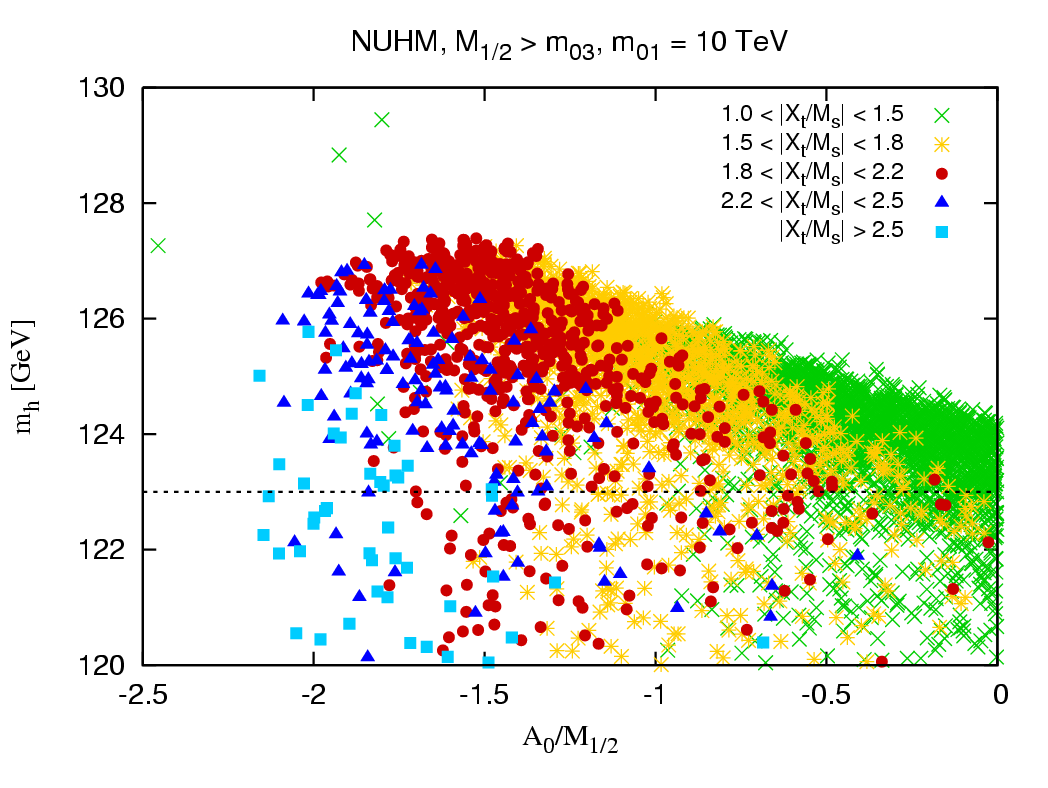} 
  \includegraphics[width=75mm]{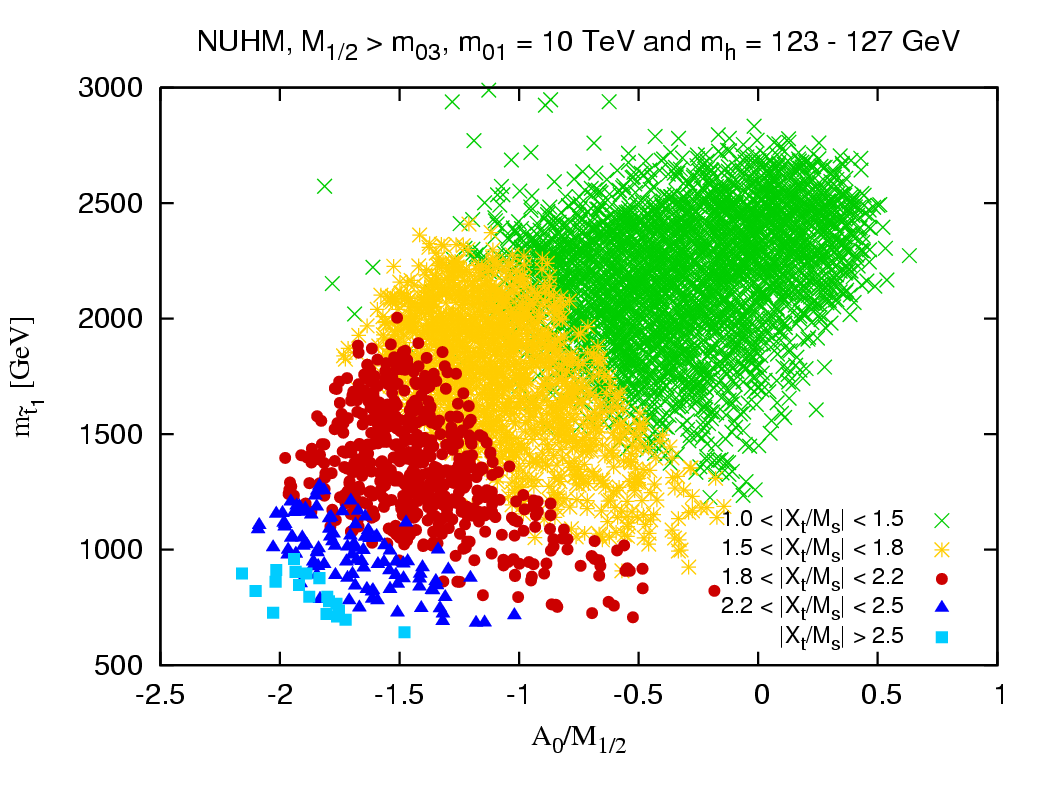}  
  \caption{In the case of very heavy first and second generation sfermions (here $m_{01}=10$~TeV) and light stops, sbottoms and staus ($m_{03}<M_{1/2}<2$~TeV), $M_S$ is driven to smaller values and maximal mixing occurs even for $|A_0|<M_{1/2}$. We find $m_h>123~(124)$~GeV with $m_{\tilde t_1}< 1$~TeV for ratios as low as $A_0/M_{1/2}\approx-0.2~(-0.6)$.
  \label{fig:esusy}}
\end{figure}

To illustrate this case, we perform a scan over the NUHM parameter space as before, but setting the soft masses of the first two generations $m_{01}=10$~TeV. For the third generation, we assume a universal soft mass $m_{03}$,  which we let vary between $0$ and $M_{1/2}$. 
As also shown in \cite{Baer:2012uy2}, in this setup a Higgs near 125 GeV with light stops is possible even for small $|A_0|$. This is because, during RG running, $M_S$ is driven down by the first two generation squarks being very heavy~\cite{ArkaniHamed:1997ab}. Therefore maximal mixing now occurs also at smaller $A_0$, see Fig.~\ref{fig:esusy}. 
Scenarios with maximal mixing, $m_{\tilde t_1}$ below 1~TeV, and a Higgs near 125~GeV can now be found for $A_0/M_{1/2}$ ratios of around $-0.2$ to $-2$.  

\clearpage
\section{Discussion and conclusions}\label{sec:conclusions}

If the MSSM is to accommodate a Higgs boson around 125 GeV, maximal stop mixing is the only way to avoid multi-TeV stop masses. Since the Higgs sector and the top/stop sector are coupled to each other by the large top Yukawa coupling, multi-TeV stop masses generically imply multi-TeV mass parameters in the Higgs potential. A relatively small electroweak scale can then only arise through delicate cancellations, which requires considerable fine-tuning. In the context of phenomenological models which prescribe the MSSM parameters at the TeV scale, it has therefore been argued that, in the most natural remaining regions of the MSSM parameter space, the stop masses should be as small as possible (see, for instance, \cite{Barbieri:2009ev,Papucci:2011wy}). To be compatible with a 125~GeV Higgs, the stops will then have to be maximally mixed. 

However, as we have argued, one of the most appealing aspects of the MSSM is that it can be valid up to very high energies. It is not clear a priori if a given set of TeV-scale MSSM parameters can result from some healthy UV completion at a high scale, or if instead it is a point in the ``swampland''. For example, and of immediate relevance to maximal mixing, it is not possible to obtain $X_t/M_S\approx +2$ from a GUT-scale model (barring hierarchically large GUT-scale trilinear terms), because of the negative gluino contribution to $A_t$ during its RG evolution. Indeed, we have shown in some detail that in models where all GUT-scale soft parameters are of the order of the gaugino mass or smaller, the trilinear coupling must be large and negative at $M_{\rm GUT}$ for maximal mixing to result. In that case one has $X_t/M_S\approx -2$ at the electroweak scale.

In models of high-scale SUSY breaking mediation, the fine-tuning required to obtain a small electroweak scale from a large soft mass scale also involves the gluino mass and the top trilinear. The electroweak scale is very sensitive to the UV-scale $M_3$, since gluino loops strongly affect the stop masses and thus indirectly the Higgs mass parameters and the electroweak scale. This is evident e.g.~from the large coefficient of $\wh M_3^2$ in Eq.~\eqref{eq:mzsq}. The sensitivity of the electroweak scale with respect to $\wh A_t$ is less pronounced. Overall, the least fine-tuned remaining parameter regions of the MSSM are characterized by low $\wh M_3$ (or low $M_{1/2}$ if the gaugino masses are universal), correspondingly low $M_S$, and maximal mixing; see Appendix \ref{sec:FT}. Even in these regions the fine-tuning is at the permille level or worse \cite{Hall:2011aa,Ghilencea:2012gz}.

Maximal mixing does not single out any particular scenario for SUSY breaking mediation; it follows from a parameter choice in models where the UV-scale trilinear soft terms are free parameters. It would be highly interesting to identify models which actually predict large and negative $A$-terms. Among the classes of models we studied, only gauge mediation and to some extent radion mediation predict the trilinear terms, and in these cases the prediction disfavors maximal mixing.

In models where maximal mixing is allowed, it becomes an interesting question what its phenomenological consequences are. We have studied three examples, with particular attention to the parameter regions which give a Higgs boson around 125 GeV. In each case all experimental constraints can be satisfied for suitable parameter choices. 
Squarks and gluinos typically turn out to be heavy, beyond the reach of the 2011 LHC run at $\sqrt{s}=7$ TeV. They may however be within reach of the $\sqrt{s}=8$ TeV run in favorable cases. 
The $gg\to h\to\gamma\gamma$ signal strength consistently turns out to be some $10$--$40\%$ below that of the Standard Model. 

Table~\ref{tab:benchmarks} lists the properties of four representative points,
two for gaugino mediation and two for the NUHM case, with stops below 1~TeV and maximal-mixing. 
GM-1 is the point with lowest $M_S$, while GM-2 is the point with lowest $M_S$ and $\mu<500$~GeV from the gaugino mediation scan. Both these points have a stau as the lightest sparticle with a relic abundance of the order of $10^{-2}$.\footnote{For a gravitino $\tilde G$ or axino $\tilde a$ LSP, this needs to be rescaled by a factor $m_{\tilde G}/m_{\tilde\tau_1}$ or $m_{\tilde a}/m_{\tilde\tau_1}$. The resulting value together with the $\tilde\tau_1$ lifetime (which depends on the nature of the LSP) is relevant for BBN constraints.}  
NUHM-1 is a low-stop-mass point from the NUHM scan with $M_{1/2}>m_0$. It features a neutralino LSP with large higgsino component and a relic density which is too low, so that the $\tilde\chi^0_1$ would provide only about 20--30\% of the dark matter.
NUHM-2 has a large $m_0$ of order 2~TeV and $A_0\approx -2m_0$. 
It exactly matches the desired Higgs mass (125~GeV) and has a bino-like neutralino LSP with a relic density of $\Omega h^2=0.1$.
The SLHA files of these maximal mixing benchmark points are included as ancillary files in this preprint. 

\begin{table}[p]\centering
\begin{tabular}{l|c|c|c|c}
  & GM-1 & GM-2 & NUHM-1 & NUHM-2 \\
\hline\hline
  $m_0$ & 0 & 0 & 759 & 2295 \\
  $M_{1/2}$ & $772$ & $974$ & $833$ & $736$ \\
  $A_{0}$   & $-2254$ & $-2099$ & $-2336$ & $-4140$ \\
  $\tan\beta$ & $15.3$ & $10.1$ & $14.7$ & $19.4$ \\
  $\mu$ & $1343$ & $273$ & $329$ & $1049$ \\
  $m_A$ & $1981$ & $1640$ & $330$ & $678$ \\
\hline
  $m_h$ & $123$ & $123.4$ & $123.6$ & $125.2$\\
\hline
  $m_{\tilde\chi^0_1}$ & 326 & 263 & 302 & 319\\
  $m_{\tilde\chi^0_2}$ & 620 & 282 & 338 & 611\\
  $m_{\tilde\chi^\pm_1}$ & 620 & 274 & 327 & 611\\
  $m_{\tilde\tau_1}$  & 221 & 116 & 604 & 2064 \\
  $m_{\tilde e_R}$  & 425 & 255 & 685 & 2210 \\
  $m_{\tilde e_L}$  & 469 & 675 & 992 & 2387 \\
  $m_{\tilde q}$  & 1532 & 1894 & 1808 & 2681\\
  $m_{\tilde g}$  & 1712 & 2120 & 1863 & 1749 \\
  $m_{\tilde t_1}$  & 730 & 821 & 652 & 862 \\
  $m_{\tilde t_2}$  & 1306 & 1549 & 1396 & 1894\\
  $X_t/M_S$  & $-2.32$ & $-2.13$  & $-2.36$ & $-2.25$\\
\hline
  $\wh m_{H_u}$ & 1498 & 1521 & $-273$ & 566 \\
  $\wh m_{H_d}$ &  585 & 1934 & 1991 & 3030\\
\hline
  BR$_{b\to s\gamma}$ [$10^{-4}$] & $2.94$ & $2.88$ & $3.16$ & $3.18$  \\
  BR$_{B\to \mu\mu}$  [$10^{-9}$] & $3.14$ & $3.07$ & $4.60$ & $4.26$\\
  $\Delta a_\mu$ [$10^{-10}$] & $6.87$ & $2.64$ & $3.47$ & $1.07$\\ 
\hline
  $\Omega h^2$(`LSP') & $8.5\times10^{-3}$ & $6.5\times10^{-3}$ 
                  & $0.026$ & $0.101$\\
  $\sigma_{\rm SI}\,(\tilde\chi^0_1p)$ [pb]& -- & -- & $2.9\times10^{-7}$ & $3.7\times10^{-10}$\\
  $\sigma_{\rm SD}\,(\tilde\chi^0_1p)$ [pb]& -- & -- & $1.5\times10^{-4}$ & $1.4\times10^{-7}$\\
\hline\hline
\end{tabular}
\caption{Sample points with light stops, maximal mixing, and a Higgs near 125 GeV in the gaugino mediation (GM-1, GM-2) and NUHM (NUHM-1, NUHM-2) models.  
$\Omega h^2$(`LSP') is the relic abundance of the $\tilde\tau_1$ for the GM points, and of the $\tilde\chi^0_1$ for the NUHM points. 
$\sigma_{\rm SI}$ and $\sigma_{\rm SD}$ are the spin-independent and spin-dependent scattering cross sections off protons; for NUHM-1 these should be rescaled by a factor $\xi=\Omega h^2/0.1123$ for comparison with the experimental limits 
($\xi\sigma_{\rm SI}=6.7\times10^{-8}$ and $\xi\sigma_{\rm SD}=3.5\times10^{-5}$ for NUHM-1).
\label{tab:benchmarks}}
\end{table}

\section*{Acknowledgments}
This work originated from discussions at the workshop on ``Implications of a 125 GeV Higgs boson'' held at LPSC Grenoble 30 Jan -- 2 Feb 2012. Discussions with the other workshop participants, in particular with G.~B\'elanger and F.~Mahmoudi, as well as financial support from the Centre de Physique Th\'eorique de Grenoble (CPTG), are gratefully acknowledged.  
We also thank H.~Baer, J.~J\"ackel, S.~Pokorski, and especially W.~Buchm\"uller for interesting discussions, and B.~Allanach and M.~Dolan for correspondence on SOFTSUSY.

\clearpage 
\appendix
\section*{Appendix}

\section{Semi-numerical solutions of the MSSM renormalization group equations}\label{sec:RGEs}

The one-loop RGES of the MSSM can be integrated analytically when keeping only the top Yukawa coupling nonzero \cite{Ibanez:1983di}.\footnote{See e.g.~\cite{Carena:1996km, Essig:2007kh} for reviews and applications.} The procedure can be summarized as follows: As a first step, one writes down the closed-form solutions for the gauge coupling and gaugino mass RGEs, which are of course easily found and well known. The top Yukawa RGE is a Bernoulli equation whose solution, given the solutions for the gauge couplings, can be expressed as a simple integral. Then the $A_t$ RGE becomes a linear ODE with known inhomogeneous term and known coefficient functions, and is easily solved in terms of an integrating factor. Finally, having solved all of these RGEs, the coupled RGEs for third-generation scalar masses and for $m_{H_u}^2$ can be integrated in a suitable basis.

In this paper we are using a somewhat refined approach to improve precision. We keep all third-generation Yukawa couplings, and we are using two-loop RGEs for the gauge couplings, Yukawa couplings, and gaugino masses. Boundary values for the gauge and Yukawa couplings are matched to SOFTSUSY GUT-scale values, in order to properly take threshold corrections into account. With the more complicated coupled two-loop system to solve, the solutions can no longer be expressed by simple integrals, but this is unnecessary to extract the information we need. We will now briefly describe our method.

As a first step, we fix some value of $\tan\beta$ and some soft mass scale $M_S$. Using appropriate boundary values for the gauge and Yukawa couplings, we solve their RGEs numerically between $M_{\rm GUT}$ and $M_S$. What we are eventually interested are however the SUSY breaking mass parameters. Quite generally, from the structure of the RGEs and from dimensional analysis, it follows that their values at $M_S$ take the form
\be
\begin{split}
 M_a(M_S)&=\sum_b \alpha_{ab}\,\widehat M_b+\sum_x\alpha'_{ax}\wh A_x\,,\\
 A_x(M_S)&=\sum_{a}\beta_{xa}\widehat M_a+\sum_y\beta'_{xy}\wh A_y\,,\\
 m_\phi^2(M_S)&=\sum_{ab} \gamma_{\phi ab} \wh M_a\wh M_b+\sum_{xy}\gamma'_{\phi xy}\wh A_x\wh A_y+\sum_{xa}\gamma''_{\phi xa}\wh A_x\wh M_a+\sum_{\chi}\gamma'''_{\phi\chi}\wh m_\chi^2\,. 
\end{split}
\ee 
Here hatted quantities denote boundary values at $M_{\rm GUT}$ as in the main text, and the $\alpha,\beta,\gamma$ coefficients are functions of $\tan\beta$ and of $M_S$. They are obtained by numerically solving the RGEs with special boundary conditions. For instance, setting all GUT-scale masses except $\wh M_1$ to zero allows to read off the $\alpha_{a1}$, $\beta_{x1}$, and $\gamma_{\phi 11}$ coefficients from the numerical solution, and similarly for the others. The same method is also applied to the $\mu$ and $B_\mu$ RGEs.

The coefficients we find roughly agree with the one-loop running, tree-level matching value often found in the literature, with the exception of $\wh M_3$ whose influence on the low-scale soft terms we find to be somewhat reduced. Compare for instance our result for $m_Z$ at $\tan\beta=10$ and $M_S=m_Z$,
\be\label{eq:mzsq}
m_Z^2=4.5\,\wh M_3^2-0.4\,\wh M_2^2+0.3\,\wh M_2\wh M_3-0.6\,\wh A_t\wh M_3- 1.3\,\wh m_{H_{u}}^2 + 0.7\, \wh m_{Q_3}^2+ 0.8\,\wh m_{U_3}^2-2.0\,|\mu|^2+\ldots
\ee
with that of \cite{Abe:2007kf}, Eqns.~(7) and (11)
\be
m_Z^2=5.5\,\wh M_3^2-0.4\,\wh M_2^2+0.5\,\wh M_2\wh M_3-0.8\,\wh A_t\wh M_3- 1.3\,\wh m_{H_{u}}^2 + 0.7\, \wh m_{Q_3}^2+ 0.7\,\wh m_{U_3}^2-2.2\,|\mu|^2+\ldots
\ee
or that of \cite{Essig:2007kh}, Eq.~(7)
\be
m_Z^2=5.2\,\wh M_3^2-0.4\,\wh M_2^2+0.5\,\wh M_2\wh M_3-0.8\,\wh A_t\wh M_3- 1.3\,\wh m_{H_{u}}^2 + 0.7\, \wh m_{Q_3}^2+ 0.7\,\wh m_{U_3}^2-2.2\,|\mu|^2+\ldots
\ee
On the other hand, we agree within $10\%$ with the two-loop result for $m_{H_u}^2(1$ TeV$)$ of \cite{Younkin:2012ui}, Eq.~(2.1).

\section{Fine-tuning}\label{sec:FT}

To find the least fine-tuned regions of parameter space, we quote again Eq.~\eqref{mzsq}, valid for $M_S=1$ TeV and $\tan\beta=20$, 
\be\begin{split}\label{mzsq2}
m_Z^2=0.2\,A_0^2 - 0.7\, A_0  M_{1/2} + 2.9\, M_{1/2}^2 - 2.1\,|\wh\mu|^2 - 1.3\,\wh m_{H_{u}}^2 + 0.7\, \wh m_{Q_3}^2+ 0.8\,\wh m_{U_3}^2+\ldots
\end{split}
\ee
A common measure of fine-tuning \cite{Barbieri:1987fn} is derived from the logarithmic sensitivity of the electroweak scale with respect to parameter variations,
\be
C_{a}=\frac{\partial\log m_Z}{\partial\log a}\,,
\ee
where $a\in\{ M_{1/2}, A_0,\wh m_{H_u},\wh m_{H_d},\wh m_{Q_3},\wh m_{U_3},\wh m_{D_3},\wh \mu,\ldots\}$ runs over all independent dimensionful GUT-scale parameters. The fine-tuning is then estimated as
\be
\frac{1}{\text{fine-tuning}}=\max_a\,C_a\,.
\ee
The worst offender in high-scale mediation scenarios is generically the gaugino mass. This remains true even if $|A_0|$ is large enough to allow for maximal mixing: From Eq.~\eqref{mzsq} we find
\be\begin{split}
C_{ M_{1/2}}&=2.9\,\frac{ M_{1/2}^2}{m_Z^2}-0.35\,\frac{ A_0  M_{1/2}}{m_Z^2}\,,\\
C_{ A_0}&=0.2\,\frac{ A_{0}^2}{m_Z^2}-0.35\,\frac{ A_0  M_{1/2}}{m_Z^2}\,.
\end{split}
\ee
For $| A_0|\,<\,3\, M_{1/2}$, $C_{M_{1/2}}$ dominates. Furthermore, from the fine-tuning point of view, it is favorable to go to maximal mixing in order to raise the Higgs mass, rather than to raise the overall soft mass scale. For the least fine-tuned regions compatible with a Higgs mass $m_{h^0}>123$ GeV, with $M_{1/2}/m_Z\approx 10$ and maximal mixing, we find a fine-tuning measure of around a few permille. It is clear that the LHC Higgs mass results, when firmly established, will substantially raise the fine-tuning price of the MSSM.

\section{Charge- and color-breaking minima}\label{sec:CCB}

We briefly review a criterion for charge- and color-breaking minima \cite{Frere:1983ag,Derendinger:1983bz,Kounnas:1983td} in the context which is relevant for us. Consider the direction $|\widetilde U_3|=|\widetilde Q_3|=|h_u|\equiv X$ in field space, with all other VEVs vanishing. The potential energy along this direction reads
\be
V(X)=3\,y_t^2\,X^4+2\,A_t\,y_t\,X^3(m_{U_3}^2+m_{Q_3}^2+m_{H_u}^2+|\mu|^2)\,X^2+D\text{-terms}\,.
\ee 
It is minimized at
\be
X_0=-\frac{A_t}{4\,y_t}\left(1\pm\sqrt{1-\frac{8(m_{U_3}^2+m_{Q_3}^2+m_{H_u}^2+|\mu|^2)}{3\,A_t^2}}\right)\,,
\ee
and is negative at $X_0$ if $A_t$ satisfies
\be\label{Atbound}
A_t^2>3(m_{U_3}^2+m_{Q_3}^2+m_{H_u}^2+|\mu|^2)\,.
\ee 
For $m_Z\ll M_S$, the RHS of this inequality is $3(m_{U_3}^2+m_{Q_3}^2+m_{H_u}^2+|\mu|^2)\approx 6\,M_S^2$. The potential energy of the realistic electroweak vacuum is
\be
V_0=-\frac{v^4}{8}\left({g'}^2+g_2^2\right)\cos^2 2\beta\,\gtrsim -(87\text{ GeV})^4\,.
\ee
If $V(X_0)<V_0$, there is a charge- and color-breaking vacuum. It is easily checked that, for soft terms in the TeV range, the domain $V(X_0)<V_0$ is reached quickly once $A_t$ starts exceeding the bound Eq.~\eqref{Atbound}: The electroweak vacuum is quite shallow, compared to the CCB vacuum whose scale is set by the soft terms. We therefore conclude that points with  
\be
|X_t/M_S|\gtrsim\sqrt{6}
\ee 
lead to charge and color breaking.

Note that this is only a sufficient criterion, and that generally stronger constraints can be found by exploring other directions in field space \cite{Casas:1995pd}.  Furthermore, we are neglecting $D$-terms and loop corrections. Finally, we have evaluated all running quantities at the scale $M_S$ here. Demanding the absence of CCB vacua at other scales may, again, lead to more restrictive bounds \cite{Casas:1995pd}. On the other hand, a CCB vacuum may be viable phenomenologically if the lifetime of our false vacuum is long on cosmological timescales.

\clearpage

\end{document}